\documentclass[a4paper,11pt]{article}
\pdfoutput=1

\usepackage{jheppub}

\usepackage{graphicx}
\usepackage{amsmath}
\usepackage{amssymb}
\usepackage{xspace}
\usepackage[small]{subfigure}
\usepackage{slashed}
\usepackage{mdwlist}

\newcommand{\eq}[1]{eq.~\eqref{eq:#1}}
\newcommand{\eqs}[2]{eqs.~\eqref{eq:#1} and \eqref{eq:#2}}
\renewcommand{\sec}[1]{sec.~\ref{sec:#1}}

\newcommand{\subsec}[1]{sec.~\ref{subsec:#1}}
\newcommand{\subsecs}[2]{secs.~\ref{subsec:#1} and \ref{subsec:#2}}
\newcommand{\app}[1]{app.~\ref{app:#1}}
\newcommand{\fig}[1]{fig.~\ref{fig:#1}}

\newcommand{\abs}[1]{\lvert#1\rvert}
\newcommand{\Abs}[1]{\bigl\lvert#1\bigr\rvert}
\newcommand{\ABS}[1]{\biggl\lvert#1\biggr\rvert}
\newcommand{\ord}[1]{\mathcal{O}(#1)}
\newcommand{\ORd}[1]{\mathcal{O}\Bigl(#1\Bigr)}
\newcommand{\ORD}[1]{\mathcal{O}\biggl(#1\biggr)}

\newcommand{\mae}[3]{\langle#1\rvert#2\rvert#3\rangle}
\newcommand{\Mae}[3]{\bigl\langle#1\bigr\rvert#2\bigr\rvert#3\bigr\rangle}
\newcommand{\MAe}[3]{\Bigl\langle#1\Bigr\rvert#2\Bigr\rvert#3\Bigr\rangle}

\newcommand{\ket}[1]{\lvert#1\rangle}

\newcommand{\braket}[2]{\langle#1\rvert#2\rangle}

\newcommand{\df}{\mathrm{d}}

\newcommand{\img}{\mathrm{i}}
\newcommand{\MS}{\overline{\rm MS}}

\newcommand{\Li}{\textrm{Li}}

\newcommand{\bfT}{{\bf T}} 

\newcommand{\sdt}{\!\cdot\!}
\newcommand{\tr}{\textrm{tr}}

\newcommand{\lra}{\leftrightarrow}

\newcommand{\al}{\alpha}
\newcommand{\bt}{\beta}
\newcommand{\ga}{\gamma}
\newcommand{\Ga}{\Gamma}
\newcommand{\de}{\delta}
\newcommand{\eps}{\epsilon}

\newcommand{\la}{\lambda}
\newcommand{\w}{\omega}

\newcommand{\cA}{{\mathcal A}}
\newcommand{\cB}{{\mathcal B}}

\newcommand{\cL}{{\mathcal L}}

\newcommand{\Tau}{{\mathcal T}}

\newcommand{\bn}{\bar{n}}

\newcommand{\hga}{\widehat{\gamma}}

\newcommand{\nn}{\nonumber}

\newcommand{\QCD}{{\rm{QCD}}}

\newcommand{\alem}{\alpha_\mathrm{em}}

\newcommand{\Ecm}{E_\mathrm{cm}}

\newcommand{\hard}{\mathrm{hard}}
\newcommand{\nons}{\mathrm{nons}}
\newcommand{\FO}{\mathrm{FO}}
\newcommand{\cusp}{\mathrm{cusp}}

\newcommand{\zero}{{(0)}}
\newcommand{\one}{{(1)}}

\newcommand{\SCET}{\ensuremath{{\rm SCET}}\xspace}
\newcommand{\SCETa}{\ensuremath{{\rm SCET}_{\rm I}}\xspace}
\newcommand{\SCETb}{\ensuremath{{\rm SCET}_{\rm II}}\xspace}
\newcommand{\SCETp}{\ensuremath{{\rm SCET}_+}\xspace}

\allowdisplaybreaks[2]

\title{Factorization and Resummation for Generic Hierarchies between Jets}

\author[a]{Piotr Pietrulewicz,}
\author[a]{Frank J.~Tackmann,}
\author[b,c]{and Wouter J.~Waalewijn}

\affiliation[a]{Theory Group, Deutsches Elektronen-Synchrotron (DESY), D-22607 Hamburg, Germany}
\affiliation[b]{ITFA, University of Amsterdam, Science Park 904, 1018 XE, Amsterdam, The Netherlands}
\affiliation[c]{Nikhef, Theory Group, Science Park 105, 1098 XG, Amsterdam, The Netherlands}

\emailAdd{piotr.pietrulewicz@desy.de}
\emailAdd{frank.tackmann@desy.de}
\emailAdd{w.j.waalewijn@uva.nl}

\abstract{
Jets are an important probe to identify the hard interaction of interest at the LHC.
They are routinely used in Standard Model precision measurements as well as in searches for new heavy particles,
including jet substructure methods.
In processes with several jets, one typically encounters hierarchies in the jet transverse momenta
and/or dijet invariant masses. Large logarithms of the ratios of these
kinematic jet scales in the cross section are at present primarily described by parton showers.
We present a general factorization framework called \SCETp, which is an extension of Soft-Collinear Effective Theory (SCET)
and allows for a systematic higher-order resummation of such kinematic
logarithms for generic jet hierarchies. In \SCETp additional intermediate soft/collinear modes are used
to resolve jets arising from additional soft and/or collinear QCD emissions.
The resulting factorized cross sections utilize collinear splitting amplitudes and soft gluon currents and
fully capture spin and color correlations. We discuss how to systematically combine the different kinematic regimes
to obtain a complete description of the jet phase space.
To present its application in a simple context, we use the case of $e^+e^- \to 3$ jets.
We then discuss in detail the application to $N$-jet processes at hadron colliders, considering
representative classes of hierarchies from which the general case can be built.
This includes in particular multiple hierarchies that are either strongly ordered in angle or energy or not.
}

\preprint{
\begin{flushright}
DESY 15-129\\
NIKHEF 2015-024\\
January 19, 2016
\end{flushright}
}

\begin{document}

\maketitle

\section{Introduction}
\label{sec:intro}

A thorough understanding of the production of hadronic jets is crucial to take full advantage of the data from high-energy colliders. Jet processes typically involve hierarchies between the short-distance scale of the hard scattering (e.g.~the jet energies or invariant masses between jets) and the scale at which the individual jets are resolved (e.g.~the mass or angular size of a jet), leading to logarithms of the ratio of these scales in the perturbative expansion of the cross section. An accurate description of these effects is obtained by resumming the dominant logarithmic corrections to all orders in perturbation theory.

In multijet events one generically encounters additional hierarchies in the hard kinematics of the jets, namely among the jet energies and/or among the angles between jets.
At the LHC, an important class of examples are jet substructure methods to reconstruct boosted heavy objects, which essentially rely on identifying soft or collinear (sub)jets.
Another example is cascade decays of heavy new (colored) particles leading to experimental signatures with jets of widely different $p_T$. There are also cases where additional jets produced by QCD are used to tag or categorize the signal events, a prominent example being the current Higgs measurements.
Whenever such kinematic hierarchies arise among QCD-induced jets, in particular in the corresponding background processes, the enhancement of soft and collinear emissions in QCD leads to additional logarithms of the jet kinematics in the cross section. 
So far, a complete and general factorization framework for multijet processes that allows for a systematic resummation of such kinematic logarithms for generic jet hierarchies has been missing. Current predictions therefore rely on Monte Carlo parton showers and are thus mostly limited to leading logarithmic accuracy.

In this paper we develop such a factorization and resummation framework for generic jet hierarchies in hard-scattering processes with large momentum transfer by considering an extension of Soft-Collinear Effective Theory (SCET)~\cite{Bauer:2000ew, Bauer:2000yr, Bauer:2001ct, Bauer:2001yt, Bauer:2002nz, Beneke:2002ph} referred to as \SCETp. Compared to the usual soft and collinear modes in SCET, \SCETp contains additional intermediate modes that behave as soft modes (with eikonal coupling) with respect to the standard collinear modes but at the same time behave as collinear modes with respect to the overall soft modes. Their precise scaling, which is now simultaneously soft and collinear, depends on the considered measurement or observable (in analogy to how the scaling of the modes in SCET is determined by the considered observable).

In SCET individual hard QCD emissions are resolved as jets, 
while the effects of soft and collinear emissions on observables are each resolved at a single scale.
The intermediate modes in \SCETp are required to further resolve the additional scales induced by measurements or hierarchies which are not separated in SCET.%
\footnote{%
We stress that this does not imply that SCET describes such effects incorrectly. It does correctly contain these effects at each fixed order but it is not sufficient for resumming the associated additional logarithms. In fact, we will match onto SCET in the limit where the additional hierarchies disappear and the corresponding logarithms are not enhanced. This is precisely analogous to the relation between SCET and fixed-order QCD for the logarithms resummed by SCET.}
The case we discuss in detail in this paper is the explicit measurement of soft or collinear (sub)jets.
Here, also individual soft or collinear emissions are explicitly resolved, and \SCETp allows us to capture their effects on observables.

Generically, there are two types of intermediate \SCETp modes that appear which can be distinguished by their origin as follows
\begin{itemize}
   \item Collinear-soft (csoft) modes arise as soft offspring from a collinear sector of a parent SCET.
   \item Soft-collinear modes arise as collinear offspring from a soft sector of a parent SCET.
\end{itemize}
This distinction is helpful, as it automatically determines the correct Wilson-line structure and interactions of the modes with respect to the other modes present in the final \SCETp. Both types of modes can be present at different scales and in different directions. There can also be cases where the two types become degenerate.

\SCETp first appeared in ref.~\cite{Bauer:2011uc}, where its purely collinear regime described by csoft modes was constructed and used to describe the situation of two energetic jets collinear to each other.
In ref.~\cite{Procura:2014cba}, \SCETp was used to describe the situation where two resolution variables are measured simultaneously, requiring csoft modes separated from the collinear modes in either virtuality or rapidity depending on the measurements.
The purely soft regime of \SCETp involving soft-collinear modes was first considered in ref.~\cite{Larkoski:2015zka}. There it was shown that this regime is essential for the resummation of nonglobal logarithms by explicitly resolving additional soft subjets (see also ref.~\cite{Neill:2015nya}).
In ref.~\cite{Larkoski:2015kga}, the soft and collinear regimes were used to factorize and resum a two-prong jet substructure variable (defined in terms of energy-correlation functions~\cite{Larkoski:2014gra}). They also discussed a way to treat the overlap between the two regimes by removing the double counting at the level of the factorized cross section.
More recently, a \SCETp setup was applied in refs.~\cite{Becher:2015hka, Chien:2015cka} for the factorization of both global and nonglobal logarithms appearing in jet rates (see e.g.~refs.~\cite{Seymour:1997kj, Cheung:2009sg, Ellis:2010rwa, Banfi:2010pa, Kelley:2012kj, vonManteuffel:2013vja}).

\begin{figure}
  \begin{center}
  \subfigure[$m_J^2 \ll s_{12} \sim s_{13} \sim s_{23}\sim Q^2$]{\epsfig{file=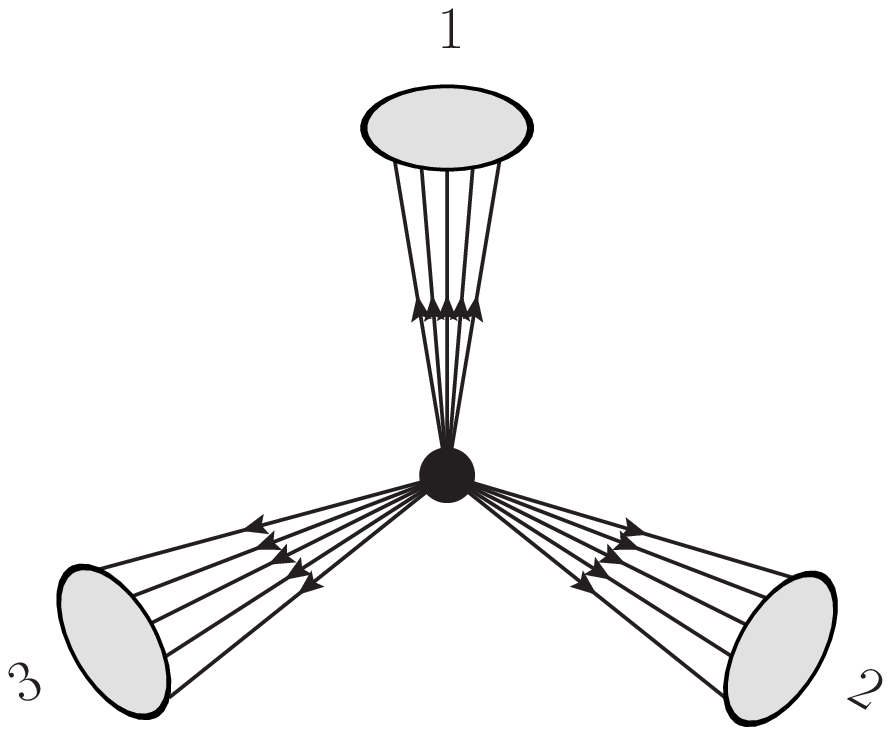,width=0.35\linewidth,clip=}}
  \subfigure[$m_J^2 \ll s_{12} \ll s_{13} \sim s_{23}\sim Q^2$]{\epsfig{file=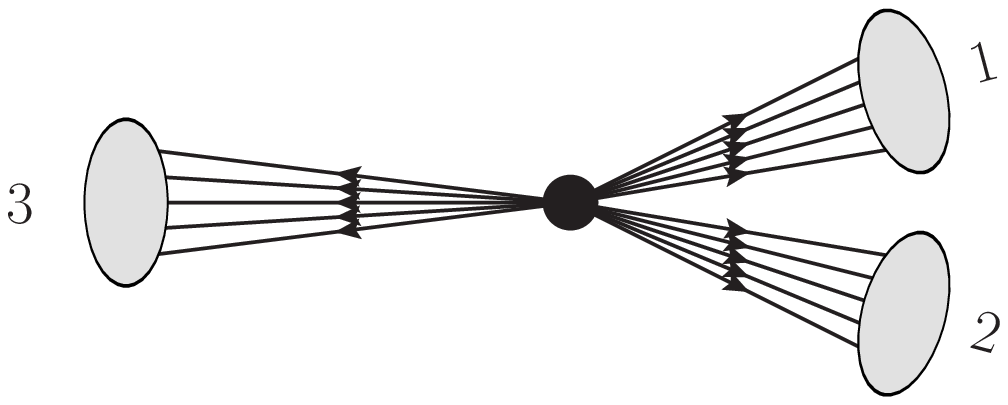,width=0.35\linewidth,clip=}}
  \subfigure[$m_J^2 \ll s_{12} \sim s_{13} \ll s_{23}\sim Q^2$]{\epsfig{file=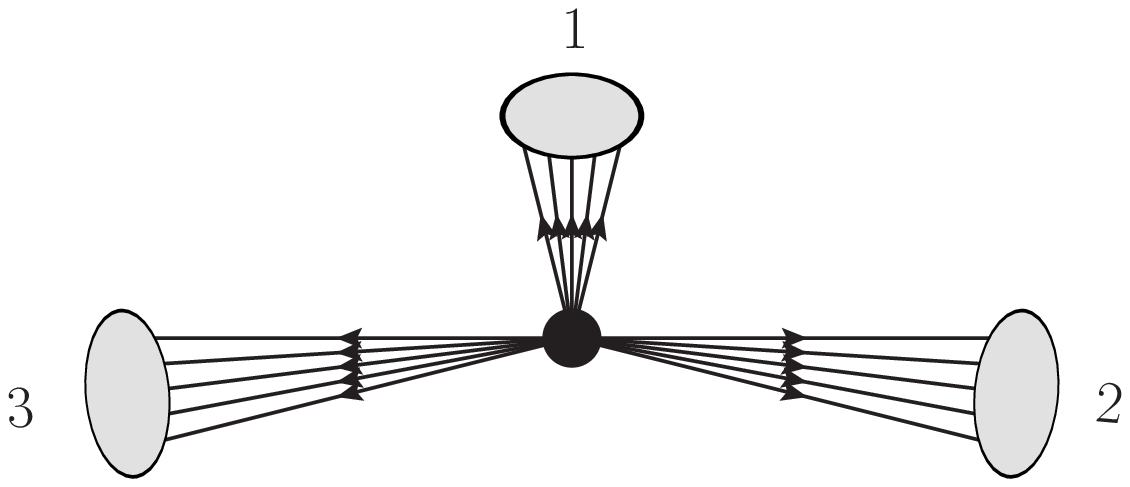,width=0.35 \linewidth,clip=}}
  \subfigure[$m_J^2 \ll s_{12} \ll s_{13} \ll s_{23}\sim Q^2$]{\epsfig{file=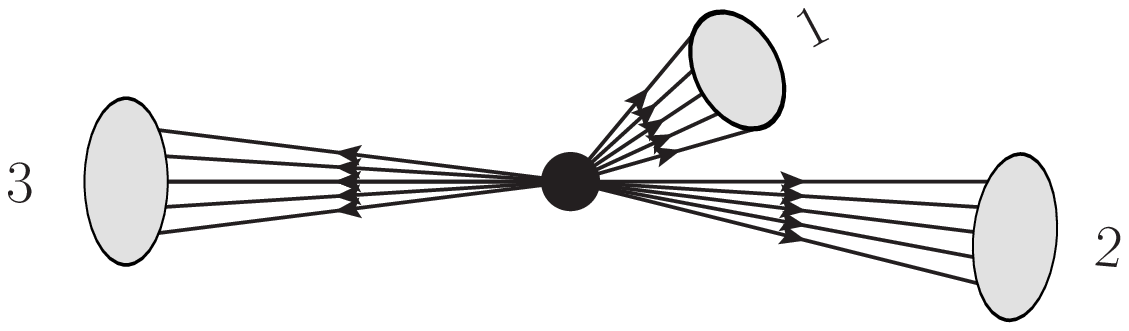,width=0.35 \linewidth,clip=}}
  \caption{Different hierarchies for three-jet events in $e^+e^-$ collisions.}
  \label{fig:3jet_configs}
  \end{center}
\end{figure}

In this paper, we give a general description of \SCETp for generic jet hierarchies. We first focus on the case of a single hierarchy. We review the purely collinear regime, following ref.~\cite{Bauer:2011uc}, which we will label as $c+$. Furthermore, we present in detail the purely soft regime (labeled $s+$) as well as the overlap between the collinear and soft regimes (labeled $cs+$), involving both csoft and soft-collinear modes.
The corresponding kinematic hierarchies for $e^+ e^- \to 3$ jets are illustrated in Fig.~\ref{fig:3jet_configs}.
Standard SCET applies to case (a) where the jets are parametrically equally hard and well separated, $s_{ij} \sim Q^2$, where $s_{ij}$ are the dijet invariant masses and $Q$ the total center-of-mass energy.
The collinear regime is shown in case (b), where two jets (labelled 1 and 2) are close to each other. It is characterized by the hierarchy $s_{12} \ll s_{13} \sim s_{23}\sim Q^2$. The soft regime is shown in case (c), where one jet (labelled 1) is less energetic than the others. It is characterized by the hierarchy $s_{12} \sim s_{13} \ll s_{23}\sim Q^2$. Finally, in the soft/collinear overlap regime, shown in case (d), one jet is softer than the others and at the same time closer to one of the hard jets, leading to the hierarchy $s_{12} \ll s_{13} \ll s_{23}\sim Q^2$.

In general, \SCETp can have multiple soft and collinear regimes (along with the corresponding overlap regimes), which is necessary to describe multiple hierarchies between several jets. We discuss in detail the application of the \SCETp formalism for a generic $N$-jet process at hadron colliders and for a number of different hierarchies. The cases we explicitly consider include
\begin{itemize*}
  \item One soft jet.
  \item Two jets collinear to each other, with or without a hierarchy in their energies.
  \item Two jets collinear to each other plus an additional soft jet.
  \item Two soft jets with or without a hierarchy in their energies.
  \item Two soft jets collinear to each other.
  \item Three jets collinear to each other with or without a hierarchy in the angles between them.
\end{itemize*}
These cases contain the nontrivial features and essential building blocks that are needed to describe arbitrary hierarchies.
In particular, we show how spin and color correlations are captured in the associated factorization theorems.

Each regime requires a different mode setup in \SCETp, so technically corresponds to a different effective field theory. We explain how they are appropriately combined and matched to the corresponding SCET in the nonhierarchical limit. This yields a complete description of the jet phase space that accounts for all possible kinematic hierarchies.

We will consider an exclusive $N$-jet cross section and require that the $N$ jets can always be distinguished from each other by imposing the parametric relation $m_J^2 \ll s_{ij}$. We assume that the corresponding jet resolution variable(s) that enforce this constraint do not exhibit any hierarchies among themselves, such that there are no parametrically large nonglobal logarithms from soft emissions.
For definiteness and simplicity, we consider $N$-jettiness~\cite{Stewart:2010tn} as our overall $N$-jet resolution variable. In refs.~\cite{Stewart:2015waa,Thaler:2015xaa}, it was shown that $N$-jettiness can be promoted into an exclusive cone jet algorithm, and with a suitable choice of $N$-jettiness measure the resulting jets are practically identical to anti-$k_T$ jets. We stress though that the general setup for the treatment of kinematic hierarchies is largely independent of the specific choice of jet resolution variable and jet algorithm. 
In the application to jet substructure the setup can get more complicated when subjets get sensitive to the jet boundary~\cite{Larkoski:2015zka,Larkoski:2015kga}.
For earlier analytic work on jet hierarchies in $e^+ e^- \to $ jets see Ref.~\cite{Catani:1992tm}.

The remainder of this paper is organized as follows: In \sec{EFT} we describe the basic \SCETp regimes ($c+$, $s+$, and $cs+$) and the structure of the resulting factorization theorems for $e^+ e^- \to N$ jets that resum the corresponding kinematic logarithms. In \sec{3jets}, we present a detailed discussion with explicit perturbative results for the case of $e^+ e^- \rightarrow 3$ jets, which is simple enough that the single hierarchies shown in \fig{3jet_configs} are sufficient to exhaust all kinematic limits.
We subsequently discuss step-by-step the generalizations required to treat a generic LHC process $pp \rightarrow N$ jets plus additional nonhadronic final states. Specifically, collinear initial-state radiation, spin and color correlations for a single kinematic hierarchy are addressed in \sec{njets}. In \sec{multihierarchy} we discuss the various cases with multiple hierarchies outlined above. We conclude in \sec{conclusions}.

\section{Overview of the effective field theory setup}\label{sec:EFT}

In this section, we discuss the general factorization framework for each regime of \SCETp, considering for simplicity $e^+e^-\to N$ jets.
We start in \subsec{standard} with reviewing the standard case without additional hierarchies, which also serves to establish our notation. The purely collinear, purely soft, and soft/collinear regimes are discussed in secs.~\ref{subsec:c+overview}, \ref{subsec:s+overview}, and \ref{subsec:cs+overview}. For now we only consider kinematic configurations with one hierarchy. The general case will be discussed in \sec{multihierarchy} in the context of $pp\to N$ jets. In \subsec{combineEFT} we show how to combine the resulting factorization theorems from the different kinematic regions. We first explicitly consider a \SCETa jet resolution observable, and we outline the modifications required for a \SCETb measurement in \subsec{scetii}.

\subsection{Standard SCET: equally separated and energetic jets}
\label{subsec:standard}

We first discuss the hard kinematics for processes with jets. The total momentum $P_i^\mu$ of the $i$th jet is given by
\begin{equation}
P_i^\mu = q_i^\mu + k_i^\mu
\,,\qquad
q_i^\mu = \w_i\,\frac{n_i^\mu}{2}
\,,\qquad
n_i^\mu = (1, \hat n_i)
\,.\end{equation}
Here, the massless reference (label) momentum $q_i^\mu$ contains the large component of the jet momentum. That is, $\w_i/2 = P_i^0 + \ord{P_i^2/P_i^0}$ corresponds to the jet energy and we take the unit vector $\hat n_i = \vec{P_i}/\abs{\vec{P_i}}$ to point along the direction of the jet. The residual momentum $k_i^\mu = P_i^\mu - q_i^\mu$ then only has small components of $\ord{P_i^2/P_i^0}$.

\begin{table}
\centering
\begin{tabular}{c|c|c}
\hline\hline
mode & $p^\mu=(+,-,\perp)$ & $p^2$ \\
\hline
collinear ($n_1,\dots,n_N$) &  $\bigl(\Tau_N,Q,\sqrt{\Tau_N Q} \bigr)$ &$ \Tau_N Q \sim m_J^2$ \\
ultrasoft & $ \bigl(\Tau_N, \Tau_N, \Tau_N \bigr)$ &$\Tau_N^2 \sim m_J^4/Q^2$\\
\hline\hline
\end{tabular}
\caption{Scaling of the modes in standard SCET for $N$ equally separated and energetic jets.}
\label{tab:scet_modes}
\end{table}

To describe the degrees of freedom of the effective field theory, it is convenient to use lightcone coordinates,
\begin{equation}
p^\mu
= \bn_i\sdt p\, \frac{n_i^\mu}{2} + n_i \sdt p\, \frac{\bn_i^\mu}{2} + p^\mu_{\perp i}
\equiv (n_i \sdt p,\bar{n}_i \sdt p,\vec p_{\perp i}) \equiv (p^+,p^-,\vec p_\perp)_i
\,,\end{equation}
where $\bar n_i^\mu = (1, - \hat n_i)$, and $p^\mu_{\perp i}$ contains the components perpendicular to $n_i^\mu$ and $\bn_i^\mu$. The subscript $i$ will be dropped if it is obvious which lightcone coordinates we are referring to.

For definiteness, we consider $N$-jettiness~\cite{Stewart:2010tn} as the \SCETa jet resolution observable, defined as
\begin{align} \label{eq:tauN}
\Tau_N &= \sum_k \min\limits_{i} \Big\{\frac{2 q_i \sdt p_k}{Q_i}\Big\}
= \sum_k \min\limits_{i} \Big\{\frac{n_i \sdt p_k}{\rho_i}\Big\}
= \sum_i \Tau^{(i)}_N
\,.\end{align}
We use a geometric measure with $Q_i = \rho_i \w_i$, where the parameter $\rho_i$ controls the size of the $i$th jet region and can in principle depend on the hard jet kinematics. It roughly corresponds to the typical jet radius $\rho_i \sim R_i^2$ and we consider it as $\rho_i\sim1$. The minimization assigns particles to the jet they are closest to, and we denote the contribution to $\Tau_N$ from the $i$th jet region by $\Tau^{(i)}_N$. Note that $Q_i \Tau^{(i)}_N$ is equal to the jet invariant mass $P_i^2$ up to power corrections.

The SCET description applies in the exclusive $N$-jet limit where all jets are sufficiently narrow and there are no additional jets from additional hard emissions. This limit corresponds to taking $\Tau_N \ll Q$. Formally, we work at leading order in the power expansion in $\lambda^2 \equiv \Tau_N/Q \sim m_J^2/Q^2$, where we use $m_J$ to denote the typical (average) jet mass. Due to the singular structure of QCD, jets typically have masses much smaller than their energy. Hence, in practice most of the events naturally have $m_J \ll Q$.

We stress that our discussion of the kinematic jet hierachies largely decouples from the precise choice of $\Tau_N$, and in principle any jet resolution observable which constrains $m_J$ (more precisely, any \SCETa-type variable) can be utilized. Furthermore, the precise jet algorithm that is used to find the actual jet momenta $P_i$, which then determine the $q_i$, is also not relevant to our discussion. One option is to promote \eq{tauN} itself to a jet algorithm by further minimizing the value of $\Tau_N$ over all possible jet directions $n_i$~\cite{Stewart:2010tn}. This is the basis of the recently introduced XCone jet algorithm~\cite{Stewart:2015waa, Thaler:2015xaa}. Any other jet algorithm that yields the same jet directions $n_i$ up to power corrections can be used, which includes the usual $k_T$-type clustering algorithms.

We denote the large pairwise invariant mass between two jets with
\begin{equation}
s_{ij}= 2 q_i \cdot q_j = \w_i \w_j\, \frac{n_i\cdot n_j}{2}
\,.\end{equation}
We order the jets such that
\begin{equation}
t \equiv s_{12} = \min_{i \neq j} \{s_{ij}\}
\,,\qquad
\w_1 < \w_2
\,,\end{equation}
and we define
\begin{equation}
u = \max_k s_{1k}
\,,\qquad
Q^2 = (q_1 + \dotsb + q_N)^2
\,.\end{equation}
So, $t$ is the smallest dijet invariant mass, and $u$ measures the softness of jet 1.
For $e^+ e^- \to 3$ jets, $u = s_{13}$ is just the intermediate dijet invariant mass.

The situation where all jets are equally energetic and well separated corresponds to $\w_i \sim Q$ and $n_i \sdt n_j \sim 1$ and therefore $t\sim u\sim s_{ij} \sim Q^2$. It is described by the usual SCET framework, since all dijet invariant masses are of the same order so there are no additional hierarchies between physical scales. In contrast, the \SCETp regimes illustrated in \fig{3jet_configs} and discussed in the following subsections are characterized by $t \ll u \sim Q^2$ ($c+$ regime), $t \sim u \ll Q^2$ ($s+$ regime), and $t \ll u \ll Q^2$ ($cs+$ regime).

The degrees of freedom in \SCETa consist of collinear modes for every jet direction and ultrasoft (usoft) modes interacting with these. The parametric scaling of these modes is summarized in table~\ref{tab:scet_modes}. The collinear modes for the different jet directions cannot interact with each other in the effective theory, while the interactions with the usoft modes decouple at leading power in $\Tau_N/Q \sim m_J^2/Q^2$ via the BPS field redefinition~\cite{Bauer:2001yt}. This leads to the following SCET Lagrangian for $N$-jet production 
\begin{align} \label{eq:L_SCET}
\mathcal{L}_{\SCET}= \sum^N_{i=1} \mathcal{L}_{n_i}  + \mathcal{L}_{us} + \cL_\SCET^\hard
\,.\end{align}
The Lagrangians $\mathcal{L}_{n_i}$ and $\mathcal{L}_{us}$ describe the dynamics of the $n_i$-collinear and usoft sectors, respectively, and only contain interactions among the fields within each sector. Their explicit expressions can be found in refs.~\cite{Bauer:2000yr, Bauer:2001ct, Bauer:2001yt}. The hard-scattering Lagrangian $\cL_\SCET^\hard$ consists of leading-power SCET operators, built from collinear fields and usoft Wilson lines, and their Wilson coefficients. It arises from matching the hard-scattering processes in QCD onto SCET, where fluctuations with a virtuality above the scale $\mu \sim Q$ are integrated out.

The factorization theorem for the differential cross section following from \eq{L_SCET} has the following structure~\cite{Fleming:2007qr, Bauer:2008dt, Ellis:2010rwa, Stewart:2010tn}
\begin{align}\label{eq:sigma_N}
\df \sigma_{\SCET}\sim  \vec{C}^\dagger_N \times \bigg[\prod_{i=1}^N J_i \otimes \widehat{S}_N\bigg] \times \vec{C}_N = {\rm tr} \bigg[ \widehat{H}_{N} \times \prod_{i=1}^N J_i  \otimes \widehat{S}_{N}\bigg] \, .
\end{align}
The Wilson coefficients $\vec{C}_N$ arise from $\cL_\SCET^\hard$ and encode the short-distance physics of the hard-scattering process. They determine the hard function $\widehat{H}_{N}=\vec{C}_N \vec{C}^\dagger_N$. The jet functions $J_i$ incorporate the dynamics of the collinear radiation that leads to the formation of jets, which takes place at the scale $\mu \sim m_J$. Finally, the cross talk between the jets via usoft radiation is described by the soft function $\widehat{S}_N$ at the scale $\mu \sim m_J^2/Q$. Here, $\vec{C}_N $ is a vector and $\widehat{S}_N$ and $\widehat{H}$ are matrices in the color space of the $N$ external hard partons. The jet functions $J_i$ are scalars in color space, i.e.~color diagonal, and can therefore be pulled outside the color trace. The precise form of the jet and soft functions and the structure of the convolution between them is determined by the $N$-jet resolution variable. Since each function in the cross section \eq{sigma_N} only involves a single scale, the logarithms of $\Tau_N/Q\sim m_J^2/Q^2$ can be systematically resummed by evaluating each function at its natural scale and evolving them to a common scale using their renormalization group evolution (RGE).

\subsection{$c+$ regime: two collinear jets}
\label{subsec:c+overview}

We now consider the kinematic situation where the first two jets come close to each other, but remain energetic, i.e.,
\begin{equation}
n_1 \sdt n_2 \ll 1
\,,\qquad
n_i \sdt n_j \sim 1
\,,\qquad
\w_i \sim Q
\qquad\Rightarrow\qquad
t \ll u \sim Q^2
\,.\end{equation}
Thus, all of the dijet invariant masses remain equally large except for $t = s_{12} \ll Q^2$. This additional hierarchy introduces large logarithms of $t/Q^2 \sim n_1\sdt n_2$ in the hard and soft functions in \eq{sigma_N}. The \SCETp theory that resums these logarithms (which we now regard as the $c+$ regime of \SCETp) was introduced in ref.~\cite{Bauer:2011uc}.%
\footnote{The refactorization of the hard sector was already discussed earlier in refs.~\cite{Bauer:2006mk, Baumgart:2010qf}.}
We briefly recall it here and refer to ref.~\cite{Bauer:2011uc} for a detailed derivation. It was applied in refs.~\cite{Larkoski:2015kga, Larkoski:2015uaa} in the context of jet substructure.

The relevant modes in the $c+$ regime are given in table~\ref{tab:modes_c+}. Due to the measurement of $t$ there are additional collinear-soft (csoft) modes. Compared to the usoft modes, they have a higher angular resolution allowing them to resolve the two nearby jets separated by the angle of order $|\vec{p}_\perp|/p^- \sim \sqrt{t}/Q$.  Hence, they interact with the usoft modes as collinear modes with lightcone direction $n_t$. At the same time, they interact with the two nearby jets 1 and 2 (the $n_1$-collinear and $n_2$-collinear sectors) as soft modes. In particular, at their own collinearity scale the directions $n_1$ and $n_2$ belong to the same equivalence class as $n_t$. The requirement that their plus component is constrained by the \SCETa jet resolution measurement implies $p^+ \sim \Tau_N \sim m_J^2/Q$ which then fully determines their scaling as given in table~\ref{tab:modes_c+}.

\begin{figure}
\centering
\includegraphics[scale=0.5]{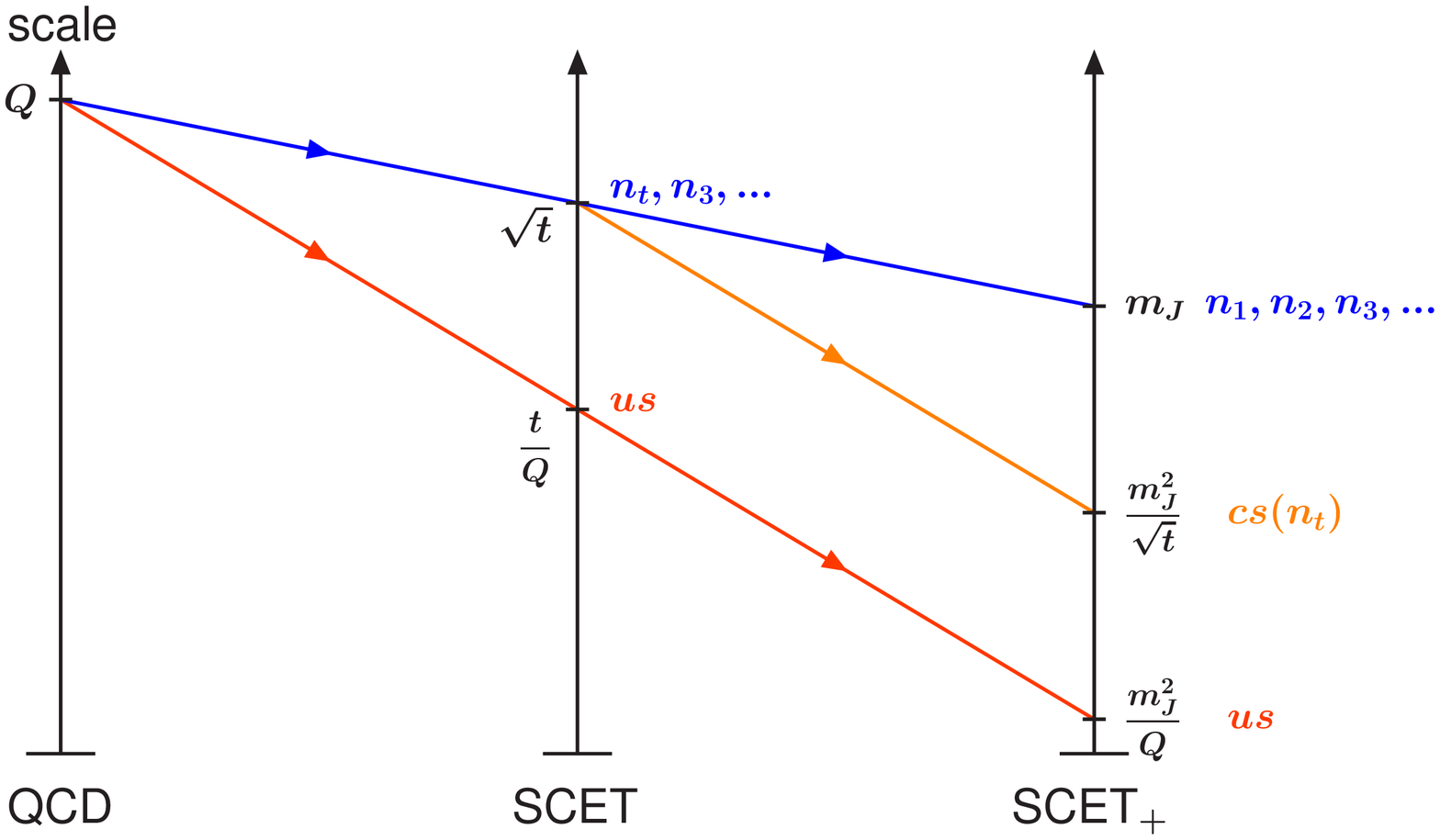}%
\caption{Illustration of the multistage matching procedure for the $c+$ regime of \SCETp with $t\ll u\sim Q^2$. The modes and their virtuality scale are indicated.}
\label{fig:scet_c+_matching}
\end{figure}

To disentangle all physical scales, we perform the two-step matching shown in Fig.~\ref{fig:scet_c+_matching}.
We first match QCD onto standard SCET with $N-1$ collinear sectors $n_t,n_3,\dots,n_N$ with corresponding invariant mass fluctuations $\sim \sqrt{t}$ and an associated usoft sector at the scale $t/Q$. At this point, the two nearby jets are not separately resolved yet and contained in the $n_t$-collinear sector. After decoupling the collinear and usoft modes, this theory is matched onto \SCETp. For the collinear sectors of jets 3 to $N$ as well as for the usoft sector only the virtuality scale is lowered to $m_J$ and $m_J^2/Q$, respectively. The $n_t$-collinear sector of the parent SCET with scaling $p^\mu_{n_t} \sim (t/Q,Q,\sqrt{t})$ is matched onto the two collinear sectors for jets 1 and 2 and the csoft mode. This step involves nontrivial matching coefficients, related to the collinear splitting amplitudes. They appear when matching the hard-scattering Lagrangian of the parent SCET onto the final $\cL_{c+}^\hard$ of \SCETp. As shown in ref.~\cite{Bauer:2011uc}, the interactions between the two collinear modes and the csoft modes can be decoupled via a further BPS field redefinition. This leads to the leading-power Lagrangian, which has again no interactions between different sectors,
\begin{align} \label{eq:Lc+}
\mathcal{L}_{c+}= \sum^N_{i=1} \mathcal{L}_{n_i} + \mathcal{L}_{n_t} + \mathcal{L}_{us} + \cL_{c+}^\hard
\,.\end{align}
Here, $\mathcal{L}_{n_t}$ is the Lagrangian for the csoft modes and is identical to the Lagrangian for collinear modes $\mathcal{L}_{n_i}$, except for the different scaling of the label momenta and associated scaling of the csoft gauge fields. It is important that the csoft fields are defined with a zero-bin subtraction~\cite{Manohar:2006nz} to avoid double counting with the usoft fields in analogy to the collinear fields. In addition, the $n_1$ and $n_2$-collinear modes are now defined with an appropriate zero-bin subtraction with respect to both csoft and usoft modes.

\begin{table}
\centering
\begin{tabular}{c|c|c}
\hline \hline
mode & $p^\mu=(+,-,\perp)$ & $p^2$ \\
\hline
collinear ($n_1,\dots,n_N$) &  $\bigl(\Tau_N,Q,\sqrt{\Tau_N Q} \bigr)$ &$ \Tau_N Q \sim m_J^2$ \\
collinear-soft ($n_t$) &  $\bigl(\Tau_N, \Tau_N\, Q^2/t, \Tau_N\,Q/\sqrt{t} \bigr)$ & $\Tau_N^2\, Q^2/t \sim m_J^4/t$ \\
ultrasoft & $ \bigl(\Tau_N, \Tau_N, \Tau_N \bigr)$ &$\Tau_N^2 \sim m_J^4/Q^2$\\
\hline \hline
\end{tabular}
\caption{Scaling of the relevant modes in the $c+$ regime of \SCETp. For the collinear-soft mode, $n_1$ and $n_2$ belong to the same equivalence class as $n_t$.}
\label{tab:modes_c+}
\end{table}

The factorization theorem for the differential cross section following from \eq{Lc+} has the structure~\cite{Bauer:2011uc}
\begin{align}\label{eq:sigma_N+}
\df \sigma_{c+} &\sim  \vec{C}^\dagger_{N-1} \, C^*_c \times \biggl[\prod_{i=1}^N J_i \otimes S_c\otimes \widehat{S}_{N-1}\biggr] \times C_c \,\vec{C}_{N-1} 
\nn \\
&=  {\rm tr} \biggl[ \widehat{H}_{N-1} \times H_{c}  \times \prod_{i=1}^N J_i  \otimes S_c \otimes \widehat{S}_{N-1}\biggr]
\,.\end{align}
Compared to \eq{sigma_N}, the hard coefficient $\vec C_N$ got factorized into $\vec{C}_{N-1}$ for $N-1$ hard external partons at the scale $\mu \sim Q$ (arising from the first matching step in \fig{scet_c+_matching}) and a collinear splitting coefficient $C_c$ describing the splitting of the $n_t$-collinear sector into the $n_1$- and $n_2$-collinear sectors at the scale $\mu \sim \sqrt{t}$ (arising from the second matching step in \fig{scet_c+_matching}). The corresponding hard functions are $\widehat{H}_{N-1}=\vec{C}_{N-1} \vec{C}^\dagger_{N-1}$ and $H_{c}=|C_c|^2$. The soft function $\widehat{S}_N$ got factorized into a usoft function $\widehat{S}_{N-1}$ at the scale $\mu \sim m_J^2/Q$ that only resolves the $N-1$ well-separated jets, and a csoft function $S_c$ that describes the csoft radiation between the two nearby jets at the scale $\mu\sim m_J^2/\sqrt{t}$. Note that $H_c$ and $S_c$ have a trivial color structure, since they are related to a $1 \rightarrow 2$ collinear splitting for which the relevant color space is one dimensional. In other words, in the collinear limit the full $N$-parton color space separates into the subspace for $N-1$ partons and the subspace for the collinear $1\to 2$ splitting.

\subsection{$s+$ regime: one soft jet}
\label{subsec:s+overview}

Next, we consider the kinematic situation that the first jet becomes less energetic, while all jets remain well separated from each other, i.e.,
\begin{equation}
\w_1 \ll Q
\,,\qquad
\w_{i\geq 2} \sim Q
\,,\qquad
n_i \sdt n_j \sim 1
\qquad\Rightarrow\qquad
t\sim u\ll Q^2
\,.\end{equation}
In this case, all dijet invariant masses involving the first soft jet are all of the same order $s_{1i} \sim u \ll Q^2$. This additional hierarchy leads to large logarithms of $u/Q^2$ in \eq{sigma_N}, appearing this time only in the hard function. There are no large logarithms in the soft function as it only depends on the angles between the jet directions, which do not exhibit any hierarchy. Hence, the appropriate EFT setup, which we identify with the $s+$ regime of \SCETp, only refactorizes the hard function. This type of setup was also considered in refs.~\cite{Larkoski:2015zka, Larkoski:2015kga} to calculate energy-correlation functions describing jet substructure. Note however, that their conjectured factorization theorem for the general $N$-jet case does not correctly account for color correlations.

The relevant modes in the $s+$ regime are given in table~\ref{tab:modes_s+}. In addition to the usual collinear modes for the energetic jet sectors $2, \ldots, N$ and the overall usoft modes, we have a soft-collinear mode with momentum scaling $p^\mu_{1} \sim \w_1(\lambda^2, 1,\lambda)$ that is responsible for the collinear dynamics of the soft jet. Its overall scaling is fixed by the kinematic constraint $s_{1i} \sim \w_1 Q \sim u$ and the constraint imposed by the measurement of the jet resolution variable requiring that $p_1^+ = \w_1 \lambda^2 \sim \Tau_N \sim m_J^2/Q$.%
\footnote{Here it is important that we are using a \SCETa jet resolution variable like $N$-jettiness, which fixes the size of small lightcone component $p_1^+$.}
Since we still have $\Tau_N \ll u/Q \sim \w_1$, this soft-collinear mode cannot couple to any of the other well-separated collinear modes.
Hence, it is just a collinear mode with a smaller energy and consequently a smaller invariant mass, $m^2_{1} \sim m_J^2 \,u/Q^2 \ll m_J^2$.

\begin{figure}
\centering
\includegraphics[scale=0.5]{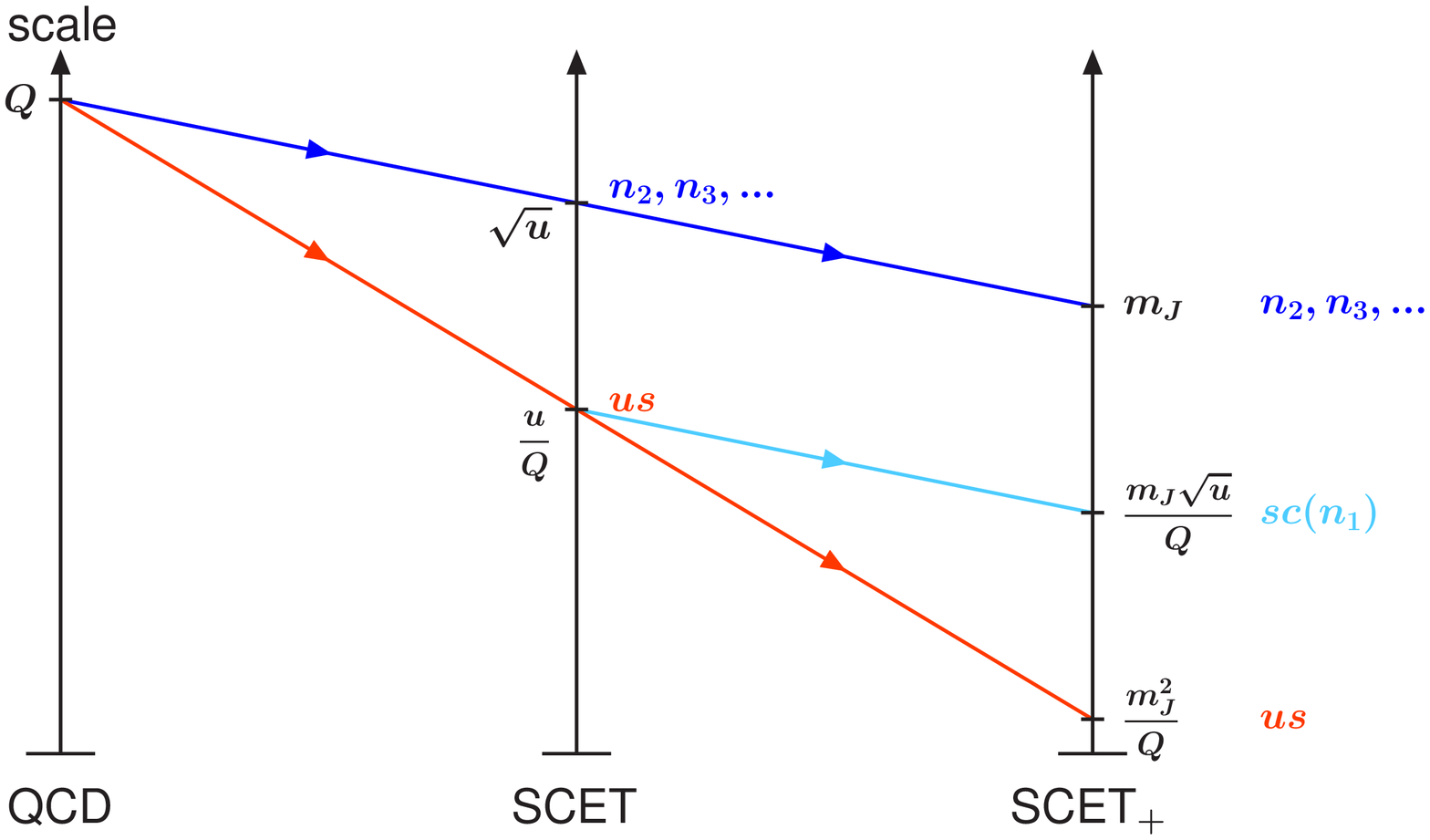}%
\caption{Illustration of the multistage matching procedure for the $s+$ regime of \SCETp with $t\sim u\ll Q^2$. The modes and their virtuality scale are indicated.}
\label{fig:scet_s+_matching}
\end{figure}

To match onto the $s+$ regime, we perform the two-step matching shown in \fig{scet_s+_matching}. We first match QCD onto standard SCET with $N-1$ collinear sectors $n_2, \ldots, n_N$ at the scale $\sqrt{u}$ and a corresponding usoft sector at the scale $u/Q$. At this point, the soft jet is still unresolved and contained in the usoft sector. After decoupling the collinear and usoft modes, we match this theory onto \SCETp. The virtuality of the collinear sectors is simply lowered to $m_J$. The decoupled usoft sector with momentum scaling $p^\mu \sim u/Q (1,1,1)$ is matched onto the soft-collinear mode for the now resolved jet 1 and the usoft sector at the lower scale $m_J^2/Q$. This involves nontrivial matching coefficients related to the soft gluon current, which appear when matching the hard-scattering Lagrangian from the parent SCET onto the $\cL_{s+}^\hard$ of \SCETp. The soft-collinear and usoft sectors can be decoupled via a second BPS field redefinition in the soft-collinear sector. Since the parent usoft sector is equivalent to full QCD at a lower scale, this decoupling proceeds completely analogous to the usual matching from QCD to SCET. The final leading-power Lagrangian has again all sectors completely decoupled,
\begin{align} \label{eq:Ls+}
\mathcal{L}_{s+} = \mathcal{L}_{n_1} + \sum^N_{i=2} \mathcal{L}_{n_i} + \mathcal{L}_{us} + \cL_{s+}^\hard
\,.\end{align}
The Lagrangian $\mathcal{L}_{n_1}$ for the soft-collinear mode is given by the usual collinear Lagrangian, but with a different power counting for the label momenta.

\begin{table}
\centering
\begin{tabular}{c|c|c}
\hline \hline
mode & $p^\mu=(+,-,\perp)$ & $p^2$ \\
\hline
collinear ($n_2,\dots,n_N$) &  $\bigl(\Tau_N,Q,\sqrt{\Tau_NQ} \bigr)$ &$ \Tau_N Q \sim m_J^2$ \\
soft-collinear ($n_1$) &  $\bigl(\Tau_N, u/Q, \sqrt{\Tau_N u/Q} \bigr)$ & $\Tau_N\, u/Q \sim m_J^2\, u/Q^2$ \\
ultrasoft & $ \bigl(\Tau_N, \Tau_N, \Tau_N \bigr)$ &$\Tau_N^2 \sim m_J^4/Q^2$\\
\hline \hline
\end{tabular}
\caption{Scaling of the relevant modes in the $s+$ regime of \SCETp.}
\label{tab:modes_s+}
\end{table}

The factorization theorem following from \eq{Ls+} has the structure
\begin{align}\label{eq:sigma_Ns}
\df \sigma_{s+}  \sim \vec{C}^\dagger_{N-1} \,  \widehat{C}^{\dagger}_{s} \times \bigg[\prod_{i=1}^N J_i \otimes \widehat{S}_N\bigg] \times \widehat{C}_{s} \,\vec{C}_{N-1} ={\rm tr} \bigg[ \widehat{C}_{s} \, \widehat{H}_{N-1} \, \widehat{C}_{s}^\dagger  \,\prod_{i=1}^N J_i \otimes \widehat{S}_N \bigg] 
\,. \end{align}
Compared to \eq{sigma_N}, the hard coefficient $\vec C_N$ got factorized into $\vec{C}_{N-1}$ for $N-1$ hard external partons at the scale $\mu \sim Q$ (arising from the first matching step in \fig{scet_s+_matching}) and a soft splitting coefficient $\widehat{C}_{s}$ describing the splitting of the parent usoft sector in SCET into the $n_1$-soft-collinear and the usoft sector in \SCETp at the scale $\mu \sim t/Q$ (arising from the second matching step in \fig{scet_s+_matching}). The $\widehat{C}_{s}$ is a matrix in color space that promotes the hard coefficient $\vec C_{N-1}$ from the $(N-1)$-parton color space to the full $N$-parton color space in which the soft function $\widehat{S}_N$ acts. Note that at leading power in $u/Q^2$ the soft jet is initiated by a gluon, $J_1=J_g$, since only gluon emissions are enhanced in the soft limit, and the natural scale for its jet function is $\mu\sim m_J \times \sqrt{u}/Q$.

\subsection{$cs+$ regime: one soft jet collinear to another jet}
\label{subsec:cs+overview}

We now consider the kinematic situation where the first two jets come close to each other and at the same time the first jet becomes soft. The remaining jets stay equally separated and energetic, i.e.,
\begin{equation}
n_1 \sdt n_2 \ll 1
\,,\qquad
\w_1 \ll Q
\,,\qquad
n_i \sdt n_j \sim 1
\,,\qquad
\w_{i\geq 2} \sim Q
\qquad\Rightarrow\qquad
t \ll u \ll Q^2
\,.\end{equation}
Hence, this case is characterized by the combination of the collinear and soft hierarchies in the dijet invariant masses, $t = s_{12} \ll u \sim s_{1i\geq3} \ll Q^2$, while all remaining $s_{jk} \sim Q^2$. Treating this case in either the $s+$ or $c+$ regimes with the corresponding generic scales would leave large logarithms of either $u/Q^2$ or $t/u$ in the hard and/or soft functions. The resummation of both types of large logarithms is achieved in the $cs+$ regime of \SCETp, which has not been discussed in the literature before. This EFT setup combines the expansion in the softness of jet 1 and the angle between jets 1 and 2, and is the theory connecting the $c+$ and $s+$ regimes. As we will see below, this kinematic situation can effectively be described within the $c+$ regime by an appropriate choice of resummation scales in the hard sector that takes into account the softness of jet 1. This has been exploited in ref.~\cite{Larkoski:2015kga}. It is nevertheless important to explicitly consider the $cs+$ regime in order to fully separate all scales and to show that all logarithms are resummed correctly in this way. This also shows that this kinematic situation cannot be described within the $s+$ regime, which lacks the required refactorization of the soft sector. The $cs+$ regime is also useful to account for the overlap between the $s+$ and $c+$ regimes, see \subsec{combineEFT}, and to be able to handle more complicated overlapping hierarchies.

The relevant modes in the $cs+$ regime are summarized in table~\ref{tab:modes_cs+}. Besides the usual collinear modes with the labels $n_2,\dots,n_N$ and the usoft modes, we have a soft-collinear mode in the $n_1$ direction that describes the collinear dynamics of the soft jet, and a csoft mode that is responsible for the cross talk between the two nearby jets 1 and 2. As for the soft case, the scaling of the soft-collinear mode is determined by $u \sim Q \w_1$ and $p_1^+ \sim \Tau_N \sim m_J^2/Q$. And as for the collinear case, to be able to resolve the two nearby jets 1 and 2, the csoft mode is boosted in the lightcone direction $n_t$ with angular resolution scale $|\vec p_{t\perp}|/p_t^- \sim \sqrt{n_1 \sdt n_2} \sim \sqrt{t/u}$. The constraint from the jet resolution measurement, $p_1^+ \sim m_J^2/Q$, then fixes its scaling.

\begin{table}
\centering
\begin{tabular}{c|c|c}
\hline \hline
mode & $p^\mu=(+,-,\perp)$ & $p^2$ \\
\hline
collinear ($n_2,\dots,n_N$) &  $\bigl(\Tau_N,Q,\sqrt{\Tau_NQ} \bigr)$ &$ \Tau_N Q \sim m_J^2$ \\
soft-collinear ($n_1$) &  $\bigl(\Tau_N,u/Q, \sqrt{\Tau_N u/Q} \bigr)$ & $\Tau_N\, u/Q \sim m_J^2\, u/Q^2$ \\
collinear-soft ($n_t$) &  $\bigl(\Tau_N, \Tau_N\, u/t, \Tau_N\sqrt{u/t}\bigr)$ & $\Tau_N^2\, u/t \sim m_J^4\, u/(Q^2t)$ \\
ultrasoft & $ \bigl(\Tau_N, \Tau_N, \Tau_N \bigr)$ &$\Tau_N^2 \sim m_J^4/Q^2$\\
\hline \hline
\end{tabular}
\caption{Scaling of the relevant modes in the $cs+$ regime of \SCETp.}
\label{tab:modes_cs+}
\end{table}

\begin{figure}
\centering
\includegraphics[scale=0.5]{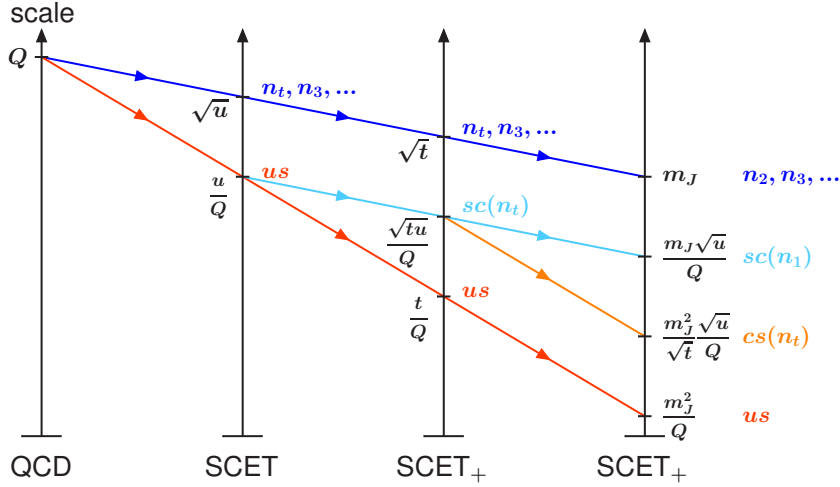}%
\caption{Illustration of the multistage matching procedure for the $cs+$ regime of \SCETp with $t \ll u \ll Q^2$. The parent SCET is matched onto an intermediate \SCETp with a single soft-collinear sector. In the final matching step, this is further matched onto separate soft-collinear and csoft modes.}
\label{fig:scet_cs+_matching}
\end{figure}

We now perform the three-step matching procedure shown in \fig{scet_cs+_matching}. We first match QCD onto SCET with $N-1$ collinear modes $n_t, n_3, \ldots, n_N$ and usoft modes with virtuality scales $\sqrt{u}$ and $u/Q$, respectively.
Next, we match onto an intermediate \SCETp with standard collinear and usoft modes at the lower virtuality scales $\sqrt{t}$ and $t/Q$, respectively, and a soft-collinear sector in the $n_t$ direction at the lower scale $\sqrt{tu}/Q$ with momentum scaling $p^\mu \sim (t/Q,u/Q,\sqrt{tu}/Q)$, which can resolve the angular size of the $n_t$-collinear sector. As before, the collinear, soft-collinear, and usoft sectors are decoupled by appropriate BPS field redefinitions. At this point, the soft jet is not yet resolved and still contained in the soft-collinear sector. This means that there is no nontrivial hard matching coefficient in this step, and as we will see in \sec{3jets}, the matching of the operators in the hard-scattering Lagrangian happens entirely at the level of soft Wilson lines. This also means that one could in principle directly construct this \SCETp and match onto it from QCD (see e.g.~refs.~\cite{Bauer:2011uc, Procura:2014cba}).

In the last step in \fig{scet_cs+_matching}, we match the intermediate \SCETp with $N-1$ collinear sectors onto the final $cs+$ theory. Here, the virtualities of the collinear and usoft modes are simply lowered, with the $n_t$-collinear mode now being refined to the final $n_2$-collinear mode. The parent decoupled soft-collinear sector is matched onto the final $n_1$-soft-collinear mode for the now resolved jet 1, and the final csoft mode in the $n_t$ direction. (Hence, taking into account its full ancestry, the final csoft mode here could be referred to as a soft-collinear-soft mode.) The corresponding matching coefficients are now related to the soft limit of the collinear splitting amplitudes or equivalently the collinear limit of the soft gluon current. Analogous to the $c+$ regime, the csoft and soft-collinear modes are decoupled by a BPS field redefinition. Note that the consistency of \fig{scet_cs+_matching} can be verified by taking the limit $t \to u$ or $u \to Q^2$ for which it reduces to the matching for the $s+$ and $c+$ regimes, respectively.

The final fully decoupled leading-power Lagrangian is given by
\begin{align} \label{eq:Lcs+}
\mathcal{L}_{cs+} = \mathcal{L}_{n_1} + \sum^N_{i=2} \mathcal{L}_{n_i} + \mathcal{L}_{n_t} + \mathcal{L}_{us} + \cL_{cs+}^\hard
\,,\end{align}
where both $\mathcal{L}_{n_1}$ and $\mathcal{L}_{n_t}$ are collinear Lagrangians with the appropriate scaling of their label momenta.

The factorization theorem resulting from \eq{Lcs+} has the structure
\begin{align}\label{eq:dsigma_cs}
\df \sigma_{cs+} &\sim\vec{C}^\dagger_{N-1} \, C^*_{cs} \times \biggl[\prod_{i=1}^N J_i \otimes S_c \otimes \widehat{S}_{N-1}\biggr] \times  C_{cs} \,\vec{C}_{N-1}  
\nn \\
&= {\rm tr} \biggl[ \widehat{H}_{N-1} \times H_{cs}  \times\prod_{i=1}^N J_i \otimes S_c \otimes \widehat{S}_{N-1} \biggr] 
\,.\end{align}
As in \eqs{sigma_N+}{sigma_Ns}, the hard coefficient $\vec{C}_{N-1}$ describes the production of $N-1$ hard partons at the scale $\mu\sim Q$. The coefficient $C_{cs}$ now describes the soft-collinear splitting at the scale $\mu \sim \sqrt{tu}/Q$. Compared to the $c+$ regime in \eq{sigma_N+}, $C_{cs}$ corresponds to the soft limit of the collinear splitting coefficient $C_c$, whose scale got lowered from $\sqrt{t}\to \sqrt{t}\times\sqrt{u}/Q$. Similarly, the scale of the csoft function $S_c$ is now lowered to $\mu \sim m_J^2/\sqrt{t} \times \sqrt{u}/Q$. Compared to $s+$ regime in \eq{sigma_Ns}, $C_{cs}$ corresponds to taking the collinear limit of the soft splitting coefficient $\widehat C_{s}$, lowering its scale from $u/Q \to u/Q \times \sqrt{t/u}$ and making it color diagonal. In addition, the soft sector got refactorized as in the $c+$ regime leading to $S_c$ at the scale $\mu \sim m_J^2/Q \times \sqrt{u/t}$. As in the $s+$ regime, the soft jet 1 is always initiated by a gluon with the natural scale for its jet function being $\mu\sim m_J \times \sqrt{u}/Q$.

\subsection{Combining all regimes}
\label{subsec:combineEFT}

\begin{figure}
\centering
\includegraphics[width=0.6\linewidth]{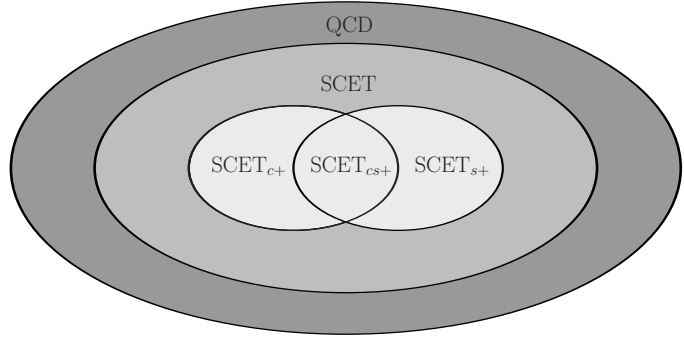}%
\caption{Schematic overview of the fixed-order content of the different theories discussed here. Approaching the $cs+$ regime in the center, more and more logarithms get large and are resummed, at the expense of additional expansions. The missing fixed-order corrections from these expansions can be incorporated by adding back the relevant nonlogarithmic fixed-order differences between the theories.}
\label{fig:scet_theories}
\end{figure}

We now discuss how to combine the factorization theorems for the different \SCETp regimes as well as the nonhierarchical SCET limit to obtain a complete description for any $t,u,Q^2 \gg m_J^2$. This will be generalized to the full $N$-jet phase space with arbitrary hierarchies among the $s_{ij} \gg m_J^2$ in \sec{njets}. The goal is to be able to resum all logarithms of any ratios of $s_{ij}$ and at the same time to reproduce the correct fixed-order result whenever there are no longer large hierarchies.

Each of the factorization theorems in eqs.~\eqref{eq:sigma_N},~\eqref{eq:sigma_N+},~\eqref{eq:sigma_Ns}, and~\eqref{eq:dsigma_cs} has been power expanded in the ratio of scales whose logarithms are being resummed. They thus receive power corrections in the corresponding scale ratio, which become $\ord{1}$ in the nonhierarchical limit where that scale ratio is no longer small. To obtain a smooth and complete description, we basically need to add to the resummed result in a given kinematic region the relevant missing nonlogarithmic (``nonsingular'') corrections at fixed order, such that we reproduce the full fixed-order result everywhere. In addition, to ensure a smooth transition across different kinematic regions it is also important to smoothly turn off the resummation in any nonhierarchical limit. This can be achieved through a suitable choice of resummation profile scales~\cite{Ligeti:2008ac, Abbate:2010xh}.

A Venn diagram of the fixed-order content of the different theories is shown in \fig{scet_theories}, from which the nonsingular corrections can be directly read off. The basic idea is to start from the inner most hierarchical (most expanded) case and go outwards step by step matching to the fixed-order content of the next less hierarchical (less expanded) case until we reach the outermost full QCD result. For $e^+ e^- \to 3$ jets this procedure will be discussed in some detail in \subsec{3jet_combination}.

We start from the $cs+$ result which resums all kinematic logarithms in the $t\ll u\ll Q^2$ limit and add nonsingular corrections to match it to the $c+$ and $s+$ results, which yields the combined \SCETp cross section,
\begin{align}\label{eq:nonsing}
\df \sigma_+ &= \df \sigma_{cs+} + \df \sigma_{c+}^\nons + \df \sigma_{s+}^\nons
\,, \nn \\
\df \sigma_{c+}^\nons &= \df \sigma_{c+} - \bigl[\df \sigma_{cs+} \bigr]_{\FO(u \ll Q^2)}
\,, \nn \\
\df \sigma_{s+}^\nons &= \df \sigma_{s+} - \bigl[ \df \sigma_{cs+} \bigr]_{\FO(t \ll u)}
\,.\end{align}
The $\FO(...)$ notation indicates that the hierarchy specified in brackets is not resummed but taken at fixed order. For example, for $[\df \sigma_{cs+}]_{\FO(u \ll Q^2)}$ the logarithms of $t/u$ are resummed, while the logarithms of $u/Q^2$ are not resummed, and are instead expanded to the same fixed order as they are present in $\df\sigma_{c+}$.
Hence, in $\df \sigma_{c+}^\nons$ the logarithms of $t/Q^2$ are still resummed, while the fixed-order corrections that are singular in $u/Q^2$ cancel between the two terms, such that $\df \sigma_{c+}^\nons$ is a power correction in $u/Q^2$.

Having obtained $\df\sigma_+$, we can further add the nonsingular corrections from SCET in the limit $t\sim u\sim Q^2$ and eventually the nonsingular corrections from full QCD relevant in the limit $m_J\sim Q$,
\begin{align} \label{eq:nonsing2}
\df \sigma &= \df \sigma_{+} + \df \sigma_{\SCET}^\nons +\df \sigma_{\QCD}^\nons
\,, \nn \\
\df \sigma_{\SCET}^\nons
&= \df \sigma_{\SCET} - \bigl[ \df \sigma_{+} \bigr]_{\FO(t \ll u \ll Q^2)}
\nn \\
&= \df \sigma_{\SCET} - \bigl[\df \sigma_{c+} + \df \sigma_{s+} -\df \sigma_{cs+} \bigr]_{\FO(t \ll u \ll Q^2)}
\,, \nn \\
\df \sigma_{\QCD}^\nons &=\df \sigma_{\QCD} - \bigl[ \df \sigma_{\SCET} \bigr]_{\FO(m_J \ll Q)}
\,.\end{align}

Note that in our approach, the overlap between the $c+$ and $s+$ regimes is automatically taken care off via the separate $cs+$ regime.
In ref.~\cite{Larkoski:2015kga} this overlap is removed manually by subtracting it at the level of the factorized cross section, which yields technically the same result.

\subsection{\SCETb observables}
\label{subsec:scetii}

Here we briefly discuss \SCETp for \SCETb-type jet resolution variables, which constrain the transverse momenta within the jets rather than their invariant mass or small lightcone momenta. A simple example is $N$-jettiness with the broadening measure,
\begin{align}
\Tau^\perp_N = \sum_k \min\limits_{i} \Bigl\{\frac{2 \abs{\vec{q}_i \times \vec{p}_k}}{Q_i}\Bigr\}
= \sum_k \min\limits_{i} \Bigl\{ \frac{\abs{\vec{p}_{k\,\perp_i}}}{\rho_i} \Bigr\}
\,, \end{align}
where $\perp_i$ denotes the component perpendicular to the direction $n_i$ of the $i$th jet.
Other examples are the XCone measures with angular exponent $\beta = 1$~\cite{Stewart:2015waa}.
Measures of this type have been utilized for jet substructure studies using $N$-subjettiness~\cite{Thaler:2010tr, Thaler:2011gf}.
These observables are in principle sensitive to the precise definition of the jet axes $n_i$ due to the fact that the recoil from soft emissions cannot be neglected. To keep the factorization theorem simple, one can employ the recoil-insensitive broadening axes~\cite{Larkoski:2014uqa}.

The distinct feature of \SCETb-type observables is that all modes in the effective theory, i.e.~collinear, soft, csoft and soft-collinear, have the same virtuality. This directly follows from the fact that the measurement of $\Tau^\perp_N$ constraints their $p_\perp$ components, which sets the scale of their virtuality $p^2 \sim p_\perp^2$. The scaling of the relevant modes in the different regimes is summarized in tables~\ref{tab:scet_modes_II}, \ref{tab:modes_II_c+}, \ref{tab:modes_II_s+}, and \ref{tab:modes_II_cs+}.
The different modes are now parametrically separated in rapidity $\sim p^-/p^+$ rather than virtuality, and the corresponding logarithms can be summed by using the rapidity renormalization group evolution~\cite{Chiu:2011qc, Chiu:2012ir}.

In the $c+$ regime, there are again csoft modes mediating between the two nearby jets 1 and 2. As before, their scaling is determined by the requirement that they have a resolution angle $|\vec{p}_\perp|/p^- \sim \sqrt{t}/Q$ and the measurement constraint $p_\perp \sim \Tau_N^\perp$. In the $s+$ regime, the scaling of the soft-collinear modes $p^\mu_{1} \sim \w_1(\lambda^2, 1, \lambda)$ is fixed by the facts that $\w_1 Q \sim u$ and $\w_1 \lambda \sim p_\perp$. Finally, the $cs+$ regime again combines the features of the $c+$ and $s+$ regimes.

The structure of the corresponding factorization theorems is analogous to those in eqs.~\eqref{eq:sigma_N},~\eqref{eq:sigma_N+},~\eqref{eq:sigma_Ns}, and~\eqref{eq:dsigma_cs}. The essential difference is that the convolutions between soft and jet functions are now in transverse momentum variables, and involve the resummation of rapidity logarithms. Since the matching steps are insensitive to the details of the jet measurement, all the arising Wilson coefficients $\vec C_N$, $\vec C_{N-1}$, $C_c$, $\widehat{C}_s$, and $C_{cs}$ are the same as for a \SCETa-type observable. The factorized cross sections for the different regimes can be combined to describe the complete phase space by accounting for the nonsingular corrections as discussed in \subsec{combineEFT}.

\begin{table}
\centering
\begin{tabular}{c|c}
\hline\hline
mode & $p^\mu=(+,-,\perp)$  \\
\hline
collinear ($n_1,\dots,n_N$) &  $\bigl(p_\perp^2/Q ,Q,p_\perp\bigr)$ \\
soft & $ \bigl(p_\perp,p_\perp,p_\perp \bigr)$  \\
\hline\hline
\end{tabular}
\caption{Scaling of the modes in \SCETb for the standard case with $N$ equally separated and energetic jets. The virtuality of all modes is $p^2 \sim p_\perp^2$ and therefore not displayed.}
\label{tab:scet_modes_II}
\end{table}

\begin{table}
\centering
\begin{tabular}{c|c}
\hline\hline
mode & $p^\mu=(+,-,\perp)$ \\ 
\hline
collinear ($n_1,\dots,n_N$) &  $\bigl(p_\perp^2/Q ,Q,p_\perp\bigr)$ \\
collinear-csoft ($n_t$) &  $\bigl(p_\perp\,\sqrt{t}/Q,p_\perp\, Q/\sqrt{t},p_\perp\bigr)$ \\
soft & $ \bigl(p_\perp,p_\perp,p_\perp \bigr)$  \\
\hline\hline
\end{tabular}  
\caption{Scaling of the relevant modes in the $c+$ regime of \SCETp for a \SCETb observable.}
\label{tab:modes_II_c+}
\end{table}

\begin{table}
\centering
\begin{tabular}{c|c}
\hline\hline
mode & $p^\mu=(+,-,\perp)$  \\
\hline
collinear ($n_2,\dots,n_N$) &  $\bigl(p_\perp^2/Q ,Q,p_\perp\bigr)$ \\
soft-collinear ($n_1$) &  $\bigl(p_\perp^2\, Q/u, u/Q ,p_\perp\bigr)$  \\
soft & $ \bigl(p_\perp,p_\perp,p_\perp \bigr)$  \\
\hline\hline
\end{tabular}
\caption{Scaling of the relevant modes in the $s+$ regime of \SCETp for a \SCETb observable.}
\label{tab:modes_II_s+}
\end{table}

\begin{table}
\centering
\begin{tabular}{c|c}
\hline\hline
mode & $p^\mu=(+,-,\perp)$ \\ 
\hline
collinear ($n_2,\dots,n_N$) &  $\bigl(p_\perp^2/Q ,Q,p_\perp\bigr)$ \\
soft-collinear ($n_1$) &  $\bigl(p_\perp^2\, Q/u, u/Q ,p_\perp\bigr)$  \\
collinear-csoft ($n_t$) &  $\bigl(p_\perp\,\sqrt{t/u},p_\perp\, \sqrt{u/t},p_\perp\bigr)$ \\
soft & $ \bigl(p_\perp,p_\perp,p_\perp \bigr)$  \\
\hline\hline
\end{tabular}  
\caption{Scaling of the relevant modes in the $cs+$ regime of \SCETp for a \SCETb observable.}
\label{tab:modes_II_cs+}
\end{table}

\section{\boldmath $e^+ e^- \to 3$ jets}
\label{sec:3jets}

In this section, we discuss in detail all kinematic regimes for $e^+ e^- \to 3$ jets, considering each hierarchy in turn. The jets are again ordered according to the kinematics such that
\begin{equation}
t\equiv s_{12} < u \equiv s_{13} < s \equiv s_{23} \sim Q^2
\,,\qquad
Q^2 = s + t + u
\,.\end{equation}
As jet resolution variable we use 3-jettiness as defined in \eq{tauN}. For simplicity, in this section we use the geometric measure with $\rho_i = 1$ so $Q_i = \w_i$ and
\begin{align} \label{eq:tau3}
\Tau_3 &= \sum_k \min \bigl\{n_1 \sdt p_k,n_2 \sdt p_k,n_3 \sdt p_k\bigr\}
= \sum_i \Tau^{(i)}_3
\,.\end{align}
In the exclusive $3$-jet limit (or more precisely at leading order in the power expansion in $\Tau_3/Q$), we can uniquely associate each jet with one of the partons in the underlying hard partonic scattering process, denoted as
\begin{equation}
e^+e^- \to \kappa_1(q_1)\, \kappa_2(q_2)\, \kappa_3(q_3)
\,,\qquad
\kappa = \{\kappa_1, \kappa_2, \kappa_3\}
\,.\end{equation}
Since we label the jets by their kinematic ordering rather than their flavor, we use $\kappa$ to denote the partonic channel, which in the present case can be any permutation of $\{g,q,\bar{q}\}$ where $q$ stands for any quark flavor.

By evaluating all functions in the factorization theorems below at their natural scales and RG evolving them to the common arbitrary scale $\mu$, all kinematic logarithms of $t/Q^2$, $u/Q^2$, and $t/u$ in their respective regimes as well as the logarithms of $\Tau_3/Q$ are resummed. The perturbative ingredients required for the resummation to NNLL are fully known. We give the one-loop results for the additional \SCETp ingredients below. The required common RGE solutions and anomalous dimensions can be found for example in the appendices of Refs.~\cite{Ligeti:2008ac, Stewart:2010qs, Berger:2010xi, Bauer:2011uc}, and are not reproduced here.

\subsection{Standard SCET regime: $t \sim u \sim Q^2$}
\label{subsec:SCET3}

We first review the notation and conventions for SCET helicity operators and the matching from QCD.
We then discuss the factorization theorem for the standard SCET case where all three jets are equally energetic and well separated.

\subsubsection{Helicity operators and matching to SCET}

We start by briefly discussing SCET helicity operators~\cite{Stewart:2012yh, Moult:2015aoa, Kolodrubetz:2016uim}, which are convenient for carrying out the matching from QCD onto SCET. In particular, they make it straightforward to construct the complete operator basis in SCET with multiple collinear sectors. We summarize the necessary definitions and some basic properties here, and refer for details to ref.~\cite{Moult:2015aoa}. A summary of the common SCET notation and conventions is given in \app{SCET}.

Collinear quark and gluon jet fields in the $n_i$-collinear sector with specified helicity are defined as
\begin{align}
\chi^\alpha_{i\pm} \equiv \frac{1 \pm \gamma_5}{2} \chi^\alpha_{n_i,-\w_i} 
\,, \quad 
\mathcal{B}^a_{i\pm} \equiv - \varepsilon_{\mp\mu}(n_i,\bar{n}_i) \,\mathcal{B}^{a\mu}_{n_i,\w_i\perp} 
\,,\end{align}
which involve the polarization vectors and spinors for massless on-shell momenta
\begin{align}\label{eq:polarization_vectors}
 \varepsilon_\pm^\mu(p,k) = \pm \frac{\mae{p\pm}{\ga^\mu}{k\pm}}{\sqrt{2} \braket{k\mp}{p\pm}}
 \,, \quad
  \ket{p\pm} =\frac{1 \pm \gamma_5}{2} \,u(p)  
\,.\end{align}
Since fermions always come in pairs, we can use currents with fixed helicity as basic building blocks of helicity operators,
\begin{align}
  J_{ij\pm}^{\bar \alpha \beta} = \pm \frac{\sqrt{2} \varepsilon_\mp^{\mu}(n_i,n_j)}{\sqrt{\omega_i \,\omega_j\vphantom{i}}} \,
  \frac{\bar \chi_{i\pm}^{\bar \alpha} \ga_\mu \chi_{j\pm}^\beta}{\braket{n_i \mp}{n_j \pm}}
\,.\end{align}
The leptonic vector current is defined analogously but does not contain any QCD Wilson lines. The normalization of the fermion currents and gluon fields are chosen such that the tree-level Feynman rules for the corresponding final state give delta functions of the label momenta $\tilde p_i^\mu=\omega_i n_i^\mu/2$,
\begin{align} \label{eq:tree_helicity}
  \mae{g_\pm^{a_1}(p_1)}{\cB_{1\pm}^{b_1}}{0} &= \de^{a_1 b_1}\, \tilde \delta(\tilde p_1 - p_1)
\,, \nn \\
  \mae{q_\pm^{\al_2}(p_2) \bar q_\mp^{\bar \al_3}(p_3)}{J_{23\pm}^{\bar \beta_2 \beta_3}}{0}
  &= \delta^{\al_2 \bar \beta_2}\, \delta^{\beta_3 \bar \alpha_3}\, \tilde \delta(\tilde p_2 - p_2)\, \tilde \delta (\tilde p_3 - p_3)
\,,\end{align}
and zero otherwise. The delta function and integral of label momenta are denoted by
\begin{align}
\tilde \delta(\tilde p_i - p) \equiv\delta_{\{n_i\},p} \, \delta(\omega_i -\bar{n}_i \sdt p) \, , \qquad \int \df \tilde p_i \equiv \sum_{\{n_i\}} \int \df \omega_i \, ,
\end{align}
with $\delta_{\{n_i\},p} = 1$ if $n_i \sdt p =\mathcal{O}(\lambda^2)$ and zero otherwise.

We now match the QCD currents onto the corresponding SCET operators, resulting in the hard-scattering Lagrangian
\begin{align} \label{eq:L_SCET3}
\cL_\SCET^\hard
&= \sum_{\lambda_g, \lambda_q, \lambda_\ell} \int \prod_{i=1}^3 \df \tilde p_i\, O_{\lambda_g(\lambda_q; \lambda_\ell)}^{a\, \bar\al \bt}(\{\tilde p_i\},\mu)\, C_{\lambda_g(\lambda_q; \lambda_\ell)}^{a\, \al \bar \beta}(\{\tilde p_i\},\mu)
\,.\end{align}
Here, the helicity labels $\lambda_g, \lambda_q, \lambda_\ell = \pm$ are summed.
For $e^+e^- \to$ 3 jets, the complete (and minimal) operator basis is given in terms of the gluon fields $\cB_{1\lambda_g}^a$ and the quark and lepton currents, $J_{23 \lambda_q}^{\bar \al\bt}$ and $J_{45 \lambda_\ell}$, as
\begin{align} \label{eq:O_3}
O_{+(+;\pm)}^{a\,\bar \al\bt} &= \cB_{1+}^a\, J_{23 +}^{\bar \al\bt}\, J_{45 \pm}
\,,\qquad &
O_{-(-;\pm)}^{a\,\bar \al\bt} &= \cB_{1-}^a\, J_{23 -}^{\bar \al\bt}\, J_{45 \pm}
\,,\nn\\
O_{+(-;\pm)}^{a\,\bar \al\bt} &= \cB_{1+}^a\, J_{23 -}^{\bar \al\bt}\, J_{45 \pm}
\,,\qquad &
O_{-(+;\pm)}^{a\,\bar \al\bt} &= \cB_{1-}^a\, J_{23 +}^{\bar \al\bt}\, J_{45 \pm}
\,.\end{align}

The operators and Wilson coefficients in \eq{L_SCET3} are written as general vectors in the color space of the external partons. For $e^+e^-\to q\bar q g$, the color decomposition is trivial because there is only one allowed color structure, $T^a_{\al\bar\beta}$, so the relevant color-conserving subspace is one dimensional. However, in anticipation of the more complicated color structure for $N$-jet production, we employ the general notation
\begin{align} \label{eq:C3}
C_{\lambda_g(\lambda_q; \lambda_\ell)}^{a\, \al \bar \beta}
&= \bar T^{a\, \al \bar \beta} \cdot  
\vec C_{\lambda_g(\lambda_q; \lambda_\ell)}
= T^a_{\al\bar \beta}\,  C_{\lambda_g(\lambda_q; \lambda_\ell)}
\,,\end{align}
where
\begin{equation} \label{eq:gqq_color}
\bar T^{a\, \al \bar \beta} = \bigl( T^a_{\al\bar \beta} \bigr)
\,,\qquad
\vec C_{\lambda_g(\lambda_q; \lambda_\ell)} = \bigl( C_{\lambda_g(\lambda_q; \lambda_\ell)} \bigr)
\end{equation}
are one-dimensional row and column vectors, respectively.
We define the corresponding conjugate vector for the Wilson coefficient by
\begin{equation} \label{eq:C3dagger}
\vec C^\dagger_{\lambda_g(\lambda_q; \lambda_\ell)}
= C_{\lambda_g(\lambda_q; \lambda_\ell)}^{*\, a\, \al \bar \beta} \bar T^{a\, \al \bar \beta} 
= \vec C^{* T}_{\lambda_g(\lambda_q; \lambda_\ell)} \cdot \widehat{T}_{g\,q\bar q}
= \bigl( C^*_{\lambda_g(\lambda_q; \lambda_\ell)} N_c C_F \bigr)
\,,\end{equation}
where the color sum matrix is given by
\begin{align}
\widehat{T}_{g\,q\bar q} = (\bar T^{a_1\, \al_2 \bar \al_3})^\dagger \bar T^{a_1\, \al_2 \bar \al_3}
 = \bigl(T^{a_1}_{\al_3 \bar \al_2} T^{a_1}_{\al_2 \bar \al_3} \bigr) = \bigl( N_c C_F \bigr) = N_c C_F \mathbf{1}
\,.\end{align}
It explicitly appears in \eq{C3dagger}, because the color basis in $\bar T^{a\,\al\bar\beta}$ is not normalized. (In the $N$-jet case, the color basis will typically also not be orthogonal, so $\widehat T$ will be a nontrivial matrix.)

The matching coefficients are given in terms of the IR-finite part of the UV-renormalized QCD amplitudes.
For example, for a specific helicity configuration, the QCD amplitude is written as
\begin{equation}
\cA(0\to g_1^+ q_2^+ \bar q_3^- \ell_4^\pm \bar\ell_5^\mp) = \img T^{a_1}_{\al_2 \bar \al_3}\,  A(1^+; 2_q^+,3_{\bar q}^-;4_\ell^\pm,5_{\bar \ell}^\mp)
\,.\end{equation}
Using dimensional regularization as UV and IR regulator, the corresponding UV-renormalized amplitude in SCET is given to all orders in $\alpha_s$ by
\begin{align} \label{eq:hel_matching}
\cA_\SCET &\equiv
\Mae{g_+^{a_1}(q_1)q_+^{\al_2}(q_2) \bar q_-^{\bar \al_3}(q_3)}{\img \cL_\SCET^\hard}{\bar \ell_\mp(-q_4) \ell_\pm(-q_5)}
\nn \\ 
&= 
 \int \prod_{i=1}^5 \df \tilde p_i\,
   \Mae{g_+^{a_1}(q_1)q_+^{\al_2}(q_2) \bar q_-^{\bar \al_3}(q_3)}{O^{a\, \bar \alpha \beta}_{+(+;\pm)}(\{\tilde p_i\})\,
   }{\bar \ell_\mp(-q_4) \ell_\pm(-q_5)}\,\img C^{a\, \bar \alpha \beta}_{+(+;\pm)} (\{\tilde p_i\})
\nn \\
&=  \img T^{a_1}_{\al_2 \bar \al_3}\, \frac{1}{Z_{g\,q\bar q} (\{q_i\})}\,C_{+(+;\pm)}(\{q_i\})
\,.\end{align}
Here we used the SCET counterterm $Z_{g\,q\bar q}$ defined by%
\footnote{In general, $\widehat Z_\kappa$ is a matrix in color space defined by
$[\vec{O}^\dagger]^{\rm bare} = \vec{O}^\dagger \widehat Z_O Z_\xi^{-n_q/2} Z_A^{-n_g/2} \equiv\vec{O}^\dagger \widehat Z_\kappa$.}
\begin{align}\label{eq:Z_3}
\bigl[O^{a\, \bar \alpha \beta}_{+(+;\pm)}\bigr]^{\rm bare}(\{\tilde p_i\})
= O^{a\, \bar \alpha \beta}_{+(+;\pm)}(\{\tilde p_i\})\, \frac{Z_{O_{g\,q\bar q}}(\{\tilde p_i\})}{Z_\xi \sqrt{Z_A}}
\equiv  O^{a\, \bar \alpha \beta}_{+(+;\pm)}(\{\tilde p_i\})\, Z_{g\,q\bar q} (\{\tilde p_i\})
\,,\end{align}
together with the fact that the matrix element of the bare operator is given by the tree-level result, since all loop graphs in the effective field theory are scaleless and vanish in pure dimensional regularization. Note that we use an outgoing convention, which is why the momentum, spin, and particle type for the incoming leptons in \eq{hel_matching} are reversed. Requiring the QCD and SCET amplitudes to be equal, the IR divergences cancel between them, implying that to all orders in $\alpha_s$ we have
\begin{align} \label{eq:C_is_A}
 C_{+(+;\pm)}(\{q_i\})  = \lim_{\eps\to 0}\bigl[ Z_{g\,q\bar q} (\{q_i\})\, A(1^+; 2_q^+,3_{\bar q}^-;4_\ell^\pm,5_{\bar \ell}^\mp) \bigr]
\,.\end{align}

\subsubsection{Factorization theorem}

The factorization theorem in SCET for $3$-jettiness is given by~\cite{Stewart:2010tn}
\begin{align}\label{eq:factorization_SCET}
 \frac{\df \sigma_{\SCET}}{\df t \,\df u \,  \prod_{i=1}^3 \df \Tau^{(i)}_3 }
&=  \sum_{\kappa} \biggl[\prod_{i=1}^3\int\! \df s_i \,  J_i(s_i,\mu)\biggr]
\tr\biggl[{\widehat H}_\kappa (t,u,Q^2,\mu)\,
   {\widehat S}_\kappa\Bigl(\Bigl\{\Tau^{(i)}_3\!-\frac{s_i}{\w_i}\Bigr\},\{n_i \sdt n_j\},\mu\Bigr) \biggr]
\nn \\ & \quad \times 
  \biggl[1 +\ORD{\frac{m_J^2}{Q^2}}\biggr]
\,.\end{align}
The partonic channel $\kappa=\{\kappa_1,\kappa_2,\kappa_3\}$ is summed over all six permutations of $\{g,q,\bar{q}\}$ and also over the desired quark flavors. Label momentum conservation
\begin{equation} \label{eq:label_conservation}
(Q, \vec{0}) = q_1^\mu + q_2^\mu + q_3^\mu = \w_1\, \frac{n_1^\mu}{2} + \w_2\, \frac{n_2^\mu}{2} + \w_3\, \frac{n_3^\mu}{2}
\end{equation}
fixes the energies of the jets, yielding
\begin{align}\label{eq:label_momenta}
\w_1=Q-\frac{s}{Q} = \frac{t+u}{Q}
\,, \qquad
\w_2=Q-\frac{u}{Q}
\,, \qquad
\w_3=Q-\frac{t}{Q}
\,.\end{align}

Since there is only one color structure for this process, the hard and soft functions in \eq{factorization_SCET} are simply one-dimensional matrices and the trace is over this trivial color space,
\begin{equation}
\widehat H_\kappa = \bigl( H_\kappa \bigr)
\,,\qquad
\widehat S_\kappa = \bigl( S_\kappa \bigr)
\,,\qquad
\tr\bigl[ \widehat H_\kappa\,\widehat S_\kappa \bigr] = H_\kappa S_\kappa
\,.\end{equation}
We employ the matrix notation to make the generalization to the $N$-jet case straightforward. We drop the hats whenever we refer to the matrix components.

The hard function $\widehat H_\kappa$ describes the physics at the hard-interaction scale $\mu_H \sim Q$. It is given in terms of the Wilson coefficients as
\begin{align}\label{eq:normalization_matching}
\widehat{H}_{\{g,q,\bar{q}\}}(t,u,Q^2,\mu)
&= \frac{1}{4Q^4 (4\pi)^3}\,
\frac{1}{4} \sum_{\lambda_\ell}
\sum_{\lambda_g,\lambda_q}\!\! \Bigl\langle \vec C_{\lambda_g(\lambda_q; \lambda_\ell)}(\{q_1,q_2,q_3,-q_{\bar \ell},-q_\ell\},\mu)
\nn \\ & \quad \times
\vec C^\dagger_{\lambda_g(\lambda_q; \lambda_\ell)}(\{q_1,q_2,q_3,-q_{\bar \ell},-q_\ell\},\mu) \Bigr\rangle_{\ell\bar\ell}
\nn \\
&= \frac{1}{4Q^4 (4\pi)^3}\,
\frac{1}{4} \sum_{\lambda_\ell}
\sum_{\lambda_g,\lambda_q} \!\! \Bigl\langle \Abs{C_{\lambda_g(\lambda_q; \lambda_\ell)}(\{q_i\}, \mu)}^2
\Bigr\rangle_{\ell\bar\ell}\, N_c C_F \mathbf{1}
\,,\end{align}
where we included the flux factor $1/(2Q^2)$, averaged over the spins of the incoming leptons, and included the prefactor of the $3$-body phase space
\begin{align}
\int\! \df\Phi_3 = \frac{1}{2 Q^2 (4\pi)^3} \int\! \df t\, \df u
\,.\end{align}
Since we do not keep track of any angular dependence between the beam directions and final-state jet axes, we have averaged over the directions of the incoming leptons indicated by $\langle \dots \rangle_{\ell\bar\ell}$.
The results for the other partonic channels can easily be obtained via crossing symmetry,
\begin{align}
\widehat{H}_{\{g,\bar{q},q\}}(t,u,Q^2,\mu) &=\widehat{H}_{\{g,q,\bar{q}\}}(t,u,Q^2,\mu)
\,, \nn \\
\widehat{H}_{\{q,g,\bar{q}\}}(t,u,Q^2,\mu) &=\widehat{H}_{\{\bar{q},g,q\}}(t,u,Q^2,\mu)=\widehat{H}_{\{g,q,\bar{q}\}}(t,Q^2-t-u,Q^2,\mu) 
\,, \nn \\
\widehat{H}_{\{q,\bar{q},g\}}(t,u,Q^2,\mu) &=\widehat{H}_{\{\bar{q},q,g\}}(t,u,Q^2,\mu)=\widehat{H}_{\{g,q,\bar{q}\}}(Q^2-t-u,u,Q^2,\mu) 
\,.\end{align}
The tree-level hard function is given by
\begin{align}
H_{\{g,q,\bar{q}\}}^{(0)}(t,u,Q^2,\mu)
&= \sigma_{0,q}\, \frac{\alpha_s(\mu) C_F}{2\pi}\, \frac{(Q^2-t)^2+(Q^2-u)^2}{Q^4 \, t \, u}
\,,\end{align}
where we have pulled out the tree-level cross section for $e^+e^- \to q \bar q$,
\begin{equation} \label{eq:sigma0q}
\sigma_{0,q}
= \frac{4\pi \alem^2\,N_c}{3 Q^2}
\biggl[ Q^2_{\ell}\, Q_q^2 + \frac{(v_q^2 + a_q^2) (v_\ell^2+a_\ell^2) - 2  Q_{\ell}\,  Q_q\, v_q v_\ell (1-m_Z^2/Q^2)}
{(1-m_Z^2/Q^2)^2 + m_Z^2 \Gamma_Z^2/Q^4} \biggr]
\,.\end{equation}
Here, $\alem$ is the electromagnetic coupling, $Q_{\ell,q}$ are the lepton and quark charges, $v_{\ell,q}$ and $a_{\ell,q}$ are the vector and axial couplings of the leptons and quarks to the $Z$ boson, and $m_Z$ and $\Ga_Z$ are the mass and width of the $Z$ boson. The one-loop hard function can be extracted from the one-loop virtual corrections for $\abs{\cA(gq\bar q)}^2$ in ref.~\cite{Ellis:1980wv}, or directly from the one-loop helicity matching coefficients given e.g. in ref.~\cite{Moult:2015aoa}.

The $J_i \equiv J_{\kappa_i}$ in \eq{factorization_SCET} are the inclusive quark and gluon jet functions in SCET, which are known to $\ord{\alpha_s^2}$~\cite{Bauer:2003pi, Fleming:2003gt, Becher:2006qw, Becher:2009th, Becher:2010pd}. They determine the contribution to the measurement from collinear radiation at the scale $\mu_J \sim \sqrt{Q\Tau_3}$. At tree level, $J_i^\zero(s_i,\mu) = \de(s_i)$.

The soft function $\widehat{S}_\kappa$ in \eq{factorization_SCET} determines the contribution to the measurement from usoft radiation at the scale $\mu_S \sim \Tau_3$. It is a matrix element containing three usoft Wilson lines in the directions $n_1$, $n_2$, and $n_3$ in the appropriate color representation. For example, for the $\{g,q,\bar{q}\}$ channel
\begin{align} \label{eq:S_3}
\widehat{S}_{\{g,q,\bar{q}\}}(\{\ell_i\},\{n_i \sdt n_j\},\mu)
&= \widehat{T}^{-1}_{g\,q\bar q}\, (\bar T^{a_1\, \al_2 \bar \al_3})^\dagger\,
S_{\{g,q,\bar{q}\}}^{a_1 \al_2 \bar \al_3\, b_1 \bar \bt_2 \bt_3}(\{\ell_i\},\{n_i \sdt n_j\},\mu) \, \bar T^{b_1\, \bt_2 \bar \bt_3} 
\nn \\
&= \frac{\mathbf{1}}{N_c C_F}\sum_{X_s} \tr\Bigl[\Mae{0}{\bar{T} \bigl[Y_{n_3}^\dagger Y_{n_1} T^a Y^\dagger_{n_1} Y_{n_2}\bigr]}{X_s}
\nn \\ & \quad \times
\Mae{X_s}{T \bigl[Y_{n_2}^\dagger Y_{n_1} T^a Y^\dagger_{n_1} Y_{n_3} \bigr]}{0} \Bigr]
\prod_{i=1}^3 \delta\big(\ell_i-n_i \sdt k^{(i)}_{s}\big) 
\,, \end{align}
where $k^{(i)}_{s}$ denotes the momentum of the soft state $X_s$ in the $i$th jet region.
At tree level, the soft function is given by $S_{\{g,q,\bar{q}\}}^\zero = \de(\ell_1) \de(\ell_2) \de(\ell_3)$.
In ref.~\cite{Jouttenus:2011wh}, the $N$-jettiness soft function for general $N$ was calculated at one loop and the all-order form of its anomalous dimension was derived. A procedure to extend this calculation to two loops has been described in ref.~\cite{Boughezal:2015eha}.

\subsection{$c+$ regime: $t \ll u \sim Q^2$}
\label{subsec:SCETc}

We now discuss the case where jets 1 and 2 come close together, which was already discussed in ref.~\cite{Bauer:2011uc}. As discussed in \subsec{c+overview}, we first match QCD onto SCET with two collinear sectors with label directions $n_t$ and $\bn_t \equiv n_3$ and virtuality $\sim \sqrt{t}$. At this scale, the two nearby jets are not yet resolved. The relevant operators in this theory are those for $e^+e^-\to q\bar q$,
\begin{equation} \label{eq:O_2}
O_{(+;\pm)}^{\bar \al\bt}
= J_{t3+}^{\bar \al\bt}\, J_{45\pm}
\,, \qquad
O_{(-;\pm)}^{\bar \al\bt}
= J_{t3-}^{\bar \al\bt}\, J_{45\pm}
\,.\end{equation}
This process also has a unique color structure,
\begin{equation} \label{eq:C_2}
C_{(\lambda_q; \lambda_\ell)}^{\al \bar \beta}
= \bar T^{\al \bar \beta} \cdot \vec C_{(\lambda_q; \lambda_\ell)}
= \de_{\al\bar \beta}\, C_{(\lambda_q; \lambda_\ell)}
\,, \qquad
\bar T^{\al \bar \beta} = \bigl(\de_{\al\bar \beta} \bigr)
\,, \quad
\vec C_{(\lambda_q; \lambda_\ell)} = \bigl( C_{(\lambda_q; \lambda_\ell)} \bigr)
\,,\end{equation}
with the corresponding color sum matrix given by
\begin{equation}
  \widehat{T}_{q\bar q} = (\bar T^{\al_1 \bar \al_2})^\dagger \bar T^{\al_1 \bar \al_2}
 = \bigl(\de_{\bar \al_1 \al_2} \de_{\al_1 \bar \al_2} \bigr) = \bigl( N_c \bigr) \equiv N_c\,\mathbf{1}
\,.\end{equation}
The matching coefficients are directly related to the IR-finite part of the $e^+e^-\to q\bar q$ amplitudes, in analogy to \eq{hel_matching}.

After decoupling the usoft degrees of freedom in the parent SCET, the two nearby jets are resolved in one of the collinear sectors at the scale $\mu \sim \sqrt{t}$. If the gluon jet is close to the quark jet, this corresponds to matching the $n_t$-collinear sector of the parent SCET onto the $n_1$-collinear, $n_2$-collinear, and $n_t$-csoft sectors of the $c+$ regime of \SCETp,
\begin{align} \label{eq:SCETc_matching}
\bar \chi_{t\lambda_q}^{\bar \al}
= (\bar \xi_{t\lambda_q} W_{n_t})^{\bar \al}
= \sum_{\lambda_g} \int\! \df\tilde p_1\, \df \tilde p_2\,
  C_{c,\lambda_q\lambda_g}^{a\,\bt \bar \ga}(\bn_t,\tilde p_1,\tilde p_2,\mu)\,
(\mathcal{X}_{n_1} \cB_{1\lambda_g})^a 
  (\bar \chi_{2\lambda_q}X^\dagger_{n_2})^{\bar \bt}  V_{n_t}^{\ga \bar\al}
\,,\end{align}
and similarly for the case of the gluon jet being close to the antiquark jet. Note that this matching preserves the helicity of the (anti)quark field. Equation~\eqref{eq:SCETc_matching} leads to matching the SCET $q\bar q$ operators in \eq{O_2} onto \SCETp $gq\bar q$ operators, which have the same helicity structure as in \eq{O_3}, but are dressed with additional csoft Wilson lines.
The csoft Wilson lines $X_{n_2}$ and $\mathcal{X}_{n_1}$ sum the emissions of csoft gluons $A_{n_t}$ from $\bar \chi_{n_2}$ and $\cB_{n_1}$. They arise from the field redefinition of the $n_{1,2}$-collinear fields decoupling them from the csoft modes. The $V_{n_t}$ Wilson line sums the csoft emissions from the remaining collinear sector(s). It can be interpreted as the csoft remnant of the collinear $W_{n_t}$ in the parent $n_t$-collinear sector. The csoft Wilson lines are defined as
\begin{align}\label{eq:Vn_Xn} 
V_{n_t}&  = {\rm \overline{P}} \, \exp\biggl[-\img g \int_0^\infty\! \df s \, \bar{n}_t \sdt A_{n_t}(s \bar{n}_t^\mu)\biggr]
= \biggl[\sum_{\rm perms} {\rm exp}\Bigl(\frac{-g}{\bar{n}_t \sdt \mathcal{P}_{n_t}} \, \bar{n}_t \sdt A_{n_t}\Bigr) \biggr]
\, , \nn \\
X_{n_2}
&= {\rm \overline{P}} \, \exp\biggl[-\img g  \int_0^\infty\! \df s \, n_2 \sdt A_{n_t}(s n_2^\mu)\biggr]
\,,\end{align}
with an analogous expression for $\mathcal{X}_{n_1}$ in the adjoint representation. The label momentum operator $\mathcal{P}_{n_t}$ in the first line acts only inside the square brackets.

Performing the color decomposition using the same color basis as in \eq{gqq_color}, we can write the matching coefficient as%
\footnote{The fact that the collinear matching coefficient $C_c^{a\,\beta\bar\gamma}$ only depends on the color space of the $1\to 2$ splitting is a direct consequence of the usoft-collinear factorization in SCET, which implies that the matching in \eq{SCETc_matching} only involves a single collinear usoft-decoupled sector. As discused in detail in ref.~\cite{Bauer:2011uc}, this also holds for the general $N$-jet case and is equivalent to the factorization of QCD amplitudes in the collinear limit in terms of universal splitting amplitudes. The results of ref.~\cite{Almelid:2015jia}, which are based on a partial $3$-loop calculation supplemented by consistency arguments in the high-energy limit, indicate that this collinear factorization of amplitudes might be violated. If this result confirmed by the complete $3$-loop calculation, it would require a more general matching condition than in \eq{SCETc_matching} involving all recoiling collinear sectors and $C_c$ would then become a general matrix from $N-1$ to $N$-parton color space. This would require the explicit non-cancellation of Glauber effects in a hard-scattering calculation at 3 loops in order to connect the different collinear sectors in the parent SCET, which would be quite unexpected.\\\emph{Note added:} After initial submission of our paper a revised version of ref.~\cite{Almelid:2015jia} appeared,
which now confirms collinear factorization at three loops, as expected from and consistent with the \SCETp matching in \eq{SCETc_matching}.}
\begin{align} \label{eq:Cc_simplify}
  C_{c,\lambda_q\lambda_g}^{a\,\bt \bar \ga}(\bn_t,\tilde p_1,\tilde p_2,\mu) =
  \bar T^{a\,\bt \bar \ga} \cdot  \vec C_{c,\lambda_q\lambda_g}
  \Bigl(2 \tilde p_1 \sdt \tilde p_2,\frac{\bn_t \sdt \tilde p_1}{\bn_t \sdt (\tilde p_1 + \tilde p_2)}, \mu\Bigr)
\,.\end{align}
We also used that reparametrization invariance~\cite{Manohar:2002fd} implies that to all orders $\vec C_c$ can only depend on
$t = 2 \tilde p_1 \sdt \tilde p_2$ and the lightcone momentum fraction $z = \bn_t \sdt \tilde p_1 / \bn_t \sdt (\tilde p_1 + \tilde p_2)$~\cite{Bauer:2011uc}.
Since it only depends on a single dimensionful scale, all large logarithms of $t$ must appear as $\ln(t/\mu^2)$, which can thus be minimized by choosing the natural scale $\mu_{H_c} \sim\sqrt{t}$.

The two nearby jets originate from collinear emissions in the $n_t$-collinear sector of the parent SCET. The total momentum $p_t^\mu = Q n_t^\mu/2 + k^\mu$ of the $n_t$-collinear sector includes a residual momentum component $k^\mu\sim \ord{t/Q}$, which is responsible for generating the small dijet invariant mass $t$. For a single collinear emission at tree level, momentum conservation reads
\begin{align}
p_t^\mu
&= (Q + k^-)\frac{n_t^\mu}{2} + k^+ \frac{\bn_t^\mu}{2} = p_1^\mu + p_2^\mu
\,,\nn\\
p_1^\mu
&= z (Q + k^-)\frac{n_t^\mu}{2} + (1 - z) k^+ \frac{\bn_t^\mu}{2} + k_\perp^\mu
\,,\nn\\
p_2^\mu
&= (1-z) (Q + k^-)\frac{n_t^\mu}{2} + z k^+ \frac{\bn_t^\mu}{2} - k_\perp^\mu
\,,\end{align}
where $k_\perp^2 = -z(1-z)(Q+k^-)k^+$ such that $p_1^2 = p_2^2 = 0$. From the point of view of \SCETp, this corresponds to the hard splitting process that determines the large jet momenta corresponding to the \SCETp label momenta. However, since SCET already contains a power expansion in $t/Q^2$, the observed jet momenta and dijet invariant masses can only be computed up to relative $\ord{t/Q^2}$ corrections. Choosing $\vec n_1 = \vec p_1 / \abs{\vec p_1}$ and $\vec n_2 = \vec p_2 / \abs{\vec p_2}$, we thus have
\begin{align} \label{eq:SCETq12}
p_1^\mu &= z Q \Bigl[1 + \ORd{\frac{t}{Q^2}}\Bigr]\,\frac{n_1^\mu}{2}
\,,\qquad
p_2^\mu = (1-z) Q \Bigl[1 + \ORd{\frac{t}{Q^2}}\Bigr] \frac{n_2^\mu}{2}
\,,\nn\\
t &= Q k^+ \Bigl[1 + \ORd{\frac{t}{Q^2}}\Bigr] = z(1-z)Q^2\,\frac{n_1 \sdt n_2}{2} \Bigl[1 + \ORd{\frac{t}{Q^2}}\Bigr]
\,.\end{align}
Once we take the final matrix element in \SCETp the measurement identifies $\tilde p_1^\mu \equiv q_1^\mu$ and $\tilde p_2^\mu \equiv q_2^\mu$ and the \SCETp label momentum conservation reads
\begin{equation}
p_t^\mu = q_1^\mu + q_2^\mu = \w_1\,\frac{n_1^\mu}{2} + \w_2\,\frac{n_2^\mu}{2}
\,,\qquad
\w_1 = zQ
\,,\qquad
\w_2 = (1-z)Q
\,,\end{equation}
where the $\ord{t/Q^2}$ corrections in \eq{SCETq12} can be absorbed into the residual components of $p_t^\mu$.

A similar discussion applies to the $\bn_t$-collinear sector,
\begin{align} \label{eq:SCETq3}
p_{\bar t}^\mu &= Q\,\frac{\bn_t^\mu}{2} + k^\mu = p_3^\mu = Q \Bigl[1 + \ORd{\frac{t}{Q^2}}\Bigr] \frac{n_3^\mu}{2}
\,,\nn\\
u &= 2 p_1\cdot p_3 = z Q^2 \Bigl[1 + \ORd{\frac{t}{Q^2}}\Bigr]
\,,\end{align}
where we chose $\vec n_3 = \vec p_3/\abs{\vec p_3}$ in the first line. The \SCETp label momentum conservation simply becomes
\begin{equation}
p_{\bar t}^\mu = q_3^\mu = \w_3\, \frac{n_3^\mu}{2}
\,,\qquad
\w_3 = Q
\,.\end{equation}
Even though this appears trivial, it is important to remember that upon matching from SCET onto \SCETp, the label momentum gets refined from $\ord{t/Q}$ to $\ord{\Tau_3}$. In particular, the parent $\bn_t$ direction corresponds to a wider equivalence class of collinear directions than the final $n_3$ direction in \SCETp, and we identified $\bn_t \equiv n_3$ from the start only for convenience.

Since $q_3$ is unaffected by the details of the nearby jets 1 and 2, we could carry out the matching in \eq{SCETc_matching} independent of the third jet. In other words, while the residual momentum conservation in the splitting of the $n_t$-collinear sector is important for determining the $n_1$ and $n_2$ directions, there is no residual momentum conservation between the $n_t$ and $\bn_t = n_3$ collinear sectors and thus also no recoil because of the power expansion in SCET.

As expected, the above label momenta correspond to the $t \ll u \sim Q^2$ limit of \eq{label_momenta}. As shown in ref.~\cite{Bauer:2011uc}, the intrinsic $\ord{t/Q^2}$ ambiguity in the $q_i$ also allows one to implement the full kinematic dependence in \eqs{label_conservation}{label_momenta} in \SCETp. This effectively incorporates some kinematic $\ord{t/Q^2}$ nonsingular corrections from the well-separated SCET regime. However, to maintain the exact consistency of the factorization theorem, one has to be careful to incorporate the corresponding recoil effect also in the parent SCET.
The above discussion shows explicitly that the power corrections of $\ord{t/Q}$ in the label momenta can be consistently dropped in the derivation of the factorization theorem, which we therefore do here.

The remaining steps are mainly related to the factorization of the measurement and are discussed in ref.~\cite{Bauer:2011uc}. The resulting factorization theorem for the $c+$ regime is given by
\begin{align}\label{eq:factorization_c+}
\frac{\df \sigma_{c+}}{\df t \,\df u  \prod_{i=1}^3 \df \Tau^{(i)}_3 }
&= \sum_{\kappa_c} \tr\biggl[ \widehat H_{c,\kappa_c}\Bigl(t,\frac{u}{Q^2},\mu\Bigr) \int\! \df k_1 \, \df k_2\, \widehat S_{c,\kappa_c}(k_1,k_2 ,\mu)\biggr]
\biggl[\prod_{i=1}^3\int\!\df s_i \,  J_i(s_i,\mu)\biggr]
\nn \\ & \quad \times 
\tr\biggl[\widehat H_{q\bar q}(Q^2,\mu) \widehat S_{q \bar q}\Bigl(\Tau^{(1)}_3\!-\!\frac{s_1}{\w_1}\!-\!\sqrt{\hat{s}_t} \, k_1,\Tau^{(2)}_3\!-\!\frac{s_2}{\w_2}\!-\! \sqrt{\hat{s}_t}  \, k_2, \Tau^{(3)}_3\!-\!\frac{s_3}{\w_3} ,\mu\Bigr) \biggr]
  \nn \\ & \quad \times
  \biggl[1+\ORd{\frac{t}{u},\frac{m_J^2}{t}}\biggr]
\,,\end{align}
where we have used
\begin{align}
\hat{s}_t = \frac{n_1 \sdt n_2}{2} = \frac{t}{\w_1 \w_2} = \frac{t}{z(1-z)Q^2}
\,,\qquad
z = \frac{u}{Q^2}
\,,\end{align}
and the kinematic ordering of the jets implies $z<1/2$. The partonic channel is now separated as
\begin{equation}
\begin{aligned}
e^+e^-\to\, &\kappa_t(q_t)\, \kappa_3(q_3)
\,,\qquad &
\kappa_2 &= \{\kappa_t, \kappa_3\}
\,,\\
&\kappa_t(q_t) \to \kappa_1(q_1)\,\kappa_2(q_2)
\,,\qquad &
\kappa_c &= \{\kappa_t; \kappa_1, \kappa_2\}
\,,\end{aligned}
\end{equation}
where $\kappa_2$ is either $\{q, \bar q\}$ or $\{\bar q, q\}$. The sum over $\kappa_c$ runs over $\{q_t; g, q\}$, $\{q_t; q, g\}$, $\{\bar q_t; g, \bar q\}$, $\{\bar q_t; \bar q, g\}$ for all desired quark flavors, which already includes the two cases for $\kappa_2$. The jet functions $J_i(s_i)$ are the same as before in \eq{factorization_SCET}, all of them having the same natural jet scale $\mu_J \sim \sqrt{Q\Tau_3}$. The two color traces in \eq{factorization_c+} are over different color spaces, which are both still trivial here. The one-loop results for the hard and soft functions have been computed in ref.~\cite{Bauer:2011uc}, and for completeness we reproduce them here. For an explicit derivation of their RGEs, anomalous dimensions, and consistency we refer to ref.~\cite{Bauer:2011uc}.

The dijet hard function $\widehat H_{q\bar q}(Q^2,\mu)$ has the natural hard scale $\mu_H \sim Q$. It is related to the matching coefficients $\vec C_{(\lambda_q; \lambda_\ell)}$ via
\begin{align} \label{eq:Hqq}
\widehat H_{q\bar q}(Q^2,\mu)
&=\frac{1}{16\pi Q^2}\,  \frac{1}{4} \sum_{\lambda_\ell} \sum_{\lambda_q}
\Bigl\langle \vec C_{(\lambda_q; \lambda_\ell)}(\{q_1, q_2,-q_{\bar \ell},-q_\ell\},\mu)\,
\vec C^\dagger_{(\lambda_q; \lambda_\ell)}(\{q_1, q_2,-q_{\bar \ell},-q_\ell\},\mu) \Bigr\rangle_{\!\ell \bar \ell}
\nn \\
&=\frac{1}{16\pi Q^2}\,  \frac{1}{4} \sum_{\lambda_\ell} \sum_{\lambda_q}
\Bigl\langle \Abs{C_{(\lambda_q; \lambda_\ell)}(\{q_1, q_2,-q_{\bar \ell},-q_\ell\},\mu)}^2 \Bigr\rangle_{\!\ell \bar \ell}
\,\,N_c\,\mathbf{1}
\,,\end{align}
where we included the flux factor $1/(2Q^2)$, averaged over the spins and relative directions of the incoming leptons, and included $1/(8\pi)$ from the two-body phase space. Up to one loop it is given by~\cite{Manohar:2003vb, Bauer:2003di} 
\begin{equation} \label{eq:H_2}
H_{q\bar q}(Q^2,\mu) = \sigma_{0,q} \biggl\{1+  \frac{\alpha_s(\mu) C_F}{2\pi} \biggl[-\ln^2 \Bigl(\frac{Q^2}{\mu^2}\Bigr)
+ 3 \ln \Bigl(\frac{Q^2}{\mu^2}\Bigr) - 8 + \frac{7\pi^2}{6}\biggr] + \ord{\al_s^2}\biggr\}
\,,\end{equation}
where the tree-level result $\sigma_{0,q}$ is given in \eq{sigma0q}.

The functions $\widehat H_{c,\kappa_c}$ contain the collinear splitting and their natural scale is $\mu_{H_c} \sim \sqrt{t}$. They are related to the matching coefficients $\vec C_{c,\lambda_q \lambda_g}$ in \eq{Cc_simplify} via
 \begin{align}\label{eq:H_c}
 \widehat H_{c,\{q;g,q\}}(t,z,\mu)
 &=\frac{1}{(4\pi)^2 Q^2 N_c}\,
\sum_{\lambda_g} \vec C_{c,\lambda_q\lambda_g}(t,z,\mu) \, \vec C^\dagger_{c,\lambda_q\lambda_g}(t,z,\mu) 
\nn \\
 &=\frac{1}{(4\pi)^2 Q^2 N_c}\, 
\sum_{\lambda_g} \Abs{C_{c,\lambda_q\lambda_g}(t,z,\mu)}^2\, N_c C_F\,\mathbf{1}
\,,\end{align}
where we averaged over the color of the initiating quark (but not its spin which is fixed) and included the prefactor from the two-body collinear phase space
\begin{align}
 \int\! \df \Phi_c = \frac{1}{(4\pi)^2 Q^2} \int\! \df t\, \df u
\,.\end{align}
Since the virtual corrections in \SCETp are scaleless and vanish in pure dimensional regularization, the $\vec C_c$ are equivalent (up to overall normalization) to the IR-finite parts of the universal collinear splitting amplitudes~\cite{Berends:1987me, Mangano:1990by, Bern:1994zx, Kosower:1999xi, Kosower:1999rx, Bern:1999ry, Sborlini:2013jba}. This is completely analogous to the discussion for the full amplitudes leading to \eq{C_is_A}. For the same reason, the $\vec C_c$ can also be computed directly from the collinear matrix elements in SCET, see \app{H+} and \eq{H3c_def} for the explicit prescription. The NLO result is given by
\begin{align}\label{eq:H3c_oneloop}
H_{c,{\{q; g,q\}}}(t,z,\mu)
&= \frac{\alpha_s(\mu) C_F}{2\pi} \, \frac{1+(1-z)^2}{z \, Q^2 t} \biggl\{1 +
\frac{\alpha_s(\mu)}{2\pi}\biggl[
-\frac{C_A}{2}\biggl(\ln^2\Bigl(\frac{t\,z}{\mu^2}\Bigr) + 2\, \Li_2(1-z) - \frac{7\pi^2}{6} \biggr)
\nn \\ & \quad
+ \Bigl(\frac{C_A}{2} - C_F\Bigr) \biggl(2\ln\Bigl(\frac{t}{\mu^2}\Bigr) \ln(1-z)+\ln^2(1-z) +2\, \Li_2(z)\biggr)
\nn \\ & \quad
+ (C_A-C_F)\, \frac{z}{1+(1-z)^2}\biggr] + \ord{\al_s^2} \biggr\}
\,,\end{align}
which can be obtained from the NLO splitting amplitudes given e.g.~in refs.~\cite{Bern:1994zx,Bern:1999ry, Sborlini:2013jba}, or alternatively from the collinear limit of the hard function for three well-separated jets,
\begin{align} 
H_{\{g,q,\bar{q}\}}(t,u,Q^2,\mu) \Big\vert_{t\ll u \sim Q^2}
= H_{q\bar{q}}(Q^2,\mu) \, H_{c,\{q; g,q\}}\Bigl(t,\frac{u}{Q^2},\mu\Bigr)\Bigl[1+ \ORd{\frac{t}{u}}\Bigr]
\,.\end{align}
The results for the other quark-initiated channels are related via
\begin{align}\label{eq:H_c_crossing}
\widehat H_{c,\{\bar q; g,\bar{q}\}}(t,z,\mu)
= \widehat H_{c,\{q;q,g\}}(t,1-z,\mu)
= \widehat H_{c,\{\bar{q};\bar q,g\}}(t,1-z,\mu)
= \widehat H_{c,\{q;g,q\}}(t,z,\mu)
\,.\end{align}
Since the collinear quark splitting amplitudes are independent of $\lambda_q$, all spin correlations between the hard interaction at the scale $\mu \sim Q$ and the splitting process $q \to qg$ or $\bar{q} \to \bar{q} g$ at the scale $\mu \sim \sqrt{t}$ drop out in \eq{factorization_c+}, and this is why we could sum over $\lambda_q$ in \eq{Hqq}. For collinear gluon splittings $g \to gg$ and $g \to q\bar{q}$, $\vec{C}_c$ does depend on the helicity of the initiating gluon, so that spin correlations between the hard sectors need to be taken into account (see \subsec{scet_c}). These would be relevant here if we were to also consider $e^+e^-\to gg$.

The csoft function for the splitting channels $\kappa_c={\{q; g,q\}}$ and ${\{\bar q; g,\bar{q}\}}$ is defined as
\begin{align}\label{eq:csoftfct}
\widehat S_{c,\{q;g,q\}}(k_1,k_2,\mu)
&= \widehat{T}_{g\,q\bar q}^{-1}\,
 (\bar T^{a_1\, \al_2 \bar \al_t})^\dagger S_{c,\{q;g,q\}}^{a_1 \al_2 \bar \al_t\, b_1 \bar \bt_2 \bt_t}(k_1,k_2,\mu) \bar T^{b_1\, \bt_2 \bar \bt_t}
\nn \\
&= \frac{\mathbf{1}}{N_c C_F} \sum_{X_{cs}} \tr\Bigl[
   \Mae{0}{\bar{T}\bigl[ V^\dagger_{n_t}  X_{n_1} T^a X_{n_1}^{\dagger} X_{n_2}\bigr]}{X_{cs}}
\\ \nn & \quad \times
   \Mae{X_{cs}}{ T\bigl[ X^\dagger_{n_2} X_{n_1} T^a X_{n_1}^{\dagger} V_{n_t} \bigr]}{0} \Bigr]
\delta\biggl(k_1 - \frac{n_1 \sdt k^{(1)}_{cs}}{\sqrt{\hat{s}_t}} \biggr)
\delta\biggl(k_2 - \frac{n_2 \sdt k^{(2)}_{cs}}{\sqrt{\hat{s}_t}} \biggr)
\,,\end{align}
which we decomposed in the color basis of \eq{Cc_simplify}. The $k_{cs}^{(i)}$ denote the momentum of the csoft state $X_{cs}$ in the $i$th jet region. The csoft function in general depends on the directions $n_1$, $n_2$ (through the measurement and the $X_{n_i}$) and $\bn_t$ (through $V_{n_t}$). Using reparametrization invariance one can show~\cite{Bauer:2011uc} that the only parametric scale the csoft function can depend on is $k_i = n_i\cdot k_{cs}^{(i)}/\sqrt{\hat s_t}$.%
\footnote{Compared to ref.~\cite{Bauer:2011uc}, we have rescaled the argument $k_i$ of $S_c$ by $\sqrt{\hat{s}_t}$, such that the explicit dependence on $\hat s_t$ drops out in $\hat S_c$, as reparametrization invariance implies, and instead appears in the factorization theorem in \eq{factorization_c+} through the convolution argument of the usoft function.}
Its natural scale is thus $\mu_{S_c}\sim k_i \sim \Tau_3/\sqrt{\hat s_t} \sim \Tau_3 Q/\sqrt{t}$.
The one-loop result is
\begin{align}\label{eq:Sc_oneloop}
S_{c,\{q;g,q\}}(k_1,k_2,\mu)
&= S_{c,\{\bar q; g,\bar{q}\}}(k_1,k_2,\mu)
\\ 
&= \delta(k_1) \delta(k_2) + \frac{\alpha_s(\mu)}{4\pi} \biggl\{ C_A \biggl[
-\frac{8}{\mu}\cL_1\Bigl(\frac{k_1}{\mu}\Bigr) + \frac{\pi^2}{6} \delta(k_1) \biggr] \delta(k_2)
 \nn \\ & \quad
 + 4C_F \biggl[\frac{1}{\mu}\cL_1\Bigl(\frac{k_1}{\mu}\Bigr) \delta(k_2)
 - \delta(k_1) \frac{1}{\mu} \cL_1\Bigl(\frac{k_2}{\mu}\Bigr) + \frac{\pi^2}{6} \,\delta(k_1) \delta(k_2) \biggr] \biggr\} + \ord{\al_s^2}
\,,\nn \end{align}
where the plus distributions are defined as usual,
\begin{align}
\cL_n(x)=\Bigl[\frac{\theta(x)\ln^n(x)}{x}\Bigr]_+
\,.\end{align}
The results for the splitting channels $\kappa_c={\{q;q,g\}}$ and ${\{\bar q; \bar{q},g\}}$ are obtained by interchanging $k_1 \leftrightarrow k_2$.
The csoft function is universal and only depends on the color representation of the partons involved in the splitting. The result for general color structures is given in ref.~\cite{Bauer:2011uc}.

Finally, the ultrasoft function involves two soft Wilson lines with directions $n_t$ and $\bar{n}_t$, and is given by
\begin{align} \label{eq:S2_def}
\widehat S_{q \bar q}(\ell_1,\ell_2,\ell_3,\mu)
&= \widehat{T}^{-1}_{q \bar q}\,
 (\bar T^{\al_1 \bar \al_2})^\dagger\, S_{q\bar q}^{\al_1 \bar \al_2\, \bar \bt_1 \bt_2}(\ell_1,\ell_2,\ell_3,\mu)\, \bar T^{\bt_1 \bar \bt_2} 
\nn \\
&= \frac{\mathbf{1}}{N_c} \sum_{X_s} \tr\Bigl[ \Mae{0}{\bar{T} \bigl[Y_{\bar{n}_t}^\dagger Y_{n_t} \bigr]}{X_s}
\Mae{X_s}{T \bigl[Y_{n_t}^\dagger Y_{\bar{n}_t} \bigr]}{0} \Bigr]
\nn \\ & \quad \times 
\delta\bigl(\ell_1-n_t \sdt k_s^{(1)}\bigr) \delta\bigl(\ell_2-n_t \sdt k_s^{(2)}\bigr) \delta\bigl(\ell_3-\bar n_t \sdt k_s^{(3)}\bigr)
\,.\end{align}
In contrast to the usual hemisphere soft function, it measures the momentum $k_s^{(i)}$ of the soft state $X_s$ in all 3 jet regions. However, since the usoft modes cannot resolve the nearby jets 1 and 2, it has no information on the angle between them (or equivalently $\hat s_t$), and the separation into the contributions $k_s^{(1)}$ and $k_s^{(2)}$ essentially happens by splitting the $n_t$-hemisphere in half. Reparametrization invariance then implies that $\widehat S_{q \bar q}$ is independent of $\hat s_t$ to all orders in $\al_s$, so its natural scale is as usual $\mu_S \sim \Tau_3$. Up to one loop it is given by
\begin{align}\label{eq:S2_oneloop}
S_{q \bar q}(\ell_1,\ell_2,\ell_3,\mu)
&= \delta(\ell_1) \delta(\ell_2) \delta(\ell_3)
+\frac{\alpha_s(\mu) C_F}{4\pi} \biggl[
\frac{\pi^2}{3}\delta(\ell_1) \delta(\ell_2) \delta(\ell_3)
- \frac{4}{\mu} \cL_1\Bigl(\frac{\ell_1}{\mu}\Bigr) \delta(\ell_2) \delta(\ell_3)
\nn \\ & \quad
-4\, \delta(\ell_1)\frac{1}{\mu}\cL_1\Bigl(\frac{\ell_2}{\mu}\Bigr) \delta(\ell_3)
-8\, \delta(\ell_1) \delta(\ell_2) \frac{1}{\mu} \cL_1\Bigl(\frac{\ell_3}{\mu}\Bigr) \biggr] + \ord{\al_s^2}
\,. \end{align}

\subsection{$s+$ regime: $t \sim u \ll Q^2$}
\label{subsec:scets}

We now discuss the case where the first jet becomes soft, following the two-step matching described in \subsec{s+overview}.
Since the soft jet is not resolved at large invariant mass fluctuations, the first matching takes place from full QCD onto SCET with two collinear sectors with label directions $n_t \equiv n_2$ and $\bn_t \equiv n_3$ and virtuality $\sim\sqrt{u}$. This step is the same as for the $c+$ case, leading to the dijet hard function $\widehat H_{q \bar q}(Q^2,\mu)$ in \eq{H_2}.

After decoupling the collinear and usoft sectors in the parent SCET, the third jet is resolved in the usoft sector at the scale $u/Q$. This corresponds to matching the usoft sector of the parent SCET onto the $n_1$-soft-collinear and usoft sectors of \SCETp,
\begin{align}\label{eq:Cs_match}
(Y_{n_2}^\dagger)^{\al \bar \bt} (Y_{n_3})^{\ga \bar \de}
= \sum_{\lambda_g} \int\! \df\tilde p_1\, C_s^{a \bt'\! \bar \bt \ga \bar \ga'}(n_2, n_3, \tilde p_1,\mu)\,
  (\mathcal{Y}_{n_1} B_{1\lambda_g})^a (Y_{n_2}^\dagger)^{\al \bar \bt'} (Y_{n_3})^{\ga' \bar \de}
\,.\end{align}
The $Y_n$ on the left-hand are the usoft Wilson lines arising in the hard scattering operator in \eq{O_2} from the decoupling [see \eq{BPS}]. The color indices $\al$ and $\bar \de$ are contracted with the collinear fields, while $\bar \bt$ and $\ga$ are contracted with the matching coefficient $C^{\bt \bar\ga}_{(\lambda_q;\lambda_\ell)}$ from \eq{C_2}. The virtuality of the usoft fields in the Wilson lines is lowered to $\Tau_3$ on the right-hand side.
Here $\mathcal{Y}_{n_1}$ is an adjoint usoft Wilson along the $n_1$ direction. Equation~\eqref{eq:Cs_match} leads to matching the SCET $q\bar q$ operators in \eq{O_2} onto the $gq\bar q$ operators in \eq{O_3}.
Due to parity invariance, the matching is independent of the gluon helicity $\lambda_g$ (up to an irrelevant phase).

Inserting the color bases for the hard Wilson coefficient and reducing the arguments of $C_s$ with reparametrization invariance, we have
\begin{align} \label{eq:Cs_simplify}
C_s^{a \bt'\! \bar \bt \ga \bar \ga'} (n_2, n_3, \tilde p_1,\mu)\, \bar T^{\bt \bar \ga}
  &= \bar T^{a \bt'\! \bar \ga'}\cdot \widehat C_{s}\Bigl(\frac{2n_2 \sdt\tilde p_1\, n_3 \sdt \tilde p_1}{n_2 \sdt n_3},\mu\Bigr)
\,.\end{align}
Thus, $\widehat C_s$ is a matrix from the $q\bar q$ color space to the $g q \bar q$ color space (which is still one dimensional in this case).

The matching onto the \SCETp $n_2$-collinear and $n_3$-collinear sectors is equivalent to that for the $n_3$-collinear sector in the $c+$ regime in \eq{SCETq3}, except that the power expansion in the parent SCET is now in $u/Q^2$. The first jet now originates from soft emissions. For a single soft emission at tree level, the jet momenta and dijet invariant masses in the parent SCET are
\begin{align}
p_1^\mu &= p_s^\mu = k^-\,\frac{n_t^\mu}{2} + k^+\,\frac{\bn_t^\mu}{2} + k_\perp^\mu
\,,\nn\\
p_2^\mu &= p_t^\mu = Q\, \frac{n_t^\mu}{2} + k_2^\mu
\,,\qquad &
p_3^\mu &= p_{\bar t}^\mu = Q\, \frac{\bn_t^\mu}{2} + k_3^\mu
\,,\nn\\
t &= 2p_1\cdot p_2 = k^+ Q \Bigl[1 + \ORd{\frac{u}{Q^2}} \Bigr]
\,,\qquad &
u &= 2p_1\cdot p_3 = k^- Q \Bigl[1 + \ORd{\frac{u}{Q^2}} \Bigr]
\,,\nn\\
s &= 2p_2\cdot p_3 = Q^2 \Bigl[1 + \ORd{\frac{u}{Q^2}} \Bigr]
\,.\end{align}
Choosing $\vec n_i = \vec p_i / \abs{\vec p_i}$, the \SCETp label momentum conservation is thus given by
\begin{align}
p_s^\mu &= q_1^\mu = \w_1\,\frac{n_1^\mu}{2}
\,,\qquad
\w_1 = k^- + k^+ = \frac{t+u}{Q}
\,, \nn \\
p_t^\mu &= q_2^\mu = \w_2\, \frac{n_2^\mu}{2}
\,,\qquad
\w_2 = Q
\,,\qquad &
p_{\bar t}^\mu &= q_3^\mu = \w_3\, \frac{n_3^\mu}{2}
\,,\qquad
\w_3 = Q
\,.\end{align}
Note that this reproduces the $t\sim u \ll Q^2$ limit of \eq{label_momenta}, as it should.

After this two-step hard matching, the derivation of the factorization theorem is identical to that in \subsec{SCET3}, since the remaining low-energy interactions communicating via residual momenta are the same. We obtain
\begin{align}\label{eq:factorization_s+}
\frac{\df \sigma_{s+}}{\df t \,\df u \,\prod_{i=1}^3 \df \Tau^{(i)}_3}
 &= \frac{2}{(4\pi)^2 Q^2} \sum_q \int\!\df s_1 \, \df s_2 \,  \df s_3 \,  J_g(s_1,\mu)\, J_q(s_2,\mu)\, J_{\bar{q}}(s_3,\mu)
 \nn \\ & \quad \times
\tr\biggl[ \widehat C_{s}\Bigl(\frac{t\, u}{Q^2},\mu\Bigr)  \widehat H_{q \bar q}(Q^2,\mu) \, \widehat C_{s}^\dagger \Bigl(\frac{t\, u}{Q^2},\mu\Bigr)
 \nn \\ & \quad \times
\widehat{S}_{\{g,q,\bar q\}}\Bigl(\Bigl\{\Tau^{(i)}_3\!-\frac{s_i}{\w_i}\Bigr\},\{n_i \sdt n_j\} ,\mu\Bigr)
\biggr]
\biggl[1 + \ORd{\frac{u}{Q^2},\frac{m_J^2}{u}}\biggr]
\,.\end{align}
The overall factor of $2$ comes from summing over the two partonic channels $\{g,q,\bar{q}\}$ and $\{g,\bar{q},q\}$ that are nonvanishing in the soft limit, and which give identical contributions. The sum over $q$ runs over the desired quark flavors. We have explicitly included the factor from the soft emission phase space
\begin{align}
\int\! \df \Phi_s = \frac{1}{(4\pi)^2 Q^2} \int\! \df t\, \df u
\,.\end{align}

In \eq{factorization_s+}, $\widehat C_s$ describes the soft large-angle splitting and is evaluated at
\begin{equation}
\frac{2n_2 \sdt q_1\, n_3 \sdt q_1}{n_2 \sdt n_3}
= \frac{2q_2 \sdt q_1\, q_3 \sdt q_1}{q_2 \sdt q_3}
= \frac{t\, u}{Q^2}
\,.\end{equation}
Since this is the only scale it depends on, it contains no large logarithms when evaluated at its natural scale $\mu_{H_s} \sim \sqrt{tu}/Q$. Because $\widehat{H}_{q\bar q}$ has a trivial color structure, we can combine $\widehat C_s$ and $\widehat C_s^\dagger$ into a single hard function 
\begin{align}\label{eq:Hs_normalization}
\widehat H_s \Bigl(\frac{t\, u}{Q^2},\mu\Bigr)
&\!= \frac{1}{(4\pi)^2 Q^2}\,
\widehat C_{s}\Bigl(\frac{t\, u}{Q^2},\mu\Bigr)  \widehat C_{s}^\dagger\Bigl(\frac{t\, u}{Q^2},\mu\Bigr)
\!= \frac{1}{(4\pi)^2 Q^2}\,
\widehat C_{s}\Bigl(\frac{t\, u}{Q^2},\mu\Bigr) \widehat T_{q \bar q}^{-1}\, \widehat C_{s}^{*T}\Bigl(\frac{t\, u}{Q^2},\mu\Bigr) \widehat T_{g\,q \bar q}
\,.\end{align}
The matching coefficients $\widehat C_s$  can be calculated directly from soft matrix elements in SCET since the virtual corrections in \SCETp are scaleless and vanish in pure dimensional regularization, see \eq{H3s_def} for the explicit prescription. They are also equivalent (up to overall normalization) to the soft gluon current~\cite{Mangano:1990by,Bern:1998sc,Bern:1999ry,Catani:2000pi}.
The result for $H_s$ up to one loop reads (see \app{Hs})
\begin{align}\label{eq:H3s_oneloop}
H_{s}\Bigl(\frac{t u}{Q^2},\mu\Bigr)
&=  \frac{\alpha_s(\mu) C_F}{\pi} \,\frac{1}{t \,u}
\biggl\{1- \frac{\alpha_s(\mu) C_A}{4 \pi}\biggl[\ln^2\Bigl(\frac{t\, u}{Q^2  \mu^2}\Bigr) - \frac{5\pi^2}{6}\biggr]
+ \ord{\al_s^2} \biggr\}
\,.\end{align}
We provide the two-loop expression in \eq{H3_twoloop} obtained from the corresponding computations of the soft gluon current in refs.~\cite{Duhr:2013msa, Li:2013lsa}.
Alternatively, $H_s$ can be obtained from the soft limit of the hard coefficient for three well-separated jets,
\begin{align} \label{eq:H_s_consistency}
H_{\{g,q,\bar{q}\}}(t,u,Q^2,\mu) \Big\vert_{t\sim u\ll Q^2}\!\!
&= H_{q \bar q}(Q^2,\mu) \, H_{s}\Bigl(\frac{t\, u}{Q^2},\mu\Bigr) \Bigl[1 + \ORd{\frac{u}{Q^2}}\Bigr]
.\end{align}

The remaining ingredients of the factorization theorem are the same as for the case of three well-separated jets in \eq{factorization_SCET}, except that the invariant mass of the soft gluon jet is smaller than for the quark jets, and the corresponding natural scale for the gluon jet function is now $\mu_{J_1} \sim \sqrt{\Tau_3 u/Q} \sim m_J \sqrt{u}/Q \ll m_J$.

Based on the $\mu$-independence of the factorization theorems in \eqs{factorization_SCET}{factorization_s+}, we can derive the all-order form for the anomalous dimension of $\widehat C_s(\mu)$.  Since the soft and jet functions are identical in both cases, it is sufficient to require consistency in the hard sector, namely
\begin{equation} \label{eq:Cs_consistency}
\frac{\df}{\df \ln\mu}\, C_{gq\bar q}(t,u,Q^2,\mu)\Big\vert_{t\sim u\ll Q^2}
 = \frac{\df}{\df \ln\mu} \biggl[C_s\Bigl(\frac{t\, u}{Q^2}, \mu\Bigr) C_{q \bar q}(Q^2,\mu) \biggr]
 \,.\end{equation}
Defining the anomalous dimensions for each of the coefficients as
\begin{equation}
\frac{\df}{\df \ln\mu}\, C_x(\ldots, \mu) = \gamma_{C_x}(\ldots, \mu)\, C_x(\ldots, \mu)
\,,\end{equation}
\eq{Cs_consistency} requires that
\begin{align}
\gamma_{C_s}\Bigl(\frac{t\, u}{Q^2},\mu\Bigr) =
\gamma_{C_{gq\bar q}}(t, u, Q^2, \mu) \Big\vert_{t\sim u\ll Q^2}
- \gamma_{C_{q \bar q}}(Q^2,\mu)
\,.\end{align}
The all-order structure of the anomalous dimensions for $C_{gq \bar q}$ and $C_{q \bar q}$ (without assuming Casimir scaling) reads~\cite{Manohar:2003vb, Chiu:2008vv, Gardi:2009qi}
\begin{align}\label{eq:gqq_anom}
\gamma_{C_{gq\bar q}}(t, u, Q^2, \mu)
&= \frac{\Gamma_\cusp^g[\alpha_s(\mu)]}{2} \ln\Bigl[\frac{(t+\img0)(u+\img0)}{\mu^2(-s-\img0)} \Bigr]
+ \Gamma_\cusp^q[\alpha_s(\mu)] \ln\Bigl(\frac{-s - \img 0}{\mu^2} \Bigr)
 + \gamma_C^{gq\bar q}[\alpha_s(\mu)]
\,, \nn \\
\gamma_{C_{q\bar q}}(Q^2, \mu)
&= \Gamma_\cusp^q[\alpha_s(\mu)] \, \ln\Bigl(\frac{-Q^2 - \img 0}{\mu^2}\Bigr) + 2 \gamma_C^q[\alpha_s(\mu)]
\,,\end{align}
where $\Ga_\cusp^i(\alpha_s)$ are the quark and gluon cusp anomalous dimensions~\cite{Korchemsky:1987wg}, and $\ga_C^i(\alpha_s)$ are the noncusp anomalous dimensions and are defined by \eq{gqq_anom}. Hence, using $s = Q^2[1 + \ord{u/Q^2}]$, we have
\begin{align} \label{eq:Cs_anom}
\gamma_{C_{s}}\Bigl(\frac{t\, u}{Q^2},\mu\Bigr)
&= \frac{\Gamma^g_\cusp[\alpha_s(\mu)]}{2} \,\ln\Bigl(\frac{- t\,u/Q^2 - \img0}{\mu^2}\Bigr)
   + \gamma_C^{gq\bar q}[\alpha_s(\mu)] - 2 \gamma_C^q[\alpha_s(\mu)]
\nn \\
&= \frac{\Gamma^g_\cusp[\alpha_s(\mu)]}{2} \,\ln\Bigl(\frac{-t\,u/Q^2 - \img 0}{\mu^2}\Bigr)
   + \gamma_C^g[\alpha_s(\mu)] + \ord{\alpha_s^3}
\,.\end{align}
In the second line we used that $\gamma_C^{gq\bar q}(\alpha_s) = 2\gamma_C^q(\alpha_s) + \gamma_C^g(\alpha_s)$, which is known to hold at least up to two loops~\cite{Aybat:2006mz}, where $2\gamma_C^g(\alpha_s)$ is the noncusp anomalous dimension of the gluon form factor.
We also explicitly verified \eq{Cs_anom} to $\ord{\alpha_s^2}$ for the perturbative results in \eqs{H3s_oneloop}{H3_twoloop}, using the explicit 2-loop expressions for $\Ga^i_\cusp$ and $\ga_C^i$ given in \app{anom_dim}.

\subsection{$cs+$ regime: $t \ll u \ll Q^2$}
\label{subsec:scetcs}

Finally, we discuss the case where the first jet becomes soft and also close to the second jet, such that all kinematic scales are separated. This follows the multistage matching procedure described in \subsec{cs+overview}. In the first step, we match full QCD onto SCET with two collinear sectors of virtuality $\sim \sqrt{u}$ and label directions $n_t \equiv n_2$ and $\bar n_t \equiv n_3$. This step is the same as in the $s+$ regime.
In the next step, the parent usoft sector splits into usoft modes with lower virtuality $t/Q$ and scaling $p_{us}^\mu \sim (u/Q)\times(t/u,t/u,t/u)$, and soft-collinear modes with virtuality $\sqrt{t \,u}/Q$ and scaling $p_{sc}^\mu \sim (u/Q)\times(t/u,1,\sqrt{t/u})$ that will eventually produce the soft gluon jet.
If the gluon jet is close to the quark jet, the matching onto this intermediate \SCETp reads
\begin{align} \label{eq:SCETcs_matching}
\bar \chi^{\bar \alpha}_{t\lambda_q} (Y_{n_t}^\dagger)^{\alpha\bar \beta} (Y_{n_3})^{\gamma \bar \delta}
= \bar \chi^{\bar \alpha}_{t\lambda_q} (X^\dagger_{n_t} V_{n_t})^{\alpha \bar \alpha'} (Y_{n_t}^\dagger)^{\alpha' \bar \beta } (Y_{n_3})^{\gamma \bar \delta}
\,.\end{align}
The $X_{n_t}$ and $V_{n_t}$ Wilson lines are defined as in \eq{Vn_Xn}. They sum up $n_t$-soft-collinear gluon emissions along the $n_t$ and $\bn_t$ directions, respectively, as required by gauge invariance.
The matching in this step is purely in terms of Wilson lines and does not introduce a hard matching coefficient: although the soft-collinear modes are being separated from the usoft modes, the soft jet is not yet resolved and thus no scale setting measurement is performed.\footnote{The precise identification of this intermediate mode as either soft-collinear or csoft is not as unique as in the $c+$ and $s+$ cases, as it shares aspects of both. Due to its momentum scaling we interpret it as collinear offspring arising from the parent usoft sector and refer to it as soft-collinear. However, the Wilson line structure in \eq{SCETcs_matching} is reminiscent of the csoft modes and can also be obtained by directly matching from QCD onto this intermediate \SCETp, where the interactions between the collinear, soft-collinear, and usoft modes can be decoupled by consecutive BPS field redefinitions, as shown in Ref.~\cite{Bauer:2011uc}.}
At the scale $\sqrt{t\, u}/Q$ the final soft jet 1 gets resolved. This corresponds to matching the $n_t$-soft-collinear sector of the parent \SCETp onto the final $n_1$-soft-collinear and $n_t$-csoft sectors of the $cs+$ regime,
\begin{align}\label{eq:Ccs_match}
(X_{n_t}^\dagger V_{n_t})^{\al \bar \de} = \sum_{\lambda_g} \int\! \df\tilde p_1\, C_{cs}^{a \bt' \bar \ga'} (n_2, \bar{n}_t, \tilde p_1,\mu)
(\mathcal{X}_{n_1} B_{1\lambda_g})^a (X_{n_2}^\dagger)^{\al \bar \bt'} (V_{n_t})^{\ga' \bar \de}
\,.\end{align}
This is the soft Wilson-line version of the $c+$ matching in \eq{SCETc_matching}. It is also identical to the soft splitting in \eq{Cs_match} with the replacements $Y_{n_2} \to X_{n_2}$, $Y_{n_3} \to V_{n_2}$ and the simplification that the indices $\bar \bt$ and $\ga$ therein are contracted.

The discussion for the label momenta proceeds in the same way as in the $c+$ and $s+$ regimes. The final label momenta are given by
\begin{align}
\w_1=\frac{u}{Q}  \, , \qquad \w_2= \w_3=Q 
\,,\end{align}
corresponding to the $t\ll u\ll Q^2$ limit of \eq{label_momenta}.

The operators that result from these matching steps are the same as for the $c+$ case with the only difference being the different scaling for the label momenta. Thus one obtains essentially the same factorization theorem
\begin{align}\label{eq:factorization_cs+}
\frac{\df \sigma_{cs+}}{\df t \,\df u  \prod_{i=1}^3 \df \Tau^{(i)}_3 }
&=  2\sum_q \tr\biggl[ \widehat H_{cs}\Big(\frac{t\,u}{Q^2},\mu\Big) \!\int\! \df k_1 \, \df k_2\, \widehat S_{cs}(k_1,k_2 ,\mu)\biggr]
\nn \\ & \quad \times
\int\! \df s_1 \, \df s_2 \,  \df s_3 \, J_g(s_1,\mu)\, J_q(s_2,\mu) \, J_{\bar{q}}(s_3,\mu)
\nn \\ & \quad \times
\tr\biggl[\widehat H_{q\bar q}(Q^2,\mu) \widehat S_{q \bar q}\Bigl(\Tau^{(1)}_3\!-\!\frac{s_1}{\w_1}\!-\!\sqrt{\hat{s}_t} \, k_1,\Tau^{(2)}_3\!-\!\frac{s_2}{\w_2}\!-\! \sqrt{\hat{s}_t}  \, k_2, \Tau^{(3)}_3\!-\!\frac{s_3}{\w_3} ,\mu\Bigr) \biggr]
\nn \\ & \quad \times
\biggl[1+\ORd{\frac{u}{Q^2},\frac{t}{u},\frac{m_J^2}{t}}\biggr]
\,.\end{align}
Summing the two nonvanishing channels $\{g,q,\bar{q}\}$ and $\{g,\bar{q},q\}$ gives rise to the overall factor of 2, as in the $s+$ case in \eq{factorization_s+}.

The hard function $\widehat H_{cs}$ incorporates the collinear-soft splitting at the natural scale $\mu_{H_{cs}} \sim \sqrt{t \, u}/Q$ and is related to the matching coefficient $\vec C_{cs}$ in \eq{Ccs_match}.
Here, $\vec C_{cs}$ can be obtained from soft-collinear matrix elements in \SCETp (see \eq{H3cs_def} for the explicit prescription), which are equivalent to the collinear limit of the soft splitting amplitudes or the soft limit of the collinear splitting amplitudes for $q \rightarrow qg$. The similarity between \eq{Ccs_match} and \eq{Cs_match} implies
\begin{align}
H_{cs}\Bigl(\frac{t\, u}{Q^2},\mu\Bigr) = H_{s}\Bigl(\frac{t\, u}{Q^2},\mu\Bigr)
\, .\end{align}
We stress however that this identity is special to the case of $e^+e^- \to 3$ jets and does not hold when the color space is nontrivial, as we will see in \subsec{Njet_scetcs}.

All of the remaining components in \eq{factorization_cs+} have already been discussed in \subsec{SCETc}. We have denoted the csoft function by $\widehat S_{cs} = \widehat S_{c,\{g,q,\bar{q}\}}$, whose natural scale is $\mu_{S_{cs}} \sim \Tau_3\sqrt{u/t}$. Furthermore, the natural scale of the gluon jet function is now $\mu_{J_1} \sim \sqrt{\Tau_3u/Q}$.

We now discuss the relationship between the factorization theorems in the $cs+$, $c+$, and $s+$ regimes. First, the difference with respect to \eq{factorization_c+} only concerns the hard sector. Expanding the $c+$ hard function $\widehat H_{c,\kappa_c}$ in the soft limit gives
\begin{align}
 H_{c,\kappa_c}\Bigl(t,\frac{u}{Q^2},\mu\Bigr) \Big\vert_{u\ll Q^2}
 = \delta_{\kappa_1 g}\,  H_{cs}\Bigl(\frac{t \, u}{Q^2},\mu\Bigr)  \Bigl[1 + \ORd{\frac{u}{Q^2}}\Bigr]
\,.\end{align}
This can be checked explicitly at one loop using \eqs{H3c_oneloop}{H3s_oneloop}. Second, we observe that the factorization theorems in $s+$ and $cs+$ regimes differ only in the usoft sector. The corresponding relation that needs to hold for consistency reads
\begin{align}\label{eq:consistency_soft}
S_{\{g,q,\bar{q}\}}(\ell_1,\ell_2,\ell_3,\mu) \big\vert_{t\ll u}
&= \int\! \df k_1 \, \df k_2 \, S_{q\bar q}\bigl(\ell_1-\sqrt{\hat{s}_t} \,k_1,\ell_2-\sqrt{\hat{s}_t}\,k_2,\ell_3,\mu\bigr)
\nn \\ & \quad 
 \times S_{c,\{g,q,\bar{q}\}}(k_1,k_2,\mu) \, \Bigl[1 + \ORd{\frac{t}{u}}\Bigr]
\,.\end{align}
(In the general case, when the color space for $\widehat S_{q\bar q}$ and $\widehat S_{c,\{g,q,\bar{q}\}}$ are nontrivial, this relation would involve a tensor product in these color spaces.)
The full $3$-jettiness soft function appearing in $\sigma_{s+}$ can be calculated fully analytically in the collinear limit, which yields the result~\cite{Bauer:2011uc}
\begin{align}\label{eq:S3expand_oneloop}
& S_{\{g,q,\bar q\}}(\ell_1,\ell_2,\ell_3,\mu)\big|_{t \ll u}
\nn \\ & \quad
= \delta(\ell_1)\, \delta(\ell_2)\, \delta(\ell_3) +\frac{\alpha_s(\mu)}{4\pi}\biggl\{ C_A\biggl[ \frac{\pi^2}{6} \, \delta(\ell_1)\, \delta(\ell_2)\, \delta(\ell_3)
- \frac{8}{\sqrt{\hat{s}_t} \, \mu}\cL_1\Bigl(\frac{\ell_1}{\sqrt{\hat{s}_t} \, \mu}\Bigr) \delta(\ell_2)\, \delta(\ell_3) \biggr]
\nn \\  & \qquad
+ C_F\biggl[\pi^2 \delta(\ell_1) \delta(\ell_2) \delta(\ell_3)
+ \frac{4}{\sqrt{\hat{s}_t} \, \mu}\cL_1\Bigl(\frac{\ell_1}{\sqrt{\hat{s}_t} \, \mu}\Bigr) \delta(\ell_2) \delta(\ell_3)
- \frac{4}{\sqrt{\hat{s}_t} \, \mu}\cL_1\Bigl(\frac{\ell_2}{\sqrt{\hat{s}_t} \, \mu}\Bigr) \delta(\ell_1) \delta(\ell_3)
 \nn \\ & \hspace{11ex}
 - \frac{4}{\mu}\cL_1\Bigl(\frac{\ell_1}{\mu}\Bigr) \delta(\ell_2) \delta(\ell_3)
 - \frac{4}{\mu}\cL_1\Bigl(\frac{\ell_2}{\mu}\Bigr) \delta(\ell_1) \delta(\ell_3)
 - \frac{8}{\mu}\cL_1\Bigl(\frac{\ell_3}{\mu}\Bigr) \delta(\ell_1) \delta(\ell_2)\biggr]\biggr\}
.\end{align}
Using eqs.~(\ref{eq:Sc_oneloop}),~(\ref{eq:S2_oneloop}) and~(\ref{eq:S3expand_oneloop}), the consistency relation in \eq{consistency_soft} can be explicitly verified at one loop (see also ref.~\cite{Bauer:2011uc}).

\subsection{Combining all regimes}
\label{subsec:3jet_combination}

As outlined in \subsec{combineEFT}, to obtain a complete description across the full $3$-jet phase space, we need to combine the resummed results obtained in the different regimes.

We start from the $cs+$ regime, where we have the maximal amount of hierarchies between the dijet invariant mass scales that can arise for $e^+ e^- \to 3$ jets, and which allows us to resum large logarithms in the kinematic ratios $u/Q$, $t/u$, and jet resolution variables $\Tau_3^{(i)}$. We then systematically add the nonsingular power corrections to take into account the correct fixed-order contributions in the less hierarchical situations where one or more scales are parametrically the same. The cross section for arbitrary $t, u, Q^2 \gg m_J^2$ is thus written as
\begin{align}
\df \sigma \equiv \frac{\df \sigma}{\df t\, \df u \prod_{i=1}^3 \df \Tau^{(i)}_3}
= \df \sigma_{cs+}
+\df \sigma_{c+}^\nons
+ \df \sigma_{s+}^\nons
+ \df \sigma_{\SCET}^\nons
+ \df \sigma_{\QCD}^\nons
\,,\end{align}
where the denominator is suppressed for convenience.

As shown in \eqs{nonsing}{nonsing2}, the nonsingular corrections for a given scale hierarchy are given by the difference of the corresponding full and resummed cross sections, where the latter has the logarithms of the scale hierarchy expanded to the same fixed order in perturbation theory as they are present in the full cross section. By including the fixed-order terms in the relevant hard, beam, jet, and soft functions to the same required order, corresponding to the often utilized N$^k$LL$'$ order counting, the fixed-order expansion to N$^k$LO can be conveniently obtained simply by turning off the resummation in the relevant scale hierarchy.

The nonsingular correction to connect $\df\sigma_{cs+}$ in \eq{factorization_cs+} to $\df\sigma_{c+}$ in \eq{factorization_c+} is given by
\begin{equation}\label{eq:factorization_ns_c+}
\df \sigma_{c+}^\nons
= \df \sigma_{c+} -  \df \sigma_{cs+} \big|_{\mu^{(cs)}=\mu^{(c)}}
\,.\end{equation}
Here, the natural scales in $\df\sigma_{cs+}$ are set equal to the ones used in $\sigma_{c+}$, i.e.~$\mu_{H_{cs}}=\mu_{H_c}$, $\mu^{(cs)}_{J_1}=\mu_J$ and $\mu^{(cs)}_{S_{c}}=\mu^{(c)}_{S_c}$. This turns off the additional resummation in $u/Q^2$ in the $cs+$ regime with respect to the $c+$ regime, and instead includes the corresponding logarithms in $u/Q^2$ at fixed order.

Similarly, the nonsingular correction connecting $\df\sigma_{cs+}$ to $\df\sigma_{s+}$ in \eq{factorization_s+} is
\begin{equation}\label{eq:factorization_ns_s+}
\df \sigma_{s+}^\nons
= \df \sigma_{s+} - \df \sigma_{cs+} \big|_{\mu^{(cs)}=\mu^{(s)}}
\,,\end{equation}
where the natural scales in $\df\sigma_{cs+}$ are now set equal to the ones used in $\df\sigma_{s+}$, i.e.~$\mu_{H_{cs}}=\mu_{H_s}$, $\mu^{(cs)}_{J_{1}} = \mu^{(s)}_{J_1}$ and $\mu^{(cs)}_{S_c}=\mu_{S}$, which turns off the additional resummation in $t/u$ in $\df\sigma_{cs+}$.

The nonsingular correction between SCET and \SCETp is given in terms of the cross sections in eqs.~\eqref{eq:factorization_SCET},~\eqref{eq:factorization_c+},~\eqref{eq:factorization_s+}, and~\eqref{eq:factorization_cs+} as
\begin{equation}\label{eq:factorization_ns_SCET3}
 \df \sigma_{\SCET}^\nons
 = \df \sigma_{\SCET}
 - \bigl[ \df \sigma_{c+} + \df \sigma_{s+} - \df \sigma_{cs+} \bigr]_{\mu_+ =\mu_{\SCET}}
\,, \end{equation}
where all the additional scales in the \SCETp cross sections are set to the corresponding ones in SCET, i.e.~$\mu_{H_{c}} = \mu_{H_{s}} = \mu_{H_{cs}}=\mu_{H}$, $\mu_{J_1}=\mu_{J}$, and $\mu_{S_c}=\mu_{S}$, which turns off all additional resummation in $t/Q^2$ in $\df\sigma_{c+}$ and $\sqrt{t u}/Q^2$ in $\df\sigma_{s+}$. The term $\df\sigma_{cs+}$ arises with an opposite sign [see \eq{nonsing2}] and removes the double counting between the $c+$ and $s+$ regimes.

Finally, the nonsingular correction between SCET and full QCD is given by
\begin{equation}\label{eq:factorization_ns_QCD}
\df \sigma_{\QCD}^\nons
= \df \sigma_{\QCD} - \df \sigma_{\SCET}|_{\mu_{\SCET}=\mu_{\rm FO}}
\,,\end{equation}
where all the resummation scales in SCET are set to a common fixed order scale, i.e.~$\mu_{H}=\mu_J=\mu_S=\mu_{\rm FO}\sim Q$, so that the resummation in $\Tau_3/Q$ is turned off.

\section{\boldmath $pp \to N$ jets}
\label{sec:njets}

In this section, we extend the discussion of \sec{3jets} to the general case of $pp \to N$ jets. We address in particular
collinear initial-state splittings and additional complications related to color, spin, and kinematics.
We consider adding one kinematic hierarchy to the standard SCET case of equally energetic and well-separated jets, and discuss the \SCETp factorization for the cases of a jet close to a beam, a soft jet, and a soft jet close to another jet, corresponding to the simplest $c+$, $s+$, and $cs+$ regimes. The $N$-jet phase space implies a proliferation of hard kinematic scales allowing for the possibility of multiple hierarchies between the jets, which may be independent, strongly ordered or correlated. These are discussed in \sec{multihierarchy}.

\subsection{$N$-jet kinematics}

We start by discussing the kinematics for $pp \to N$ jets. The initial-state partons that enter the hard interaction are labeled with $a$ and $b$, such that the hard scattering process is
\begin{align} \label{eq:njet_process}
\kappa_a(q_a) \kappa_b(q_b)  \to \kappa_1(q_1) \kappa_2(q_2) \cdots \kappa_N(q_N) + L(q_L)
\,,\qquad
\kappa=\{\kappa_a,\kappa_b;\kappa_1, \dots, \kappa_N\}
\,.\end{align}
In cases where a parton's helicity becomes relevant we will include a helicity label such as $\kappa_i^{\lambda_i}$.
The large label momenta $q_i$ for the final-state jets are defined as discussed in \subsec{standard}.
We also allow for an additional color-singlet final state $L$ with total momentum $q_L^\mu$, which is suppressed in $\kappa$. It does not affect the factorization setup for the QCD final state, apart from contributing to the overall label momentum conservation
\begin{equation} \label{eq:Njetlabelmomentum}
q_a^\mu + q_b^\mu = q_1^\mu + \dotsb + q_N^\mu + q_L^\mu
\,.\end{equation}
The label momenta for the initial partons are defined as
\begin{align} \label{eq:q_ab}
  q_{a,b}^\mu = \frac12\, x_{a,b} E_{\rm cm} (1, \pm \hat n) = \frac12\, \w_{a,b}  (1, \pm \hat n)
\,.\end{align}
They are given in terms of the hadronic center of mass energy $E_{\rm cm}$, the momentum fractions $x_a$, $x_b$ and the unit vector $\hat n$ pointing along the beam axis. Alternatively, they may be written in terms of $\w_{a,b} = Q e^{\pm Y}$, where $Q$ is the invariant mass and $Y$ the total rapidity of the hard partonic system which are determined from \eq{Njetlabelmomentum}.

To keep the notation concise, we collectively denote by $\Phi_N$ the full dependence on the kinematics, helicities, and partonic channel of the hard process. In particular, we abbreviate the hard Wilson coefficients as
\begin{equation}
\vec C(\Phi_N) \equiv \vec C_{\lambda_a\dotsb\lambda_N}(\{q_i, \Phi_L\}, \mu)
\,.\end{equation}
Correspondingly, the fully-differential Born phase space measure is denoted by $\df \Phi_N$ and given by
\begin{align}
\int\! \df \Phi_{N}
 \equiv \frac{1}{2 E^2_{\rm cm}}  \int \frac{\df x_a}{x_a} \, \frac{\df x_b}{x_b} \int \frac{\df q_L^2}{2\pi} \,\df \Phi_{N+1}(q_a+q_b;q_1,\dots, q_N,q_L) \,\df \Phi_L(q_L) \sum_{\kappa} \sum_{\{\lambda_i\}}
\,,\end{align}
where $\df\Phi_{N+1}(q_a+q_b; q_1, ...)$ and $\df \Phi_L(q_L)$ denote the standard Lorentz-invariant phase space for $N\!+\!1$ final-state momenta and for the nonhadronic final state. We also included the flux factor $1/(2\Ecm^2)$, the integral over momentum fractions, the sum over partonic channels $\kappa$, including the desired quark flavors, and the sum over helicities.

As jet resolution variable we use again $N$-jettiness in \eq{tauN} with the general geometric measure $Q_i = \rho_i \w_i$.
We write the $N$-jet cross section with additional kinematic constraints $X$ on the jets as
\begin{align}\label{eq:dsigma_Nbody}
\frac{\df \sigma (X)}{\df \Tau_N} = \int\! \df \Phi_{N} \, \frac{\df \sigma (\Phi_N)}{\df \Tau_N} \, X(\Phi_{N})
\,.\end{align}
In the following, we will discuss the results for $\df \sigma (\Phi_{N})/\df \Tau_N$, which is fully differential in $\Phi_N$. For simplicity we only consider the cross section differential in the total sum $\Tau_N$ of the contributions from each jet and beam region as in \eq{tauN}.

We define the invariant masses $s_{ij}$, taking into account that $a$ and $b$ are incoming, as
\begin{align}
  s_{ab} = (q_a+q_b)^2
  \,, \quad
  s_{ak} = (q_a - q_k)^2
  \,, \quad
  s_{bk} = (q_b - q_k)^2
  \,, \quad
  s_{kl} = (q_k + q_l)^2
\,,\end{align}
where $k, l \neq \{a,b\}$. We also define the generalized angular measures
\begin{align}
\hat s_{ij} = \frac{2q_i \sdt q_j}{Q_i Q_j} = \frac{|s_{ij}|}{Q_i Q_j} = \frac{n_i \sdt n_j}{2 \rho_i \rho_j}
\,.\end{align}

\subsection{Standard SCET regime}
\label{subsec:scet}

The $N$-jettiness factorization for $pp$ collisions was derived for active-parton
cross sections in refs.~\cite{Stewart:2009yx, Stewart:2010tn, Jouttenus:2011wh};
the factorization for generalized measures was discussed in ref.~\cite{Stewart:2015waa}.%
\footnote{In this paper, we only consider factorization for the active-parton scattering cross sections, initiated by incoming quarks or gluons. This avoids the complications associated with the spectator partons present for incoming hadrons.
In a MC context, this corresponds to the primary hard interaction without additional multi-parton interactions.
The associated factorization formulae for inclusive event shapes like $N$-jettiness do not include contributions from perturbative Glauber gluon exchange that start at ${\cal O}(\alpha_s^4)$~\cite{Gaunt:2014ska, Zeng:2015iba}. These terms can be incorporated using the Glauber operator framework of Ref.~\cite{Rothstein:2016bsq}, but do not affect the additional factorization in \SCETp, which we are primarily interested in here.}
The factorized $N$-jettiness cross section for $N$ energetic well-separated jets, $s_{ij} \sim Q^2$, is given by
\begin{align} \label{eq:factorization_Njet_SCET}
\frac{\df \sigma_{\SCET} (\Phi_{N})}{\df \Tau_N}
&= s_{\kappa} \int\! \df s_a\, \df s_b \, B_a(s_a,x_a,\mu) \, B_b(s_b,x_b,\mu) \,\biggl[\prod_{k=1}^N \int\! \df s_k \, J_k(s_k,\mu)\biggr]
\\ & \quad \times
\vec{C}^\dagger(\Phi_{N},\mu) \,
 \widehat{S}_\kappa\biggl(\Tau_N- \sum_{i=a}^N\frac{s_i}{Q_i},\{\hat s_{ij} \}, \mu\biggr)\,  \vec{C}(\Phi_{N},\mu)
 \biggl[1 +\ORd{\frac{m_J^2}{Q^2}}\biggr]
\,.\nn\end{align}
Here, $s_\kappa$ denotes the symmetry factor for the partonic channel $\kappa$, which also accounts for color averaging for incoming partons. 
The beam functions $B_{a,b}$ are the counterparts of the jet functions $J_k$ for initial states, and depend on the transverse virtualities $s_{a,b}$ and momentum fractions $x_{a,b}$.
They describe the collinear initial-state radiation contributing to the measurement of $\Tau_N$ and incorporate the nonperturbative parton distribution functions~\cite{Stewart:2009yx, Stewart:2010qs}. 

The hard matching coefficient $\vec{C}(\Phi_N)$ is now a vector and the soft function $\widehat{S}_\kappa$ a matrix in the nontrivial color space for the $N+2$ colored partons participating in the hard interaction described by \eq{njet_process}.
As discussed in detail in ref.~\cite{Moult:2015aoa}, the hard matching coefficients $\vec C(\Phi_N)$ are directly related to the IR-finite parts of the color-stripped QCD helicity amplitudes in dimensional regularization [analogous to \eq{C_is_A}]. Making its color decomposition in terms of a color basis $T_k^{\alpha_a \dots \alpha_N}$ explicit,\footnote{Here the same indices $\alpha_i$ are used for both fundamental and adjoint representations.}
\begin{equation}
C^{\alpha_a\dots\alpha_N}
 = \sum_k T_k^{\alpha_a \dots \alpha_N}\, C^k
 \equiv \bar T^{\alpha_a \dots \alpha_N} \cdot \vec{C}
\,.\end{equation}
The conjugate vector $\vec C^\dagger$ is given by
\begin{equation}
\vec{C}^\dagger
= C^{*\,\alpha_a\dots\alpha_N} \bar T^{\alpha_a \dots \alpha_N}
= \vec C^{*T} \cdot \widehat T_\kappa
\,.\end{equation}
with the color sum matrix
\begin{align}
\widehat T_\kappa
= (\bar T^{\alpha_a, \dots, \alpha_N}_i)^\dagger\, \bar T^{\alpha_a, \dots, \alpha_N}
\,.\end{align}
The typically utilized color bases are not orthogonal in which case $\widehat T_\kappa$ is a nontrivial matrix.

The color decomposition of the soft function is given by\footnote{For a degenerate (i.e.~nonminimal) color basis  $\widehat{T}^{-1}$ is a generalized inverse matrix, i.e.~$\widehat{T}\, \widehat{T}^{-1}\,\widehat{T} = \widehat{T}$.}
\begin{align}\label{eq:soft_matrix}
\widehat{S}_\kappa = \widehat{T}_\kappa^{-1}(\bar{T}^{\alpha_a\dots\alpha_N})^\dagger  S_\kappa^{\alpha_a \dots \alpha_N\,\beta_a \dots \beta_N} \, \bar{T}^{\beta_a\dots\beta_N}
\,,\end{align}
which at tree level reduces to the identity
\begin{align}
\widehat{S}_\kappa^{(0)}(\ell)
= \widehat{T}_\kappa^{-1}(\bar{T}^{\alpha_a\dots\alpha_N})^\dagger \, \delta(\ell)\, \delta_{\alpha_a \beta_a} \cdots \delta_{\alpha_N \beta_N}\, \bar{T}^{\beta_a\dots\beta_N}
= {\bf 1}\, \delta(\ell)
\,.\end{align}

To describe the kinematic jet hierarchies in the general $N$-jet case, we always assume that the corresponding
nonhierarchical limit where all jets are equally separated and equally energetic is described in standard SCET
by \eq{factorization_Njet_SCET}. We note that while parametrically this corresponds to counting all
$s_{ij}\sim Q^2$, the relevant numerical value for the hard matching scale $\mu_H$, at which the matching coefficients
$\vec{C}(\Phi_{N},\mu)$ are calculated, typically differs by $\ord{1}$ factors from the total partonic invariant mass $Q$.
For example, a good choice for the hard scale for $V+$ jet would be $\mu_H \simeq p_T^J$ (see e.g.~ref.~\cite{Jouttenus:2013hs}).
This means we can describe any processes where there is an underlying hard scattering taking
place with a hard momentum transfer $\sim Q$ into the final state, to which we then add a number of additional
soft or collinear jets, as discussed in the following subsections and \sec{multihierarchy}.

An important situation that falls outside the above general class of processes is the case of purely collinear
forward ($t$-channel) scattering, such as $pp\to 2$ jets with both jets collinear to one of the beams, for which there is no
hard momentum transfer $\sim Q$. This would corresponds to a parametric regime $s_{12} \gg |s_{1a}|, |s_{2b}| ,q_L^2 $ and requires a fundamentally different factorization theorem already in SCET (see ref.~\cite{Rothstein:2016bsq}). A framework that allows to resum
the single logarithms in this multi-Regge limit, i.e.~energetic forward jets with large rapidity separation and small transverse momenta in $t$-channel scattering, has been discussed e.g.~in ref.~\cite{Andersen:2009nu}.
The soft version of this would be a purely soft scattering,
i.e., $pp\to N$ jets in the limit $N \gg 1$ such that all final-state jets are parametrically soft compared
to the total partonic beam energy. In this case there is again effectively no hard interaction with a
hard momentum transfer to the final state, requiring a different description already in SCET.

\subsection{$c+$ regime}
\label{subsec:scet_c}

We continue with the case where two jets are close to each other or one jet is close to one of the beams. Since both cases are very similar and the former was discussed in ref.~\cite{Bauer:2011uc}, we focus on the latter. We take the first jet to be close to the direction of beam $a$, while all other jets remain equally energetic and separated, so we have
\begin{equation}
0<-t \equiv -s_{1a} \ll |s_{ij}| \sim Q^2
\,,\qquad
z = \frac{\w_1}{\w_a}
\,, \qquad
\hat{s}_t = \hat s_{a1}
\,,\end{equation}
with $\{i,j\} \neq \{1,a\}$.
The factorization procedure follows the two-step matching in \subsec{SCETc}, which separates the process into
\begin{align} \label{eq:c+process}
\kappa_a(q_a) \to\, & \kappa_t(q_t)\, \kappa_1(q_1)
\qquad &
\kappa_c &= \{\kappa_a; \kappa_1, \kappa_t^{\lambda_t}\}
\nn \\ 
&\kappa_t(q_t)\, \kappa_b(q_b) \to \kappa_2(q_2) \cdots \kappa_N(q_N)
\qquad &
\kappa_{N-1} &= \{\kappa_t^{\lambda_t}, \kappa_b; \kappa_2, \dots, \kappa_N\}
\,, \end{align}
and analogously for $\kappa_c'$ and $\kappa'_{N-1}$ with $\lambda_t \to \lambda_t'$. A new feature compared to \subsec{SCETc} is that the helicity of $\kappa_t$ can differ between the amplitude and conjugated amplitude, which can lead to nontrivial spin correlations.

The factorization formula is given by
\begin{align}\label{eq:factorization_Njet_SCETc}
\frac{\df \sigma_{c+} (\Phi_{N})}{\df \Tau_N} 
&= s_{\kappa_{N-1}}  s_{\kappa_c} \int\! \df s_a \df s_b \, B_{a}(s_a,x_a,\mu) \, B_{b}(s_b,x_b,\mu)\
\biggl[\prod_{k=1}^N \int\! \df s_k \, J_{k}(s_k,\mu) \biggr]
\nn \\ &\quad \times
\int\!\df k \sum_{\lambda_t,\lambda_t'} \vec C^\dagger_{c,\lambda_t'}(\Phi_c,\mu) \,\widehat S_{c,\kappa_c}(k, \rho_a, \rho_1, \mu) \, \vec C_{c,\lambda_t}(\Phi_c,\mu)
\nn \\ & \quad \times
\vec{C}^\dagger_{\lambda_t'}(\Phi_{N-1},\mu)
  \,\widehat{S}_{\kappa_{N-1}} \biggl(\Tau_N- \sum_{i=a}^N\frac{s_i}{Q_i} - \sqrt{\hat{s}_t} \,k,\{\hat s_{ij}\}, \mu\biggr)\,\vec{C}_{\lambda_t}(\Phi_{N-1},\mu)
\nn \\ &\quad \times
\biggl[1 + \ORd{\frac{m_J^2}{t},\frac{t}{Q^2}}\biggr]
\,,\end{align}
where $s_{\kappa_{N-1}}$ and $s_{\kappa_c}$ are the symmetry factors for each hard interaction process,
\begin{align}\label{eq:averaging_factors}
  s_{\kappa_c} s_{\kappa_{N-1}} = \frac{1}{N_c \de_{\kappa_t q} + (N_c^2-1) \de_{\kappa_t g}}\,  s_{\kappa}
\,.\end{align}
Here, $\Phi_c \equiv \{\kappa_c; \lambda_a, \lambda_1, \lambda_t; t, z, \varphi \}$ contains all information on the collinear splitting, whose phase space can be parametrized by the variables $t$, $z$, and an azimuthal angle $\varphi$. There is no phase-space factor, because the measurement is fully differential and the phase space factorizes $\df \Phi_N = \df \Phi_{N-1}\, \df \Phi_c$ in the collinear limit~\cite{Giele:1993dj}.

The short-distance scattering process $\kappa_{N-1}$ is described by the Wilson coefficient $\vec C(\Phi_{N-1})$, whose natural scale is $\mu_H \sim Q$, and which is a vector in the color space of the $N+1$ colored particles in $\kappa_{N-1}$. The corresponding soft function $\widehat{S}_{\kappa_{N-1}}$ is built out of Wilson lines in the directions of these $N\!+\!1$ partons. It depends on the angles $\{\hat{s}_{ij}\}$ of all well-resolved directions and in addition also on the measures $\rho_a$ and $\rho_1$ of the closeby jet and beam that determine the separation between their regions.

The matching coefficient $\vec C_c(\Phi_c)$ describes the (universal) $\kappa_a \to \kappa_1 \kappa_t$ splitting at the natural scale $\mu_{H_c} \sim \sqrt{-t}$. Although the color space for $\vec C_c$ is trivial [see \eq{Cc_simplify}],%
\footnote{Charge conjugation invariance prohibits the $d^{abc}$ color structure to all orders in $\alpha_s$. See also the footnote above \eq{Cc_simplify}.}
\begin{align} 
  \bar T^{a\,\bt \bar \ga} = \bigl(T^a_{\bt \bar \ga}\bigr)
  \,, \qquad
  \bar T^{abc} = \bigl(\img f^{abc} \bigr)
  \,, \qquad
  \vec C_c = \bigl( C_c \bigr)
\,,\end{align}
we keep our notation more general in anticipation of \sec{multihierarchy}. 
The matching coefficients $\vec C_{c,\kappa_c}$ can be obtained from the final-state collinear splitting amplitudes (see e.g.~ref.~\cite{Bern:1994zx}) using crossing. At tree level (dropping an irrelevant overall phase), we have
\begin{alignat}{2}\label{eq:C_c}
 C^\zero_{c,\{\bar q^-;g^+,\bar q^-_t\}}(t,z,\varphi,\mu)
 & = -\sqrt{2}g\,\frac{\sqrt{1-z}}{\sqrt{z}\, \langle a1 \rangle} 
 &&  = \sqrt{2}g\,\frac{1}{\sqrt{-t}}\, \frac{\sqrt{1-z}}{\sqrt{z}}\, e^{-\img \varphi} 
 \,, \nn \\
 C^\zero_{c,\{\bar q^-;g^-,\bar q^-_t\}}(t,z,\varphi,\mu)
 & = \sqrt{2}g\,\frac{1}{\sqrt{z(1-z)}\,[a1]}
 && = -\sqrt{2}g\,\frac{1}{\sqrt{-t}}\, \frac{1}{\sqrt{z(1-z)}}\, e^{\img \varphi} 
 \,, \nn \\
 C^\zero_{c,\{\bar q^-;\bar q^-,g^+_t\}}(t,z,\varphi,\mu)
 &= -\sqrt{2}g\,\frac{z}{(1-z)\langle a1\rangle}
 &&= \sqrt{2}g\,\frac{1}{\sqrt{-t}}\, \frac{z}{1-z}\, e^{-\img \varphi}
 \,, \nn \\
 C^\zero_{c,\{\bar q^-;\bar q^-,g^-_t\}}(t,z,\varphi,\mu)
 &= \sqrt{2}g\, \frac{1}{(1-z)[ a1 ]}
 &&= -\sqrt{2}g\,\frac{1}{\sqrt{-t}}\, \frac{1}{1-z}\, e^{\img \varphi}
 \,, \nn \\
 C^\zero_{c,\{g^-;g^+,g^-_t\}}(t,z,\varphi,\mu)
 &= -\sqrt{2}g\, \frac{1-z}{\sqrt{z}\, \langle a1 \rangle} 
 &&= \sqrt{2}g\,\frac{1}{\sqrt{-t}}\, \frac{1-z}{\sqrt{z}}\, e^{-\img \varphi}
 \,, \nn \\
 C^\zero_{c,\{g^-;g^-,g^-_t\}}(t,z,\varphi,\mu)
 &= \sqrt{2}g\, \frac{1}{\sqrt{z}(1-z)[a1]}
 &&= -\sqrt{2}g\,\frac{1}{\sqrt{-t}}\, \frac{1}{\sqrt{z}(1-z)}\, e^{\img \varphi}
 \,, \nn \\
 C^\zero_{c,\{g^+;g^+,g^-_t\}}(t,z,\varphi,\mu)
 &= \sqrt{2}g\,\frac{z^{3/2}}{(1-z)[a1]}
 &&= -\sqrt{2}g\, \frac{1}{\sqrt{-t}}\, \frac{z^{3/2}}{1-z}\, e^{\img \varphi}
 \,, \nn \\
 C^\zero_{c,\{g^+;g^-,g^-_t\}}(t,z,\varphi,\mu) &= 0
\,,\end{alignat}
where the subscript $t$ labels the off-shell parton $\kappa_t$.
We have written these both in terms of spinor products $\langle ij \rangle, [ij]$ (see e.g.~refs.~\cite{Dixon:1996wi, Dixon:2013uaa} for a review) of the first two partons in $\kappa_c$ and as function of $t$ and $\varphi$. For convenience we adopt a spinor convention here such that $\varphi$ is both the azimuthal angle and the phase.
All other channels can be obtained from parity and charge conjugation invariance, where parity flips all helicities and sends $\varphi \to \pi - \varphi$ and charge conjugation changes $q \lra \bar q$.

The parton type of $\kappa_t$ is completely fixed by $\kappa_a$ and $\kappa_1$ but its helicity is not, inducing correlations between $\vec C_{\kappa_{N-1}}$ and $\vec C_{c,\kappa_c}$. The helicity $\lambda_t$ of $\kappa_t$ in the amplitude and $\lambda_t'$ in the conjugate amplitude need not be the same and are summed over in \eq{factorization_Njet_SCETc}. This interference shows up when $\kappa_t$ is a gluon and introduces a dependence on the azimuthal angle $\varphi$ in $\vec C_c$. For example,
\begin{align} \label{eq:azim_dep}
  \vec C^\zero_{c,{\{\bar q^-;\bar q^-,g^+_t}\}} \, \vec C^{\dagger \zero}_{c,{\{\bar q^-;\bar q^-,g^-_t}\}}(t,z,\varphi,\mu) + \mathrm{h.c.}
  = 4g^2 C_F N_c\, \mathbf{1}\,\frac{1}{t}\, \frac{z}{(1-z)^2}\, \cos 2\varphi
\,,\end{align}
leading to a nonvanishing dependence on the azimuthal angle. It is straightforward to verify that for $\kappa_1=g$ in the soft limit $z \to 0$, $\vec C_{c,\lambda_t} \vec C^\dagger_{c,\lambda_t'}$ is independent of the gluon helicity and the azimuthal angle $\varphi$.

The csoft function $\widehat S_{c,\kappa_c}$ is fully determined by the $1 \to 2$ splitting and thus given in terms of \eq{csoftfct} by projecting onto the global $\Tau_N$ measurement,
\begin{align} \label{eq:csoftfctpp}
\widehat S_{c,\kappa_c}(k,\rho_a,\rho_1,\mu)
&=  \sum_{X_{cs}} \big\langle 0 \big\vert  \bar{T}[V^{\dagger \ga_t \al_t}_{n_t}  X^{\dagger \ga_a \al_a}_{n_a} X^{ \ga_{1} \al_{1}}_{n_1}] \big\vert X_{cs}\big\rangle \big\langle X_{cs}\big\vert T[X_{n_1}^{\dagger\bt_{1} \ga_{1}} X_{n_a}^{\bt_a \ga_a}  V_{n_t}^{\bt_t \ga_t}]\big\vert 0 \big\rangle \nn \\
& \quad \times  
\widehat{T}_{\kappa_c}^{-1}\, (\bar T^{\al_t \al_{a} \al_1})^\dagger \bar T^{\bt_t \bt_{a} \bt_1} 
 \delta\biggl(k - \frac{n_a \sdt k^{(a)}_{cs}}{\rho_a \sqrt{\hat{s}_t}} - \frac{n_1 \sdt k^{(1)}_{cs}}{\rho_1 \sqrt{\hat{s}_t}} \biggr)
\,. \end{align}
The Wilson lines $X_n$ and $V_n$ can be either in the fundamental or adjoint representation, as determined by $\kappa_c$. We emphasize that $\widehat S_{c,\kappa_c}$ now also depends on $\rho_a/\rho_1$ because we no longer assume $\rho_i=1$.
The different causal structure of the Wilson lines $X_{n_a}$ and $V_{n_t}$ (which enters through the $\img 0$-prescription of the eikonal propagator~\cite{Arnesen:2005nk}) does not affect the perturbative results, at least up to two-loop order~\cite{Kang:2015moa}.
Note that in contrast to $C_c$, both $\widehat{S}_{\kappa_{N-1}}$ and $\widehat{S}_{c,\kappa_c}$ are spin independent.

\subsection{$s+$ regime}
\label{subsec:scet_s}

We now consider the case where the first jet becomes soft but separated from the remaining energetic jets,
\begin{equation}
\w_1 \ll \w_{i\neq 1} \sim Q
\,,\qquad
\hat s_{ij} \sim 1
\qquad\Rightarrow\qquad
u \sim |s_{1k}| \ll |s_{kl}| \sim Q^2
\end{equation}
with $k,l \neq 1$. We now use $u$ as the generic soft dijet invariant mass scale. The full process separates into the hard interaction and soft splitting
\begin{align}
\kappa_a(q_a) \kappa_b(q_b) &\to \kappa_2(q_2) \cdots \kappa_N(q_N)
\qquad &
\kappa_{N-1} &= \{\kappa_a, \kappa_b; \kappa_2, \ldots, \kappa_N\}
\nn \\
&\to \kappa_1(q_1)\, \kappa_2(q_2) \cdots \kappa_N(q_N)
\qquad &
\kappa_N &= \{\kappa_a, \kappa_b; \kappa_1, \kappa_2, \ldots, \kappa_N\}
\,.\end{align}
The factorization procedure follows the same steps as in \subsec{scets}, but now involves the associated nontrivial color spaces, leading to the factorized cross section
\begin{align}\label{eq:factorization_Njet_SCETs}
\frac{\df \sigma_{s+} (\Phi_{N})}{\df \Tau_N} 
&=  s_{\kappa_{N-1}}\int\! \df s_a \df s_b \, B_{a}(s_a,x_a,\mu) \, B_{b}(s_b,x_b,\mu) \int\! \df s_1 \, J_g(s_1,\mu) \biggl[\prod_{k=2}^N \int\! \df s_k \, J_{k}(s_k,\mu)\biggr]
\nn \\ & \quad \times \,
 \vec{C}^\dagger(\Phi_{N-1},\mu) \,  \widehat{C}^\dagger_{s,\kappa}(\w_1,\{n_{i}\},\mu) \,
 \widehat{S}_\kappa\biggl(\Tau_N -\sum_{i=a}^N\frac{s_i}{Q_i} , \{\hat s_{ij}\}, \mu\biggr) \,
\nn \\ & \quad \times \widehat{C}_{s,\kappa}(\w_1,\{n_{i}\},\mu)  \,\vec{C}(\Phi_{N-1},\mu)
\biggl[1 +\ORd{\frac{m_J^2}{u},\frac{u}{Q^2}}\biggr]
\,,\end{align}
where $s_{\kappa_{N-1}}=s_{\kappa_N}$.
The hard matching coefficient $\vec{C}_{\kappa_{N-1}}$ is the same as in \eq{factorization_Njet_SCETc}, and $\widehat{S}_\kappa$ is the same soft function as in \eq{factorization_Njet_SCET}.

The soft jet is gluon initiated ($\kappa_1=g$) and generated by the soft splitting amplitude $\widehat{C}_{s,\kappa}$, which is
now a matrix converting the $(N+1)$-parton color space of $\vec{C}_{\kappa_{N-1}}$ to the $(N+2)$-parton color space that $\widehat{S}_\kappa$ acts on. It depends on the momentum $q_1$ of the soft parton as well as the directions of the hard partons in $\Phi_{N-1}$ but not on their helicities. Its natural scale is $\mu_{H_s} \sim \w_1 \sim u/Q$. At tree level, it is given by (see e.g.~ref.~\cite{Catani:2000pi} and references therein)
\begin{align}\label{eq:Cs+0}
\widehat{C}_{s,\kappa}^\zero(\w_1,\{n_{i}\},\mu) =  g(\mu) \sum_{i\neq1} \bfT_i^{a_1}  \, \frac{\varepsilon_{\lambda_1} \!\cdot q_i}{q_1 \cdot q_i}
\,,\end{align}
where $a_1$ is the color and $\varepsilon^\mu_{\lambda_1}$ the polarization vector of the gluon $\kappa_1$ with helicity $\lambda_1$.  The polarization vector leads to an angle-dependent phase, which, however, upon squaring drops out in the cross section. (This is no longer true in more complicated cases that require us to sum over the polarization at the amplitude level as in \eq{factorization_Njet_sc}.) Using the explicit spinor representation of the polarization vectors in \eq{polarization_vectors} in terms of an auxiliary momentum vector $k^\mu$, the tree-level matching coefficient reads e.g.~for $\lambda_1 =+$ 
\begin{align}\label{eq:Cs_spinor}
\widehat{C}_{s,\kappa}^\zero(\w_1,\{n_{i}\},\mu)
& =  \sqrt{2} g(\mu) \sum_{i\neq1} \bfT_i^{a_1}   \frac{[1|\slashed{q}_i| k\rangle}{\langle 1 i\rangle [i1] \langle k 1\rangle}
= \frac{\sqrt{2} g(\mu)}{N+1}\sum_{i \neq j\neq1} \bfT_i^{a_1}   \frac{\langle j i\rangle}{\langle 1 i\rangle \langle j 1\rangle}
\,,\end{align}
where in the second step we averaged over the $N+1$ different choices of $k^\mu = q_j^\mu$ with $j \neq 1$ for symmetry reasons and used $[1|\slashed{q}_i| j\rangle=[1i]\langle ij\rangle$.

At the cross section level, the kinematic dependence from $\widehat C_s^\dagger \dotsb \widehat C_s$ arises through the familiar soft factors
\begin{align} \label{eq:softkin}
\frac{s_{ij}}{s_{1i}\,s_{1j}}
= \frac{1}{\w_1^2} \,\frac{2n_i \sdt n_j}{n_1 \sdt n_i \,n_1 \sdt n_j}
= -\frac{1}{2} \sum_{\lambda_1}\frac{\varepsilon_{\lambda_1} \!\cdot q_i}{q_1 \cdot q_j} \, \frac{\varepsilon_{\lambda_1} \!\cdot q_j}{q_1 \cdot q_j}
\,.\end{align}
However, we emphasize that in contrast to \subsec{scets} the matching coefficients $\widehat{C}_{s,\kappa}$ and $\widehat{C}^\dagger_{s,\kappa}$ in \eq{factorization_Njet_SCETs} cannot be combined into a hard function matrix in color space.
Apart from this, our result in \eq{factorization_Njet_SCETs} for the factorized cross section in the soft jet limit agrees with the conjecture for it made in ref.~\cite{Larkoski:2015zka}.

The color charge operator $\bfT_i^{a_1}$ in \eqs{Cs+0}{Cs_spinor} transforms the color space from $(N+1)$-parton to $(N+2)$-parton color space,
\begin{align}
\bfT_i^{a_1} \, \bar T_{\kappa_{N-1}}^{\alpha_a \alpha_b \alpha_2 \dots\beta_i \dots\alpha_N}
&= t^{a_1}_{\alpha_i \beta_i} \bar T_{\kappa_{N-1}}^{\alpha_a \alpha_b \alpha_2 \dots\beta_i \dots\alpha_N}
\,,\end{align}
where the $\alpha_i$ and $\beta_i$ can be in the (anti)fundamental or adjoint representations, while $a_1$ is always in the adjoint representation since $\kappa_1=g$. The dependence on the partonic channel $\kappa$ enters through the representation of $t^{a_1}_{\al_i \beta_i}$
\begin{align}
t^{a_1}_{\al_i \beta_i} = 
\left\{\begin{array}{lr}
        T^{a_1}_{\al_i \beta_i} & \quad \text{if } \kappa_i = q \, ,\\
        - T^{a_1}_{\beta_i \al_i} & \quad  \text{if } \kappa_i = \bar q  \, ,\\
        \img f^{\al_i a_1 \beta_i} & \quad  \text{if }\kappa_i = g \, ,
        \end{array}\right.
\end{align}
where $T^{a_1}_{\al_i \beta_i}$ denote the usual $SU(3)$ generators and $f^{\al_i a_1 \beta_i}$ the structure constants. 

At higher orders in perturbation theory, the color of the emitted soft gluon is correlated to several external legs, resulting in a more involved structure for $\widehat{C}_{s,\kappa}$. At one loop~\cite{Catani:2000pi}
\begin{equation}\label{eq:Cs+1}
\widehat{C}_{s,\kappa}^\one(\w_1,\{n_{i}\},\mu) 
= g(\mu) \frac{\al_s(\mu)}{8\pi} \!\!\sum_{i \neq j\neq1}\!\!  \img f^{ba_1 c} \, \bfT^{b}_i \,  \bfT^{c}_j \Bigl(\frac{\varepsilon_{\lambda_1}\! \cdot q_i}{q_1 \cdot q_i} - \frac{\varepsilon_{\lambda_1}\! \cdot q_j}{q_1 \cdot q_j}\Bigr)\biggl[\ln^2\Bigl(\frac{- s_{1i}\, s_{1j} -\img0}{s_{ij}\, \mu^2}\Bigr) + \frac{\pi^2}{6}\biggr]
.\end{equation}
This result can also be obtained directly from the calculation in \app{H3s} by retaining the general color charge operators.
Note that crossing momenta does not affect the overall sign of the argument of the logarithm in \eq{Cs+1}. Using \eq{softkin} for the argument of the logarithm, we can see that the natural scale for $\widehat C_{s,\kappa}$ is indeed $\mu_{H_s}\sim \w_1$ since $n_i\cdot n_j\sim 1$.

The general form of the anomalous dimension for $\widehat{C}_{s,\kappa}$ can be derived from RG consistency analogous to \subsec{scets}. The additional factorization in the $s+$ regime with respect to \SCET concerns only the hard matching, $\vec{C}(\Phi_N, \mu)|_{u\ll Q^2} = \widehat{C}_{s,\kappa}(\w_1, \{n_i\}, \mu) \vec{C}(\Phi_{N-1},\mu)$, which requires the $\mu$ dependence to satisfy
\begin{equation} \label{eq:consistency}
\frac{\df}{\df\ln\mu}\,\widehat{C}_{s,\kappa}(\mu)
= \hga_{C_{\kappa_N}}(\{s_{ij}\}, \mu)\big|_{u\ll Q^2}\, \widehat{C}_{s,\kappa}(\mu)
- \widehat{C}_{s,\kappa}(\mu)\, \hga_{C_{\kappa_{N-1}}}(\{s_{ij}\}, \mu) 
\,.\end{equation}
The general all-order structure of the hard anomalous dimension follows from the $\mu$ independence of the cross section (see e.g.~refs.~\cite{Chiu:2008vv, Gardi:2009qi, Becher:2009qa}) and is given by
\begin{align} \label{eq:gamma_C_N}
\hga_{C_\kappa}(\{s_{ij}\}, \mu)
&= - \Gamma_\cusp[\alpha_s(\mu)] \sum_{i<j}{\bfT_i \cdot \bfT_j} \ln\Bigl(\frac{-s_{ij}- \img 0}{\mu^2} \Bigr)
+ \hga_{C_\kappa}[\alpha_s(\mu)]
\,,\nn\\
\hga_{C_\kappa}(\alpha_s)
&= {\bf 1} \sum_i \gamma_C^{\kappa_i}(\alpha_s) + \ord{\alpha_s^3}
\,,\end{align}
where $\bfT_i \cdot \bfT_j \equiv \sum_a \bfT_i^a  \bfT_j^a$. The noncusp anomalous dimension $\hga_{C_\kappa}(\alpha_s)$ is proportional to the identity operator and independent of the $s_{ij}$ up to two loops~\cite{Aybat:2006mz} but not beyond~\cite{Almelid:2015jia}. Combining \eqs{consistency}{gamma_C_N} yields
\begin{align} \label{eq:gamma_C_s}
\frac{\df}{\df \ln\mu}\,\widehat{C}_{s,\kappa}(\mu)
&= - \Gamma_\cusp[\alpha_s(\mu)] \biggl\{ \sum_{i< j \neq 1}\!\! \ln\Bigl(\frac{-s_{ij}- \img0}{\mu^2} \Bigr) \bigl[\bfT_i \cdot \bfT_j, \widehat{C}_{s,\kappa}(\mu) \bigr]
\\ \nn & \quad
+ \sum_{i \neq 1} \ln\Bigl(\frac{-s_{1i}-\img0}{\mu^2} \Bigr) \bfT_1 \cdot \bfT_i\, \widehat{C}_{s,\kappa}(\mu) \biggr\}
+ \Bigl\{\gamma_C^g[\alpha_s(\mu)] + \ord{\al_s^3}\Bigr\} \widehat{C}_{s,\kappa}(\mu)
\,.\end{align}

While the two terms proportional to $\Gamma_\cusp$ in \eq{gamma_C_s} separately depend on $s_{ij}$ and $s_{1i}$, they must combine into logarithms of $s_{1i} s_{1j}/s_{ij}$, which only depend on $q_1$ and $\{n_{i\neq 1}\}$, which imposes a constraint on the color structure of $\widehat C_{s,\kappa}$. We can check explicitly at one loop how this happens by inserting the tree-level expression \eq{Cs+0} into the right-hand side. Using the color identities
\begin{align}
  \bigl[\bfT_i \cdot \bfT_j, \bfT_k^a \bigr]
  &= \img f^{bac} \big( \de_{ik}\, \bfT_j^b \bfT_k^c
  +\de_{jk}\, \bfT_i^b \bfT_k^c \big)
  &  & \text{for } i \neq j \neq 1
  \,,\nn \\
  (\bfT_1 \sdt \bfT_i\, \widehat\bfT_k)^{a}
  &= 
  - \img f^{bac}\, \bfT_i^b\, \bfT_k^c 
    &  & \text{for } i,k \neq 1
 \,, \nn \\
 \sum_{i\neq 1} \bfT_i^a &= 0
     &  & \text{on $N\!+\!1$ parton color space}
\,,\end{align}
and the anomalous dimensions in \app{anom_dim}, we find at one-loop order
\begin{align} 
\frac{\df}{\df \ln\mu}\,\widehat{C}_{s,\kappa}
&= g(\mu)\, \frac{\al_s(\mu)}{2\pi} \biggl[
\sum_{i \neq j\neq1}  \img f^{ba_1 c} \, \bfT^{b}_i \,  \bfT^{c}_j \, \Bigl(\frac{\varepsilon_{\lambda_1}\! \cdot q_i}{q_1 \cdot q_i} - \frac{\varepsilon_{\lambda_1}\! \cdot q_j}{q_1 \cdot q_j}\Bigr) \ln\Bigl(\frac{-s_{ij}-\img0}{\mu^2}\Bigr)
\\ & \quad
+ 2\sum_{i,j\neq1}  \img f^{ba_1 c} \, \bfT^{b}_i \,  \bfT^{c}_j \, \frac{\varepsilon_{\lambda_1}\! \cdot q_j}{q_1 \cdot q_j}\, \ln\Bigl(\frac{s_{1i}-\img0}{\mu^2}\Bigr)
- \frac{\beta_0}{2} \sum_{i\neq1} \bfT_i^{a_1}  \, \frac{\varepsilon_{\lambda_1} \!\cdot q_i}{q_1 \cdot q_i}
\biggr]
\nn \\
&= -g(\mu) \frac{\al_s(\mu)}{2\pi} \biggl[\sum_{i \neq j\neq1}\!\!  \img f^{ba_1 c} \, \bfT^{b}_i \,  \bfT^{c}_j  \Bigl(\frac{\varepsilon_{\lambda_1}\! \cdot q_i}{q_1 \cdot q_i} - \frac{\varepsilon_{\lambda_1}\! \cdot q_j}{q_1 \cdot q_j}\Bigr) \ln\Bigl(\frac{-s_{1i}\, s_{1j}-\img0}{s_{ij}\, \mu^2}\Bigr)
\nn \\ & \qquad\qquad\qquad
+ \frac{\beta_0}{2} \sum_{i\neq1} \bfT_i^{a_1}  \frac{\varepsilon_{\lambda_1} \!\cdot q_i}{q_1 \cdot q_i} \biggr]
.\nn \end{align}
This also agrees with directly taking the $\mu$-derivative of $\widehat{C}_{s,\kappa}^\zero+\widehat{C}_{s,\kappa}^\one$
using \eqs{Cs+0}{Cs+1}.

\subsection{$cs+$ regime}
\label{subsec:Njet_scetcs}

Finally, we consider the case, where the first jet becomes soft and close to a beam,
\begin{equation}
0< -t=-s_{a1} \ll u \sim |s_{1i}| \ll |s_{jk}| \sim Q^2
\,,\qquad
z = \frac{\w_1}{\w_a}
\,, \qquad
\hat{s}_t = \hat s_{a1}
\end{equation}
for $i \neq a$ and $j,k \neq 1$, which is in close analogy to the case of a soft jet close to a final-state jet. The hard process now splits in the same way as in the $c+$ case in \eq{c+process}. The factorization proceeds as in \subsec{scetcs} and the result for the factorized cross section corresponds to the soft limit of \eq{factorization_Njet_SCETc} or the collinear limit of \eq{factorization_Njet_SCETs},
\begin{align}\label{eq:factorization_Njet_SCETcs}
\frac{\df \sigma_{cs+} (\Phi_{N})}{\df \Tau_N} 
&= s_{\kappa_{N-1}} s_{\kappa_c}\int\! \df s_a \df s_b \, B_a(s_a,x_a,\mu) \, B_b(s_b,x_b,\mu) \int\! \df s_1 \, J_g(s_1,\mu)
\nn \\ &\quad \times \biggl[\prod_{k=2}^N \int\! \df s_k \, J_{k}(s_k,\mu) \biggr]
\int\! \df k\,  \vec C^\dagger_{cs,\kappa_c}(t\,z,\mu)  \,\widehat S_{c,\kappa_c}(k,\rho_a,\rho_1,\mu)\, \vec C_{cs,\kappa_c}(t\,z,\mu)
\nn \\ & \quad \times 
  \vec{C}^\dagger(\Phi_{N-1},\mu) \,\widehat{S}_{\kappa_{N-1}} \biggl(\Tau_N-\sum_{i=a}^N\frac{s_i}{Q_i}-\!\sqrt{\hat{s}_t}\, k,\{\hat s_{ij}\},\mu\biggr)\, \vec{C}(\Phi_{N-1},\mu)
 \nn \\ & \quad \times
\biggl[1+\ORd{\frac{m_J^2}{t},\frac{t}{u},\frac{u}{Q^2}}\biggr]
\,. \end{align}
The hard coefficient $\vec C(\Phi_{N-1})$ as well as the csoft function $\widehat S_{c,\kappa_c}$ and the soft function $\widehat S_{\kappa_{N-1}}$ are the same as in the $c+$ regime in \eq{factorization_Njet_SCETc}.

The Wilson coefficient $\vec C_{cs,\kappa_c}$ now describes the collinear-soft splitting at its natural scale $\mu_{H_{cs}} \sim \sqrt{-t \,z}$. It can be obtained from the soft limit $z \to 0$ of the collinear matching coefficient $\vec C_{c,\kappa_c}$, e.g.~at tree-level for $\lambda_1 =+$
 \begin{align}\label{eq:Ccs_coll}
C^{(0)}_{c,\kappa_c}(t, z, \varphi, \mu) \big|_{z \to 0}
 = - \delta_{\kappa_1g}\, \de_{\lambda_t \lambda_a} \,\sqrt{2} g(\mu) \,\frac{1}{\sqrt{z} \, \langle a 1\rangle} = \delta_{\kappa_1 g}\, \de_{\lambda_t\lambda_a} \, C^{(0)}_{cs,\kappa_c}(tz, \mu) \, .
 \end{align}
The spin correlations and interference effects that were present in the $c+$ case now vanish between $\vec C(\Phi_{N-1})$ and $\vec C_{cs,\kappa_c}$ because the helicity of the initial splitting parton does not change as shown by the factor $\de_{\lambda_t\lambda_a}$ in \eq{Ccs_coll}. The associated hard function defined by\footnote{We include the color averaging factors for the sake of a common normalization with the soft jet case.}
\begin{align}
s_{\kappa} \widehat H_{cs,\kappa_c} (t \, z,\mu)= s_{\kappa_{N-1}} s_{\kappa_c} \,\vec C_{cs,\kappa_c}(t \, z,\mu) \, \vec C^\dagger_{cs,\kappa_c}(t \, z, \mu) \,
\end{align}
thus has at leading order the familiar expression
\begin{align} \label{eq:Hcs_coll}
\widehat H^{(0)}_{cs,\kappa_c}(t \, z ,\mu)
& = 8\pi\alpha_s(\mu) \frac{1}{-t \,z} \,\bfT^2_{t}
\,.\end{align}

Alternatively, $\vec C_{cs,\kappa_c}$ can be obtained from the collinear limit $s_{a1}/s_{1i} \sim t/u \to 0$ of $\widehat C_{s,\kappa}$, resulting in a dependence only on the one-dimensional color space related to the subprocess $\kappa_c$. Using \eq{Cs_spinor} for the tree-level expression of $\widehat C_{s,\kappa}$ with $\lambda_1 =+$, only terms with $i=a$ or $j =a$ contribute at leading order in the collinear limit, which yields
\begin{align}\label{eq:Ccs_soft}
\widehat{C}_{s,\kappa}^\zero  \Big|_{\abs{t} \ll u}
&= \frac{\sqrt{2} g(\mu)}{N+1} \bigg(\bfT_a^{a_1} \sum_{j\neq1,a}   \frac{\langle j a\rangle}{\langle 1 a\rangle \langle j 1\rangle} + \sum_{i\neq1,a} \bfT_i^{a_1}   \frac{\langle a i\rangle}{\langle 1 i\rangle \langle a 1\rangle}\bigg) \nn \\
 & = 
- \sqrt{2} g(\mu) \, \bfT_a^{a_1} \frac{1}{\sqrt{z} \, \langle a 1\rangle} = \bfT_a^{a_1}C^\zero_{cs,\kappa_c}
\equiv \bar T_{\kappa_c}^{a_1 \alpha_a \bt_a}\cdot \vec{C}_{cs}^\zero
\equiv \widehat{C}_{cs,\kappa_c}^\zero \, ,
\end{align}
where we used
\begin{align} \label{eq:collinear_relations}
  \langle 1 i\rangle & = \sqrt{z}  \langle a i\rangle\Bigl[1 + \ORd{\sqrt{\abs{t}/u}} \Bigr]
  && \text{for } i\neq a,1
  \,,\nn \\
 \sum_{i\neq 1} \bfT_i^{a_1} &= 0
     &  & \text{on $(N\!+\!1)$-parton color space}
.\end{align}
In the last line of \eq{Ccs_soft} we highlighted that the internal one-dimensional color space of $\vec C_{cs}$ behaves as a color matrix proportional to the color charge operator $\bfT_a^{a_1}$, going from $(N+1)$-parton to $(N+2)$-parton color space. In other words, the internal color space of the $1\to 2$ splitting comes as a tensor product with the $(N+1)$-parton color space.

To see more explicitly that the collinear limit of the $s+$ regime coincides with $cs+$ regime at the level of the factorization theorem, we first note that also the soft functions need to satisfy
\begin{equation}\label{eq:Softfct_factorized}
\widehat{S}_\kappa(\ell,\{\hat s_{ij}\},\mu) \big\vert_{\abs{t}\ll u}
= \frac{1}{\bfT_a^2} \int\! \df k\,  \widehat \bfT_{a}
\widehat{S}_{\kappa_{N-1}}\bigl(\ell-\sqrt{\hat{s}_t}\, k,\{\hat s_{ij}\},\mu\bigr) \widehat \bfT_a^\dagger \otimes
\widehat{S}_{c,\kappa_c}(k,\rho_a,\rho_1,\mu)
\,,\end{equation}
where $\widehat \bfT_a$ corresponds to the action of $\bfT_a^{a_1}$.
This agrees with eq.~(6.38) of ref.~\cite{Bauer:2011uc} (where $\widehat\bfT_t = \widehat\bfT_a$ since we are in the soft limit). The $\otimes$ indicates that $\widehat{S}_{\kappa_{N-1}}$ and $\widehat{S}_{c,\kappa_c}$ formally live in different color spaces. Thus we can write
\begin{align}
\vec{C}^\dagger_{\kappa_{N-1}}\widehat{C}^{\dagger}_{s,\kappa}\, \widehat{S}_{\kappa}\, \widehat{C}_{s,\kappa} \,\vec{C}_{\kappa_{N-1}}
\Big|_{\abs{t}\ll u}
& = \bfT_a^2 \, (\vec{C}^*_{cs,\kappa_c})^T \,  \widehat{S}_{c,\kappa_c}\, \vec{C}_{cs,\kappa_c} \times\, \vec{C}^\dagger_{\kappa_{N-1}} \widehat{S}_{\kappa_{N-1}} \vec{C}_{\kappa_{N-1}}
\\
& = \frac{s_{\kappa_{N-1}} s_{\kappa_{c}}}{s_{\kappa}}\, \vec{C}^\dagger_{cs,\kappa_c} \,  \widehat{S}_{c,\kappa_c}\, \vec{C}_{cs,\kappa_c} \times\, \vec{C}^\dagger_{\kappa_{N-1}} \,\widehat{S}_{\kappa_{N-1}} \vec{C}_{\kappa_{N-1}} \, , \nn
\end{align}
using $(\vec{C}_{cs,\kappa_c}^*)^T = \vec{C}_{cs,\kappa_c}^\dagger \widehat{T}_{\kappa_c}^{-1}$ with $\widehat{T}_{\kappa_c} = \bfT_a^2[N_c \de_{\kappa_a q} + (N_c^2-1) \de_{\kappa_a g}]$ and \eq{averaging_factors}. This demonstrates the relation between the factorization theorems in \eqs{factorization_Njet_SCETs}{factorization_Njet_SCETcs} explicitly.

Finally, we also give the one-loop result for the hard function $\widehat H_{cs,\kappa_c}$, which is in direct correspondence to the expression for $\widehat C^{(1)}_{s,\kappa}$ in \eq{Cs+1},\footnote{The result is independent of $\lambda_1$. Since we consider fixed helicity, \eq{Hcs_oneloop} differs by a factor of $1/2$ with respect to \eq{H3s_oneloop}, in addition to phase space factors that we have not included here.}
\begin{align} \label{eq:Hcs_oneloop}
\widehat H^{(1)}_{cs,\kappa_c}(t\,z, \mu)
&= 
\Bigl[\widehat C^{(0)}_{s,\kappa} \widehat C^{(1)\dagger}_{s,\kappa} +
\widehat C^{(1)}_{s,\kappa} \widehat C^{(0)\dagger}_{s,\kappa}\Bigr]_{\abs{t}\ll u}
\nn \\
&  =  - 2\alpha^2_s(\mu) \sum_{i\neq j\neq 1,a} \img f^{b a_1 c}\, \bfT_i^{a_1} \bfT_a^{b} \bfT_j^{c}\,
\frac{s_{ia}}{s_{1i}s_{1a}}
\biggl[\ln^2\Bigl(\frac{-s_{1a} s_{1j}-\img0}{s_{aj} \mu^2}\Bigr) + \frac{\pi^2}{6}\biggr] + \text{h.c.}
\nn \\
&= 8\pi\alpha_s(\mu)\, \bfT_a^2\, \frac{1}{t\,z} \frac{\alpha_s(\mu) C_A}{4\pi} \biggl[\ln^2\Bigl(\frac{-t\,z-\img0}{\mu^2}\Bigr) - \frac{5\pi^2}{6} \biggr]
\,,\end{align}
using the relations in \eq{collinear_relations}. The fact that only in the collinear limit the soft splitting amplitude collapses onto a one-dimensional color subspace, which renders the expression for $\widehat H_{cs,\kappa_c}$ independent of all widely separated partons, represents a key difference to the case of $e^+ e^- \to 3$ jets, where $\widehat H_{3,cs}=\widehat H_{3,s}$ holds without any additional expansion.

\section{Multiple Hierarchies}
\label{sec:multihierarchy}

Up to this point we have restricted our attention to kinematic hierarchies induced by one splitting process, resulting in one soft jet or two nearby jets (or their combination). For $e^+ e^- \to 3$ jets this describes all possible kinematic configurations. However, for $pp \to N $ jets, we can encounter more complicated kinematic hierarchies. These can arise due to splitting processes at different scales, which can be independent (\subsec{indep}) or strongly ordered in energies or angles (\subsec{so}). In addition, multiple emissions can arise from the same splitting, which we discuss in \subsec{bso}.

\subsection{Independent hierarchies}
\label{subsec:indep}

First we discuss the case that jet hierarchies arise from splitting processes in separate sectors, which allows us to perform the respective matching steps independently of each other. 
If collinear splittings occur in different collinear sectors leading to independent pairs of nearby jets or beams, they are described by iterating the results in \subsec{scet_c}. The same naturally holds when some of these splittings are collinear-soft. All these cases can also be combined with a single soft jet, as we now discuss in the context of an example.

We consider a kinematic hierarchy with one soft jet (labelled as 1) and two nearby jets (labelled as 2 and 3), such that
\begin{equation}
\w_1 \ll \w_i \sim Q
\,,\qquad s_{23} \ll |s_{ij}| \sim Q^2
\,,\qquad
z = \frac{\w_2}{\w_2+\w_3}
\end{equation}
with $i \neq j \neq 1$. This situation is relevant in the context of jet substructure, when performing the first step in the resummation of the leading nonglobal logarithms in the dressed gluon approximation of ref.~\cite{Larkoski:2015zka} for a resolved pair of jets next to each other.

The soft and collinear splittings are independent of each other since the resolved collinear emission only affects the sectors originating from the $n_{23}$-collinear mode describing the parent fat jet above the scale $s_{23}$ and since the soft emission does not resolve the two nearby jets at leading order in the power counting. The partonic content of the associated subprocesses is given by
\begin{align}
\kappa_a(q_a) \kappa_b(q_b) &\to \kappa_t(q_t)\, \kappa_4(q_4)\dotsb \kappa_N(q_N)
&\kappa_{N-2} &= \{\kappa_a, \kappa_b; \kappa_t^{\lambda_t}, \kappa_4, \dots, \kappa_N\}
\nn \\
&\to \kappa_1(q_1)\, \kappa_t(q_t)\, \kappa_4(q_4)\dotsb \kappa_N(q_N)
\qquad &
\kappa_{N-1} &= \{\kappa_a, \kappa_b; \kappa_1, \kappa_t^{\lambda_t}, \kappa_4, \dots, \kappa_N\}
\nn \\ & \qquad\qquad\,\,\,
\kappa_t(q_t) \to \kappa_2(q_2) \kappa_3(q_3)
\qquad &
\kappa_c &= \{\kappa_t^{\lambda_t}; \kappa_2, \kappa_3\} 
\,. \end{align}
Applying the results from \subsecs{scet_c}{scet_s} the active-parton factorization reads
 \begin{align}\label{eq:factorization_Njet_cs}
\frac{\df \sigma_{+} (\Phi_{N})}{\df \Tau_N} 
&= s_{\kappa_{N-2}}  s_{\kappa_c} \int\! \df s_a \df s_b \, B_a(s_a,x_a,\mu) \, B_b(s_b,x_b,\mu) \int\! \df s_1 \, J_{g}(s_1,\mu)
 \biggl[\prod_{k=2}^N \int\! \df s_k \, J_k(s_k,\mu) \biggr]
\nn \\ &\quad \times
 \int\! \df k \sum_{\lambda_t,\lambda_t'} \vec C^\dagger_{c,\lambda_t'}(s_{23},z,\varphi,\mu) \,\widehat S_{c,\kappa_c}(k,\mu) \, \vec C_{c,\lambda_t}(s_{23},z,\varphi,\mu)
 \nn \\ & \quad \times
 \vec{C}^\dagger_{\lambda_t'}(\Phi_{N-2},\mu)\,
\widehat{C}^\dagger_{s,\kappa_{N-1}}(\w_1,\{n_{i}\}_{\kappa_{N-1}},\mu)
  \,\widehat{S}_{\kappa_{N-1}} \biggl(\Tau_N- \sum_{i=a}^N\frac{s_i}{Q_i} - \sqrt{\hat{s}_{23}} \,k,\{\hat{s}_{ij}\}, \mu\biggr)
\nn \\ &\quad \times
\widehat{C}_{s,\kappa_{N-1}}(\w_1,\{n_{i}\}_{\kappa_{N-1}},\mu)  \,\vec{C}_{\lambda_t}(\Phi_{N-2},\mu)\,
  \biggl[1+\ORd{\frac{m_J^2}{s_{23}},\frac{m_J^2}{Q \w_1},\frac{\w_1}{Q},\frac{s_{23}}{Q^2}}\biggr]
\,,\end{align}
We stress that the relative hierarchy between the invariant mass scales of the soft and collinear splitting, $\mu^2_{H_s} \sim s_{1i} s_{1j}/s_{ij} \sim \w_1^2$ and $\mu^2_{H_c} \sim s_{23}$, is irrelevant for setting up the factorization in this case.

\subsection{Strong ordering in angles or energies}
\label{subsec:so}

Moving on to more complicated hierarchies, we consider the case where the consecutive hierarchies are strongly ordered in their angles or energies (this condition will be relaxed in \subsec{bso}). The strong ordering enables an iterative treatment. We separately consider the case of multiple jets that are close to each other with a strong ordering in their angles and multiple soft jets with a strong ordering in their energies.
These two cases can be combined as in \subsec{indep} if they involve independent sectors.

\subsubsection{Strong ordering in angles}
\label{subsec:so+}

Let us start by considering the case where all jets are equally energetic, $\w_i \sim Q$, and $M$ jets are close to each other ordered in their angles. (The case of $M\!-\!1$ jets close to a beam is related by crossing, requiring minor modifications as in \subsec{scet_c}.)
This is described by iterating the $c+$ in \subsecs{SCETc}{scet_c}, where in each successive step the virtuality is lowered and an additional (proto)jet becomes separately resolved, building a tree of $1\to 2$ splittings. Strong ordering requires that angles are parametrically smaller as one follows any path down this tree, but angles of independent branches \emph{do not} have to be strongly ordered with respect to each other (see \subsec{indep}). This picture resembles a parton shower but is not limited to leading-logarithmic accuracy.

To illustrate this with a specific example, we take $M = 3$ with
\begin{align}
s_{12} \ll s_{123}  \ll Q^2
\,,\qquad
\w_i \sim Q
\,.\end{align}
The partonic process is separated into
\begin{align}
\kappa_a(q_a)\, \kappa_b(q_b) \to &\kappa_{123}(q_{123})\, \kappa_4(q_4)\dotsb \kappa_N(q_N)
\qquad
&\kappa_{N-2} &= \{\kappa_a, \kappa_b; \kappa_{123}^{\lambda_{123}}, \kappa_4, \dots, \kappa_N\}
\nn \\
&\kappa_{123}(q_{123}) \to \kappa_{12}(q_{12})\, \kappa_3(q_3)
\qquad
&\kappa_c &= \{\kappa_{123}^{\lambda_{123}}; \kappa_{12} ^{\lambda_{12}}, \kappa_3\}
\nn \\
&\qquad\qquad\qquad\!
\kappa_{12}(q_{12}) \to \kappa_1(q_1)\, \kappa_2(q_2)
&\tilde \kappa_c &= \{\kappa_{12} ^{\lambda_{12}}; \kappa_1, \kappa_2\}
\,.\end{align}
The helicities of $\kappa_{123}$ and $\kappa_{12}$ in the amplitude are denoted by $\lambda_{123}, \lambda_{12}$ and in the conjugate amplitude by $\lambda_{123}',\lambda_{12}'$, encoding spin correlations between the hard process at the scale $\mu \sim Q$ and the two collinear splittings.
The corresponding factorization formula is given by
\begin{align}\label{eq:factorization_Njet_cc}
\frac{\df \sigma_{+} (\Phi_{N})}{\df \Tau_N} 
&= s_{\kappa_{N-2}} s_{\kappa_c} s_{\tilde \kappa_c} \int\! \df s_a \df s_b \, B_a(s_a,x_a,\mu) \, B_b(s_b,x_b,\mu)
\biggl[\prod_{k=1}^N \int\! \df s_k \, J_{k}(s_k,\mu) \biggr]
\nn \\ &\quad \times 
\int\! \df k \sum_{\lambda_{123},\lambda_{123}'} \vec C^\dagger_{c,\lambda_{123}'}(s_{123},  z_1+z_2,\varphi,\mu) \,\widehat S_{c,\kappa_c}(k,\mu) \, \vec C_{c,\lambda_{123}}(s_{123}, z_1+z_2,\varphi,\mu)
\nn \\ &\quad \times \int\! \df \tilde k
\sum_{\lambda_{12},\lambda_{12}'} \vec C^\dagger_{c,\lambda_{12}'}\Bigl(s_{12},\frac{z_1}{z_1+z_2},\tilde \varphi,\mu\Bigr)  \,
\widehat S_{c,\tilde \kappa_c}(\tilde k,\mu)\, \vec C_{c,\lambda_{12}}\Bigl(s_{12},\frac{z_1}{z_1+z_2},\tilde \varphi,\mu\Bigr)
\nn \\ & \quad \times 
\vec{C}^\dagger_{\lambda_{123}'}(\Phi_{N-2},\mu)\,
  \widehat{S}_{\kappa_{N-2}} \biggl(\Tau_N- \sum_{i=a}^N\frac{s_i}{Q_i}-\sqrt{\hat{s}_{123}}\,k-\sqrt{\hat{s}_{12}}\,\tilde k,\{\hat s_{ij}\},\mu\biggr)
  \nn \\ & \quad \times \vec{C}_{\lambda_{123}}(\Phi_{N-2},\mu)\,
\biggl[1+\ORd{\frac{m_J^2}{s_{12}},\frac{s_{12}}{s_{123}},\frac{s_{123}}{Q^2}}\biggr]
\,,\end{align}
with
\begin{align} \label{eq:s123_def}
\hat s_{12} = \frac{s_{12}}{Q_1 Q_2}
\,, \qquad
\hat s_{123} = \frac{s_{123}}{(Q_1 + Q_2)Q_3}
\, ,\qquad 
z_i \equiv \frac{\w_i}{\w_1+\w_2+\w_3}
\,.\end{align}
The csoft function $\widehat S_{c,\tilde \kappa_c}$ communicates between the jets 1 and 2 at the natural scale $\mu \sim m_J^2/\sqrt{s_{12}}$ and is the same as in \eq{csoftfctpp}. The $\widehat S_{c,\kappa_c}$ describes the csoft radiation between the protojet (12) and jet 3 at the scale $\mu \sim m_J^2/\sqrt{s_{123}}$. It has the same Wilson line structure as in \eq{csoftfctpp} and is given by
\begin{align}
\widehat S_{c,\kappa_c}(k,\mu) = \int\! \df k_1 \, \df k_2 \, \df k_3 \,\widehat  S_{c,\kappa_c}(k_1,k_2,k_3 ,\mu) \, \delta (k-k_1-k_2-k_3) \,,\end{align}
where
\begin{align}
\widehat S_{c,\kappa_c}(k_1,k_2,k_3,\mu)
&=  \sum_{X_{cs}} \big\langle 0 \big\vert  \bar{T}[V^{\dagger \ga \al}_{n_{123}}  X^{\ga_3 \al_3}_{n_3}
X^{\tilde \ga \tilde\al }_{n_{12}}] \big\vert X_{cs}\big\rangle \big\langle X_{cs}\big\vert T[X_{n_{12}}^{\dagger\tilde \bt \tilde\ga} X_{n_3}^{\dagger\bt_3 \ga_3}  V_{n_{123}}^{\bt \ga}]\big\vert 0 \big\rangle
\nn \\ & \quad \times  
\widehat{T}_{\kappa_c}^{-1}\, (\bar T^{\al \tilde \al \al_3})^\dagger \bar T^{\bt \tilde \bt \bt_3}
\delta\Bigl(k_3 - \frac{n_3 \sdt k^{(3)}_{cs}}{\rho_3 \sqrt{\hat{s}_{123}}}\Bigr)
\prod_{i=1}^2
\delta\Bigl(k_i - \frac{n_{12} \sdt k^{(i)}_{cs}}{\rho_i \sqrt{\hat{s}_{123}}}\Bigr)
\,. \end{align}
The representations of the Wilson lines $X$ and $V$ and the color indices $\alpha_i, \beta_i, \ga_i$ are determined by $\kappa_c$.
It now resolves the contribution to the measurement of $\Tau_N^{(1)}$, $\Tau_N^{(2)}$, and $\Tau_N^{(3)}$.
Even though there is only one Wilson line in the combined $n_{12}$ direction, the measurement is separated into $k_1$ and $k_2$. This is analogous to the soft function $\widehat S_{q\bar q}$ in \eq{S2_def}, which is built out of two Wilson lines but separates the contribution to $\Tau_3$ from all three jet regions.

\subsubsection{Strong ordering in energies}

We next consider the case where the first $M$ jets are soft and strongly ordered in their energies, while all jets are well separated,
\begin{equation}
\w_1 \ll \w_2 \ll \dots \ll \w_M \ll \w_k \sim Q
\,,\qquad
\hat s_{ij} \sim 1
\,.\end{equation}
where $k\notin \{1, \ldots, M\}$.
This is described by iterating the $s+$ regime in \subsecs{scets}{scet_s} and corresponds to a sequential rather than a tree-like structure.

For example, for $M=2$ this involves the subprocesses
\begin{align}
\kappa_a(q_a) \kappa_b(q_b) &\to \kappa_3(q_3) \cdots \kappa_N(q_N)
\qquad &
\kappa_{N-2} &= \{\kappa_a, \kappa_b; \kappa_3, \ldots, \kappa_N\}
\nn \\
&\to \kappa_2(q_2)\, \kappa_3(q_3) \cdots \kappa_N(q_N)
\qquad &
\kappa_{N-1} &= \{\kappa_a, \kappa_b; \kappa_2, \kappa_3, \ldots, \kappa_N\}
\nn \\
&\to \kappa_1(q_1)\, \kappa_2(q_2)\, \kappa_2(q_2) \cdots \kappa_N(q_N)
\qquad &
\kappa_N &= \{\kappa_a, \kappa_b; \kappa_1, \kappa_2, \kappa_3, \ldots, \kappa_N\}
\,.\end{align}
The corresponding factorization formula is given by
\begin{align}\label{eq:factorization_Njet_ss}
\frac{\df \sigma_{+} (\Phi_{N})}{\df \Tau_N} 
&= s_{\kappa_{N-2}} \int\! \df s_a \df s_b \, B_a(s_a,x_a,\mu) \, B_b(s_b,x_b,\mu) \int\! \df s_1 \, \df s_2\, J_g(s_1,\mu)\, J_g(s_2,\mu)
\nn \\ & \quad \times
 \biggl[\prod_{k=3}^N \int\! \df s_k \, J_{k}(s_k,\mu)\biggr]
 \vec{C}^\dagger(\Phi_{N-2},\mu) \,\widehat{C}^\dagger_{s,\kappa_{N-1}}(\w_2,\{n_{i}\}_{\kappa_{N-1}},\mu) \,  \widehat{C}^\dagger_{s,\kappa}(\w_1,\{n_i\},\mu) \,
\nn \\ & \quad \times \widehat{S}_\kappa\biggl(\Tau_N-\sum_{i=a}^N\frac{s_i}{Q_i},\{\hat{s}_{ij}\}, \mu\biggr)
\, \widehat{C}_{s,\kappa}(\w_1,\{n_{i}\},\mu)\,   \widehat{C}_{s,\kappa_{N-1}}(\w_2,\{n_i\}_{\kappa_{N-1}},\mu)
\nn \\ & \quad \times
\vec{C}(\Phi_{N-2},\mu)\,
\biggl[1+\ORd{\frac{m_J^2}{Q \w_1},\frac{\w_1}{\w_2},\frac{\w_2}{Q}}\biggr]
\,.\end{align}
In the strongly ordered limit all soft jets are initiated by a gluon. The soft splitting coefficients $\widehat{C}_{s,\kappa}$ successively promote the color space from $N$-parton to $(N+1)$-parton and from $(N+1)$-parton to $(N+2)$-parton color space, respectively.

\subsubsection{Correlated strong ordering in angles and energies}

Finally, we also discuss the case with soft jets close to each other arising from an ordered sequence of soft and collinear splittings. As an example, we consider the situation where a soft jet further splits into two collinear jets, corresponding to the partonic subprocesses
\begin{align}
\kappa_a(q_a)\, \kappa_b(q_b) &\to \kappa_3(q_3) \dotsb \kappa_N(q_N)
\qquad&
\kappa_{N-2} &= \{\kappa_a, \kappa_b; \kappa_3,\dots, \kappa_N\}
\nn \\
&\to \kappa_{12}(q_{12})\, \kappa_3(q_3)\dotsb \kappa_N(q_N)
\qquad &
\kappa_{N-1} &= \{\kappa_a, \kappa_b; \kappa_{12}^{\lambda_{12}}, \kappa_3, \dots, \kappa_N\}
\nn \\ & \quad\,\,\,
\kappa_{12} \to \kappa_1(q_1)\, \kappa_2(q_2)
\qquad &
\kappa_c &= \{\kappa_{12}^{\lambda_{12}}; \kappa_1, \kappa_2\}
\,, \end{align}
where $\kappa_{12} = g$. This is characterized by the kinematic hierarchies
\begin{equation}
\w_1 \sim \w_2 \ll \w_{i\neq1,2} \sim Q
\,,\quad
\hat s_{12} \ll \hat s_{ij} \sim 1
\quad\Rightarrow\quad
u \sim s_{1k} \sim s_{2k} \ll Q^2
\,,\quad
s_{12} \ll \frac{s_{1k}s_{2l}}{s_{k\ell}}
\,,\end{equation}
with $k, \ell \neq 1,2$.
Combining the results in \subsecs{scet_c}{scet_s} yields the factorized cross section
\begin{align}\label{eq:factorization_Njet_sc}
\frac{\df \sigma_{+} (\Phi_{N})}{\df \Tau_N} 
&= s_{\kappa_{N-2}}  s_{\kappa_c}\int\! \df s_a \df s_b \, B_a(s_a,x_a,\mu) \, B_b(s_b,x_b,\mu)
\biggl[\prod_{k=1}^N \int\! \df s_k \, J_{k}(s_k,\mu)\biggr]
 \\ 
&\quad \times \int\! \df  k
\sum_{\lambda_{12},\lambda_{12}'} \vec C^\dagger_{c, \lambda_{12}'}(s_{12}, z, \varphi,\mu)  \,\widehat S_{c, \kappa_c}( k,\mu)\,
\vec C_{c, \lambda_{12}}(s_{12}, z, \varphi,\mu)
\nn \\
& \quad \times
 \vec{C}^\dagger(\Phi_{N-2},\mu) \,\widehat{C}^\dagger_{s,\kappa_{N-1},\lambda_{12}'}(\w_1+\w_2,\{n_{i}\}_{\kappa_{N-1}},\mu)
\nn \\ & \quad \times
\widehat{S}_{\kappa_{N-1}}\biggl(\Tau_N-\sum_{i=a}^N\frac{s_i}{Q_i}-\sqrt{\hat{s}_{12}}\, k,\{\hat{s}_{ij}\}, \mu\biggr)
\nn \\ & \quad \times
\widehat{C}_{s,\kappa_{N-1},\lambda_{12}}(\w_1+\w_2,\{n_{i}\}_{\kappa_{N-1}},\mu)\, \vec{C}(\Phi_{N-2},\mu)\,
\biggl[1+\ORd{\frac{m_J^2}{s_{12}},\frac{Q^2 s_{12}}{u^2},\frac{u}{Q^2}}\biggr]
\nn\, .\end{align}
As discussed in \subsec{scet_s} the spin interference effects between the hard process at the scale $\mu \sim Q$ and the soft splitting vanish, but in general they do not between the soft and collinear splitting processes, since the helicity of the soft gluon is not fixed. Therefore, we have explicitly denoted the dependence of $\widehat{C}_{s,\kappa_{N-1}}$ on the helicity of the soft gluon $\lambda_{12}$.

\subsection{Beyond strong ordering}
\label{subsec:bso}

Having discussed the strongly-ordered case, we now discuss situations where several jets exhibit a hierarchy with respect to the remaining energetic and well-separated jets, but not among each other, i.e.~where multiple soft or collinear jets originate from the same sector at the same scale.

\subsubsection{Multiple collinear emissions}

First we discuss the case of $M$ energetic jets being close to each other without any special ordering in the angles between them,
\begin{equation}
t \sim s_{ij} \ll s_{ik} \sim s_{k\ell} \sim Q^2
\,,\qquad
\w_i \sim \w_j \sim Q
\,,\end{equation}
where $i,j \in \{1,\dots, M\}$ and $k, \ell \notin \{1,\dots, M\}$. The corresponding partonic process is
\begin{align}
\kappa_a(q_a)\, \kappa_b(q_b) \to\, &\kappa_t(q_t)\, \kappa_{M+1}(q_{M+1}) \dotsb \kappa_N(q_N)
&\qquad
\kappa_{N-M} &= \{\kappa_a, \kappa_b; \kappa_t^{\lambda_t}, \kappa_{M+1}, \dots, \kappa_N\}
\nn \\
&\kappa_t(q_t) \to \kappa_1(q_1) \dotsb \kappa_M(q_M)
&\qquad
\kappa_c &= \{\kappa_t^{\lambda_t}; \kappa_1, \dots, \kappa_M\}
\,.\end{align}
Taking for example $M=3$, this leads to the factorized cross section
\begin{align}\label{eq:factorization_Njet_cc2}
\frac{\df \sigma_{+} (\Phi_{N})}{\df \Tau_N} 
&= s_{\kappa_{N-2}}s_{\kappa_{c}}\int\! \df s_a \df s_b \, B_a(s_a,x_a,\mu) \, B_b(s_b,x_b,\mu)
\biggl[\prod_{k=1}^N \int\! \df s_k \, J_{k}(s_k,\mu)\biggr]
\\ &\quad \times
\int\! \df k\sum_{\lambda_t,\lambda_t'} \vec C^\dagger_{cc,\lambda_t'}(\Phi_{cc},\mu)  \,\widehat S_{cc,\kappa_c}(k,\{\hat{s}_{ij}\}_{\kappa_c},\mu)\, \vec C_{cc,\lambda_t}(\Phi_{cc},\mu)
\nn \\ & \quad \times 
C^\dagger_{\lambda_t'}(\Phi_{N-2},\mu) \, \widehat{S}_{\kappa_{N-2}} \biggl(\Tau_N -\sum_{i=a}^N\frac{s_i}{Q_i}- \sqrt{\hat s_{123}}\,k,\{\hat{s}_{ij}\}_{\kappa_{N-2}},\mu\biggr)\,C^\dagger_{\lambda_t}(\Phi_{N-2},\mu)
\nn \\ & \quad
\times \biggl[1+\ORd{\frac{m_J^2}{t},\frac{t}{Q^2}}\biggr]
\,,\nn\end{align}
where we define
\begin{equation}
t = s_{123} = (q_1 + q_2 + q_3)^2
\,,\quad
\hat s_{123} = \frac{s_{123}}{Q_1 Q_2 + Q_1 Q_3 + Q_2 Q_3}
\,,\quad
z_i = \frac{\w_i}{\w_1+\w_2+\w_3}
\,.\end{equation}
An interesting new feature is that the color space of $\vec C_{cc,\kappa_c}$ and $\widehat S_{cc,\kappa_c}$ is now nontrivial.
The matching coefficient $\vec C_{cc,\kappa_c}(\Phi_c, \mu)$ describes the collinear $1 \to 3$ splitting at the scale $\mu \sim \sqrt{t}$, and we denoted with $\Phi_{cc}$ the $3$-body collinear phase space. It can be extracted at tree level from the $1\to 3$ collinear splitting amplitudes in refs.~\cite{Campbell:1997hg,Catani:1999ss,DelDuca:1999ha}. For example, for $q \to q \bar{Q} Q$ (with different quark flavors $Q\neq q$) the collinear splitting amplitudes read 
\begin{align} \label{eq:cc_example}
\vec C_{cc, \{q^+_t; q^+ \bar Q^- Q^+\}}(q_1, q_2, q_3)
&= -g^2 \frac{1}{s_{23}}
  \biggl[\frac{\sqrt{z_1 z_2 z_3}}{1-z_1} + \frac{[13](\sqrt{z_1} \langle 12 \rangle - \sqrt{z_3} \langle 23 \rangle)}{s_{123}}\biggr]
  \begin{pmatrix} 1 \\ -1/N_c \end{pmatrix}
 , \nn \\
  \vec C_{cc,\{q^-_t; q^- \bar Q^- Q^+\}}(q_1, q_2, q_3) &= \vec C_{cc,\{q^+_t; q^+ \bar Q^- Q^+\}} (q_1, q_3, q_2)
\,,\end{align}
where we used the color basis
\begin{align}
  \bar T^{\bt \bar \al  \de \bar \ga} = \bigl(\de_{\bt \bar \ga} \de_{\de \bar \al}, \ \de_{\bt \bar \al} \de_{\de \bar \ga} \bigr)
\,.\end{align}
In the strongly-ordered limit, this reduces to the product of two $1\to 2$ splitting coefficients, which will reproduce the result in \subsec{so+},
\begin{align}
C_{cc, \{q^+_t; q^+ \bar Q^- Q^+\}}^{\bt  \bar \al \de \bar \ga} \Big\vert_{s_{12} \ll s_{123}}
  = \sum_{\la_g} C_{c,\la_g \lambda_q}^{a \bt \bar \al} C_{c,\lambda_g \lambda_Q}^{a \de \bar \ga}
\,.\end{align}
One can check this relation at tree level using the explicit results for the collinear splitting amplitudes in ref.~\cite{Bern:1994zx} (related to the ones given in \eq{C_c} via crossing) and \eq{cc_example}. Note that the leading $1/s_{23}$ term in \eq{cc_example} cancels, which requires a careful expansion up to order $1/\sqrt{s_{23} s_{123}}$, as pointed out e.g.~in refs.~\cite{Kosower:2003bh,Somogyi:2005xz}.

The csoft function $\widehat S_{cc,\kappa_c}$ characterizes the csoft radiation exchanged between the $3$ nearby jets at the scale $\mu = m_J^2/\sqrt{t}$,
\begin{align}
\widehat S_{cc,\kappa_c}(k,\{\hat s_{ij}\}_{\kappa_c},\mu)
&= 
 \widehat{T}_{\kappa_c}^{-1}\, (\bar T^{\al_t \al_1 \al_2 \al_3})^\dagger
 \widehat S_{cc,\kappa_c}^{\al_t \al_1 \al_2 \al_3\,\bt_t \bt_1 \bt_2 \bt_3}(k,\{\hat s_{ij}\}_{\kappa_c},\mu)\,
  \bar T^{\bt_t \bt_1 \bt_2 \bt_3} 
\nn \\ &= 
\sum_{X_{cs}} \Big\langle 0 \Big\vert  \bar{T}\Big[V^{\dagger \ga_t \al_t}_{n_t} \prod_{i=1}^3 X_{n_i}^{\ga_i \al_i}\Big] \Big\vert X_{cs}\Big\rangle \Big\langle X_{cs}\Big\vert T\Big[\prod_{i=1}^3 X_{n_i}^{\bt_i \ga_i} V_{n_t}^{\bt_t \ga_t}\Big]\Big\vert 0 \Big\rangle \nn \\
& \quad \times 
\widehat{T}_{\kappa_c}^{-1}\,
 (\bar T^{\al_t \al_1 \al_2 \al_3})^\dagger \bar T^{\bt_t \bt_1 \bt_2 \bt_3}
\delta\biggl(k - \sum_{i=1}^{3} \frac{n_i \sdt k^{(i)}_{cs}}{\rho_i \sqrt{\hat s_{123}}}\biggr)
\,. \end{align}
The representation of the Wilson lines $V_{n_t}$ and $X_{n_i}$ are determined by the parton $\kappa_i$. Unlike in \eqs{csoftfct}{csoftfctpp}, the function now depends on several invariants, namely the generalized angles $\hat s_{ij} = 2q_i \sdt q_j/(Q_i Q_j) = n_i \sdt n_j/(2\rho_i \rho_j)$ and the $\rho_i$ with $i\neq j \in \{1,2,3\}$. However, the typical angular scale is still $\sqrt{\hat s_{123}}$ which we pull out front.

Using that at tree level $\widehat S^{(0)}_{cc, \kappa_c} = {\bf 1} \,\de(k)$, and summing (averaging) over the outgoing (incoming) helicities, leads to the $q \to q\bar Q Q$ tree-level splitting function~\cite{Catani:1998nv}
\begin{align}
&\frac12 \sum_{\lambda=\lambda'} \vec C^{\zero\dagger}_{cc,\kappa_c}(\Phi_{cc},\mu)\, \widehat S^\zero_{cc,\kappa_c}(k,\{\hat s_{ij}\}_{\kappa_c},\mu) \,\vec C^{\zero}_{cc,\kappa_c}(\Phi_{cc},\mu)
\nn \\ & 
= \frac{2g^4 C_F T_F}{s_{23} s_{123}}\biggl[-\frac{1}{s_{23}s_{123}}\Big(2\, \frac{z_2 s_{13} \!-\! z_3 s_{12}}{z_2\!+\!z_3} \!+\! \frac{z_2\!-\!z_3}{z_2\!+\!z_3}\, s_{23}\Big)^2 \!+\! \frac{4z_1\!+\!(z_2\!-\!z_3)^2}{z_2\!+\!z_3}\!+\!z_2\!+\!z_3\!-\!\frac{s_{23}}{s_{123}}\biggr] \delta(k)
\nn \\ & 
\equiv P_{q \to q\bar Q Q} \, \delta(k)
\end{align}

\subsubsection{Multiple soft emissions}

Next we discuss the case where all jets are equally separated and the first $M$ jets are soft without any special ordering in their energies,
\begin{equation}
\w_i \sim \w_j \ll \w_k \sim Q
\,,\qquad
\hat s_{ij} \sim \hat s_{ik} \sim \hat s_{k\ell} \sim 1
\quad\Rightarrow\quad
u \sim s_{ik} \ll s_{k\ell} \sim Q^2
\,.\end{equation}
with $i, j \in \{1,\dots, M\}$ and $k, \ell \notin \{1, \dots,M\}$. The partonic process separates as
\begin{align}
\kappa_a(q_a)\, \kappa_b(q_b) &\to \kappa_{M+1}(q_{M+1}) \dotsb \kappa_N(q_N)
\qquad &
\kappa_{N-M} &= \{\kappa_a, \kappa_b; \kappa_{M+1}, \dots, \kappa_N\}
\nn \\
&\to \kappa_1(q_1) \dotsb \kappa_N(q_N)
\qquad &
\kappa_N &= \{\kappa_a, \kappa_b; \kappa_1, \ldots, \kappa_N\}
\,.\end{align}
This case involves the soft splitting amplitudes for $M$ particles and is a straightforward generalization of \eq{factorization_Njet_SCETs}. For example, for $M=2$ we get
\begin{align}\label{eq:factorization_Njet_ss2}
\frac{\df \sigma_{+} (\Phi_{N})}{\df \Tau_N} 
&= s_{\kappa_{N-2}}\int\! \df s_a \df s_b \, B_a(s_a,x_a,\mu) \, B_b(s_b,x_b,\mu) \biggl[\prod_{k=1}^N \int\! \df s_k \, J_{k}(s_k,\mu) \biggr]
\nn \\ & \quad \times
 \vec{C}^\dagger(\Phi_{N-2},\mu) \, \widehat{C}^{\dagger}_{ss,\kappa}(\w_1,\w_2,\{n_{i}\},\mu)\,
 \widehat{S}_\kappa\biggl(\Tau_N - \sum_{i=a}^N \frac{s_i}{Q_i}, \{\hat{s}_{ij}\},\mu\biggr)
\nn \\ & \quad \times \widehat{C}_{ss,\kappa}(\w_1,\w_2,\{n_{i}\},\mu) \,
\vec{C}(\Phi_{N-2},\mu)
\biggl[1 +\ORd{\frac{m_J^2}{u},\frac{u}{Q^2}}\biggr]
\, ,\end{align}
The matching coefficients $\widehat{C}_{ss,\kappa}$ have as natural scale $\mu \sim u/Q$ and are given in terms of soft splitting amplitudes. They are now matrices going from $N$-parton to $(N+2)$-parton color space. The soft jets no longer have to be gluon jets, since a soft gluon can split into a soft $q\bar q$-pair. The tree-level expressions for $\widehat{C}_{ss}$ can be obtained from refs.~\cite{Campbell:1997hg,Catani:1999ss}. For the emission of two soft gluon jets we have
\begin{align}
\widehat C_{ss,\kappa}^{(0)}
&= g^2
 \biggl\{ \sum_{i \neq j \neq 1,2} \bfT_i^{a_1}\, \frac{\varepsilon_{\lambda_1} \!\sdt q_i}{q_1 \sdt q_i}\,
 \bfT_j^{a_2}\, \frac{\varepsilon_{\lambda_2} \!\sdt q_j}{q_2 \sdt q_j}
 \nn \\ & \qquad
 + \sum_{i \neq 1,2} \biggl[ 
 \biggl( \de^{a_1 a} \bfT_i^{a_2}\, \frac{\varepsilon_{\lambda_2}\! \sdt q_i}{q_2 \sdt q_i}
 - \img f^{a_2 a_1 a}\, \frac{\varepsilon_{\lambda_2} \!\sdt q_1}{q_1 \sdt q_2}\biggr) \bfT_i^a\,
 \frac{\varepsilon_{\lambda_1} \!\sdt q_i}{(q_1+q_2) \sdt q_i}
 \nn \\ & \qquad\qquad\quad
 + \frac14\, \img f^{a a_1 a_2} \bfT_i^a \,\frac{\varepsilon_{\lambda_1} \!\sdt \varepsilon_{\lambda_2}}{q_1 \sdt q_2}
 \, \frac{(q_2-q_1)\sdt q_i}{(q_1+q_2)\sdt q_i}+ (1 \lra 2)\biggr]\biggr\}
\,.\end{align}
If the energies of the gluons are strongly ordered, this reduces to the iteration of the $\widehat C_{s,\kappa}$ as in \eq{factorization_Njet_ss}. At tree level, we have
\begin{align}
\widehat C_{ss,\kappa}^{(0)}\Big|_{s_{1k}\ll s_{2k}}
&= g^2
\biggl( \sum_{i, j \neq 1,2} \bfT_i^{a_1}\, \frac{\varepsilon_{\lambda_1} \!\sdt q_i}{q_1\sdt q_i}\,
 \bfT_j^{a_2}\, \frac{\varepsilon_{\lambda_2} \!\sdt q_j}{q_2\sdt q_j}
 + \sum_{i\neq 1,2}
 \img f^{a_2 a_1 a}\, \frac{\varepsilon_{\lambda_1} \!\sdt q_2}{q_1\sdt q_2}\, \bfT_i^a\,
 \frac{\varepsilon_{\lambda_2} \!\sdt q_i}{q_2\sdt q_i}
\biggr)
\nn \\ 
&= g^2 \sum_{i \neq 1} \bfT_i^{a_1}\, \frac{\varepsilon_{\lambda_1} \!\sdt q_i}{q_1\sdt q_i} \sum_{j \neq 1,2} 
 \bfT_j^{a_2}\, \frac{\varepsilon_{\lambda_2} \!\sdt q_j}{q_2\sdt q_j}
\nn \\ &
= \widehat C_{s,\kappa}^{(0)} \,\widehat C_{s,\kappa_{N-1}}^{(0)}
\,,\end{align}
where we used that $(\bfT_2^{a_1})_{a_2 a_2'} = \img f^{a_2 a_1 a_2'}$ and the tree-level expression in \eq{Cs+0}.

\subsubsection{Remaining cases}

The case of several soft jets close to energetic jet(s) combines the features of \eqs{factorization_Njet_cc2}{factorization_Njet_ss2}. It leads to the color structure and soft functions in \eq{factorization_Njet_cc2}. 
As the soft jets arise from collinear-soft emissions, the corresponding matching coefficient is the analogue of $\widehat{C}_{ss}$ on the set of the nearby jets, see \eq{Ccs_match}. One can also encounter the situation of several nearby jets which are partially hierarchically ordered in their energies, but not in their angles, leading to collinear and collinear-soft splittings at different invariant mass scales and a communication via a common collinear-soft function that depends on all of their directions.

The factorization formulae in this section (and their generalizations) can be combined with those describing the strongly-ordered kinematics following the same logic as in \subsecs{combineEFT}{3jet_combination}. This allows one to cover the complete jet phase space and all possible jet hierarchies and thus to systematically resum all kinematic logarithms.
For multiple jets the number of possible kinematic hierachies quickly proliferates. In practice, the number of relevant cases
can be greatly reduced by imposing restrictions on the jet kinematics one is interested in and the perturbative accuracy one
aims to achieve. For example, any hierarchy for which a fixed-order description is sufficient can be ignored and is then automatically
included via the nonsingular matching corrections.

\section{Conclusions}
\label{sec:conclusions}

Processes with multiple jets in the final state depend on several hard kinematic variables, like the jet energies and invariant masses between jets, generating large logarithms in the cross section whenever there are sizable hierarchies between the corresponding kinematic scales. This is in fact the generic situation, due to the enhancement of soft and collinear emissions in QCD. To obtain precise predictions with well-controlled perturbative uncertainties, the systematic resummation of these kinematic logarithms beyond the leading-logarithmic accuracy provided by the parton shower is needed. This is particularly relevant at the LHC, where there is plenty of phase space and a large kinematic range between the highest probed scales at  $\sim$ few TeV to the lowest jet energies at $\sim 30$ GeV.

We constructed the effective field theory framework that enables the systematic resummation of kinematic logarithms for generic jet hierarchies in multijet hard-scattering processes through RG equations. We have presented this in detail for all hierarchies for $e^+ e^- \to 3$ jets, and discussed a representative set of the general $pp \to N$ jet case, demonstrating how to handle the complications arising from the large number of possible hierarchies and due to spin and color correlations. Our framework allows for a combination of the results for the various different regimes via a sequence of nonsingular corrections that avoids double counting. Although we have mainly focused on jets defined via a \SCETa jet resolution variable like $N$-jettiness, the \SCETp framework is general and applicable also to other jet definitions and resolution variables like a \SCETb-type $p_T$-veto applied in jet binning, since in particular the factorization in the hard sector is independent of the specific jet definition.

Important applications of our framework include jet substructure analyses and jet binning.
Though the numerical implementation is beyond the scope of this work and left for future work, the necessary perturbative ingredients are generically known for the resummation up to NNLL, which requires the full one-loop matching corrections and two-loop noncusp anomalous dimensions. In particular,
our results for the exclusive $N$-jet cross sections from \SCETp can be used to systematically improve upon
the LL description of kinematic logarithms in parton showers, for example by incorporating them into
the \textsc{Geneva} Monte Carlo framework~\cite{Alioli:2012fc, Alioli:2013hqa, Alioli:2015toa} or possibly by extending the MINLO method~\cite{Hamilton:2012np, Hamilton:2012rf, Frederix:2015fyz}.
Furthermore, in ref.~\cite{Larkoski:2015zka} it was argued that nonglobal logarithms can be systematically accounted for by considering and marginalizing over increasingly resolved hierarchical multijet configurations, for which the kinematic logarithms can be resummed. Our results make it possible to explicitly carry out this procedure to higher perturbative accuracy and subleading orders.
Differential measurements with jets play an increasingly important role in collider physics and the aim of our \SCETp framework is to improve the theoretical predictions and to better control perturbative uncertainties in multijet processes.

\begin{acknowledgments}
We thank Ian Moult, Duff Neill, Matthias Ritzmann, and Maximilian Stahlhofen for discussions.
This work was supported by the German Science Foundation (DFG) through the Emmy-Noether Grant No. TA 867/1-1, the Collaborative Research Center (SFB) 676 Particles, Strings and the Early Universe. This work is part of the D-ITP consortium, a program of the Netherlands Organization for Scientific Research (NWO) that is funded by the Dutch Ministry of Education, Culture and Science (OCW). 
\end{acknowledgments}

\appendix
\section{SCET notation and conventions}
\label{app:SCET}

In this appendix we briefly summarize the common SCET notation we use. The momentum of a particle in the $n$-collinear direction is decomposed into a large label momentum $\tilde{p}_n^\mu$ with respect to the $n$-collinear direction and a small residual momentum $k^\mu$ of order $Q\lambda^2$,
\begin{align}\label{eq:scaling_momenta}
p_n^\mu=\tilde{p}_n^\mu+k^\mu
\,,\qquad
\tilde{p}_n^\mu = \bn\sdt \tilde{p}\,\frac{n^\mu}{2} + \tilde{p}_{n\perp}^\mu \sim Q(0,1,\lambda)
\,,\qquad
k^\mu \sim Q(\lambda^2,\lambda^2,\lambda^2)
\,.\end{align}
The $n$-collinear quark and gluon fields with label momentum $\tilde{p}_n^\mu$ are denoted by $\xi_{n,\tilde{p}}(x^\mu)$ and $A^\mu_{n,\tilde{p}}(x^\mu)$, where the coordinate $x^\mu$ is conjugate to the residucal momentum $k^\mu$. The label operator $\mathcal{P}_n^\mu$ picks out the label momentum of a field, $\mathcal{P}_n^\mu \, \xi_{n,\tilde{p}}=\tilde{p}^\mu \,\xi_{n,\tilde{p}}$, while derivatives acting on the fields pick out the residual momentum dependence, $\img\partial^\mu \sim k^\mu$. We are often only interested in the label $n$ for the collinear direction, $\xi_{n}$ and $A^\mu_{n}$, which implies that the momentum labels are implicitly summed over subject to overall label momentum conservation.

The operators appearing in the hard-scattering Lagrangian are constructed from fields and Wilson lines that are invariant under collinear gauge transformations~\cite{Bauer:2000yr, Bauer:2001ct}. The smallest building blocks are collinearly gauge invariant quark and gluon fields, which are defined as
\begin{align}
\chi_{n,\w}(x)
&= \Bigl[ \delta(\w - \bar{n} \sdt \mathcal{P}_n)\, W_n^\dagger(x)\, \xi_{n}(x) \Bigr]
\,,\nn\\
\mathcal{B}_{n,\w \perp}^{\mu}(x)
&= \frac{1}{g}\Bigl[ \delta(\w + \bar{n} \sdt \mathcal{P}_n) W_n^\dagger(x) \, \img D_{n\perp}^\mu(x) \,W_n(x)\Bigr]
\,.\end{align}
With these standard conventions, $\w>0$ for an incoming quark or outgoing gluon and $\w < 0$ for an outgoing antiquark or incoming gluon. The collinear covariant derivative is given by $\img D_{n\perp}^\mu=\mathcal{P}^\mu_{n\perp}+g A^\mu_{n\perp}$, and $W_n(x)$ is a Wilson line of $n$-collinear gluons in label-momentum space
\begin{align}\label{eq:wilsonline_momentumspace}
W_n(x)=\biggl[\sum_{\rm perms} {\rm exp}\Big(\frac{-g}{\bar{n} \sdt \mathcal{P}_n} \, \bar{n} \sdt A_{n}(x)\Big) \biggr]
\,.\end{align}

The usoft fields $A_{us}^\mu$ couple to the collinear fields via the usoft covariant derivative $\img D_{us}^\mu = \img\partial^\mu + g A^\mu_{us}$. These interactions in the collinear Lagrangians are eliminated by the field redefinition~\cite{Bauer:2001yt}
\begin{align} \label{eq:BPS}
\chi_{n,\w}(x)=Y_n(x)\, \chi_{n,\w}^{(0)}(x)
\,, \qquad
\cB_{n,\w\perp}^{\mu}(x) = Y_n(x)\, \cB_{n,\w \perp}^{\mu (0)}(x)\, Y_n^{\dagger}(x)
\,,\end{align}
where $Y_n(x)$ denotes the ultrasoft Wilson line along the $n$ direction,
\begin{align} \label{eq:Yn}
Y_{n}(x) &= \overline{\mathrm{P}} \,  \exp \biggl[-\img g\int_{0}^{\infty} \!\df s \, n\sdt A_{us}(s n^\mu+x^\mu) \biggr]
\,,\end{align}
and $\overline{\rm P}$ denotes anti-path-ordering. Usually we do not display the superscript ${(0)}$ explicitly on the redefined fields for notational simplicity.

\section{Anomalous dimensions}
\label{app:anom_dim}

Here we give explicit expressions for the cusp and noncusp anomalous dimensions of the hard Wilson coefficient in \eqs{Cs_anom}{gamma_C_N} and the $\beta$ function. Using the expansions
\begin{align} \label{eq:betafunction}
\beta(\alpha_s)
&= - 2 \alpha_s \sum_{n=0}^\infty \beta_n\Bigl(\frac{\alpha_s}{4\pi}\Bigr)^{n+1}
\,,\qquad
\Gamma_\cusp(\alpha_s)
= \sum_{n=0}^\infty \Gamma_n \Bigl(\frac{\alpha_s}{4\pi}\Bigr)^{n+1}
\,,\end{align}
the one-loop and two-loop coefficients in the $\overline {\rm MS}$ scheme are given by~\cite{Korchemsky:1987wg, Caswell:1974gg, Jones:1974mm}
\begin{align} \label{eq:betacusp}
\beta_0
&= \frac{11}{3}\,C_A -\frac{4}{3}\,T_F\,n_f
\,,  \qquad &
\beta_1 &= \frac{34}{3}\,C_A^2  - \Bigl(\frac{20}{3}\,C_A\, + 4 C_F\Bigr)\, T_F\,n_f
\,,\nn \\
\Gamma_0 &= 4
\,, \qquad &
\Gamma_1
&= \Bigl( \frac{268}{9} -\frac{4\pi^2}{3} \Bigr)\,C_A  - \frac{80}{9}\,T_F\, n_f
\,,\end{align}
with $\Gamma_n^q = C_F \Gamma_n$ and $\Gamma_n^g = C_A\Gamma_n$.

For the noncusp anomalous dimensions of the quark and gluon form factors
\begin{equation}
\gamma_C^i(\alpha_s)
= \sum_{n=0}^\infty \gamma_{C\,n}^i \Bigl(\frac{\alpha_s}{4\pi}\Bigr)^{n+1}
\,,\end{equation}
the coefficients are~\cite{Idilbi:2005ni, Idilbi:2006dg}
\begin{alignat}{2} \label{eq:anom_dim}
\gamma_{C \, 0}^{q}
&=-3C_F
\,, \quad
&& \gamma_{C \, 1}^{q}
= -C_F \biggl[  \biggl( \frac{41}{9}-26 \zeta_3  \biggr)C_A +\biggl( \frac{3}{2}-2\pi^2+24 \zeta_3 \biggr) C_F +\biggl( \frac{65}{18}+\frac{\pi^2}{2}   \biggr)  \beta_0  \biggr]
\,, \nn \\
\gamma_{C \, 0}^{g} &=-\beta_0
\,, \quad
&& \gamma_{C \, 1}^{g} = \biggl( -\frac{59}{9}+2\zeta_3 \biggr) C_A^2 + \biggl(-\frac{19}{9}+\frac{\pi^2}{6} \biggr) C_A \beta_0 -\beta_1
\,.\end{alignat}

\section{Hard splitting functions in \SCETp}
\label{app:H3s}

In this appendix, we explain how to directly calculate the hard splitting functions $H_c$, $H_s$, and $H_{cs}$ for $e^+ e^- \to 3$ jets discussed in \sec{3jets}. We use this specifically to compute $H_s$ at one loop and to extract the two-loop result from available results in the literature.

\subsection{Calculational prescription}
\label{app:H+}

The hard function $H_c$ can be directly computed from the matching within a single collinear sector in SCET using the fact that the associated loop diagrams in \SCETp are scaleless and vanish in pure dimensional regularization. Following a similar line of reasoning as discussed above \eq{C_is_A}, we can write $H_c$ for the collinear splitting $q \to gq$, i.e.~for the partonic channel $\kappa_c=\{q;g,q\}$, as
\begin{align}\label{eq:H3c_def}
 H_{c,\{q;g,q\}}(t,z,\mu)
 &= \ABS{\frac{Z_{g\,q\bar q,c}}{Z_{q\bar q}}}^2  \frac{1}{4 \pi N_c Q^3} \sum_{q,g}\int \! \df^4 x \, e^{\img \frac{t}{2Q} x^-}
 \tr\Bigl[\MAe{0}{\frac{\slashed{\bar{n}}}{2} \chi_{n}(x)}{g q} \Mae{g q}{\bar{\chi}_{n,Q}(0)}{0} \Bigr] \delta_{\tilde p_g^-,z Q}
 \nn \\
&= \ABS{\frac{Z_{g\,q\bar q,c}}{Z_{q\bar q}}}^2 \frac{(2\pi)^3}{N_c Q^2}\,
   \int\! \frac{\df^3 p_q}{(2\pi)^3 2p_q^0 } \int\! \frac{\df^3 p_g}{(2\pi)^3 2p_g^0}
   \sum_{\text{color,spins}}\!\!\Abs{\mathcal{M}_c\bigl(0\to g(p_g) q(p_q) \bigr)}^2
\nn \\ & \quad \times
  \de(Q - p_q^- - p_g^-)\,
  \de^2(p_{q\perp}+p_{g\perp})\,
  \de[t - Q(p_q^+ + p_g^+)] \,
  \de\Bigl(z - \frac{p_g^-}{Q}\Bigr)
\,.\end{align}
The factor $Z_{g\,q\bar q,c}$ indicates the common counterterm of the operators $O^{a \bar \alpha \beta}_{\lambda_g(\lambda_q;\lambda_\ell)}$ in the $c+$ regime with three collinear directions in analogy to \eq{Z_3}, while $Z_{q\bar q}$ is the counterterm of the $q\bar q$ operators $O^{\bar \alpha \beta}_{(\lambda_q;\lambda_\ell)}$ in \eq{O_2}. The states $\vert g \rangle$ and $\vert q \rangle$ in the first line of \eq{H3c_def} denote on-shell gluon and quark states with momenta $p_g^\mu = \tilde{p}_g^\mu +k_g^\mu$ and $p^\mu_q= \tilde{p}_q^\mu +k_q^\mu$, respectively, which we have split up into label and residual components in SCET as in \eq{scaling_momenta}. The spins and polarizations are summed over and the trace runs also over color indices. The second line of \eq{H3c_def} represents the direct computational prescription in terms of the collinear amplitude $\mathcal{M}_c\bigl(0\to g(p_g) q(p_q))$ obtained from the collinear SCET Feynman rules.

Similarly, the hard function $H_s$ can be directly computed from the usoft sector in SCET with two collinear sectors and can be written as
\begin{align}\label{eq:H3s_def}
H_{s}\Bigl(\frac{t\, u}{Q^2},\mu\Bigr)
&= \ABS{\frac{Z_{g\,q\bar q,s}}{Z_{q\bar q}}}^2  \frac{1}{N_c} \sum_{g}
\tr\Bigl[ \Mae{0}{\bar{T}\bigl[Y_{\bar{n}}^\dagger Y_{n} \bigr]}{g} \Mae{g}{T\bigl[Y_{n}^\dagger Y_{\bar{n}}\bigr]}{0} \Bigr]  \, \delta(t-Q k_g^+) \,  \delta(u-Q k_g^-)
\nn \\ 
&= \ABS{\frac{Z_{g\,q\bar q,s}}{Z_{q\bar q}}}^2  \frac{1}{N_c}
\int\! \frac{\df^3 p_g}{(2\pi)^3 2p_g^0}\,  \Abs{\mathcal{M}_s\bigl(0\to g(p_g) \bigr)}^2 \, \delta(t - Q p_g^+) \,  \delta(u - Qp_g^-) 
\,,\end{align}
where $Z_{g\,q\bar q,s}$ is the counterterm of the operators $O^{a \bar \alpha \beta}_{\lambda_g(\lambda_q;\lambda_\ell)}$ in the $s+$ regime. Here, $\vert g \rangle$ is an on-shell gluon state with momentum $p_g^\mu =k_g^\mu$ (i.e. with vanishing label momentum in the parent SCET).

Finally, the hard function $H_{cs}$ can be directly computed from the csoft sector in \SCETp with two collinear sectors and can be written as
\begin{align}\label{eq:H3cs_def}
H_{cs}\Big(\frac{t\, u}{Q^2},\mu\Big) & =  \ABS{\frac{Z_{g\,q\bar q,cs}}{Z_{q\bar q}}}^2  \frac{1}{N_c Q} \sum_{g}
\tr\Bigl[ \Mae{0}{\bar{T}[V_{n}^\dagger X_{n} ]}{g} \Mae{g}{T [X^\dagger_{n}  V_{n}]}{0} \Bigr]  \, \delta(t-Q k_g^+) \, \delta(k_g^-) \,  \delta_{\tilde{p}_g^-,u/Q}  \nn \\
&= \ABS{\frac{Z_{g\,q\bar q,cs}}{Z_{q\bar q}}}^2  \frac{1}{N_c}
\int\! \frac{\df^3 p_g}{(2\pi)^3 2p_g^0}\,  \Abs{\mathcal{M}_{cs}\bigl(0\to g(p_g) \bigr)}^2 \, \delta(t - Q p_g^+) \,  \delta(u - Qp_g^-) 
\,,\end{align}
where $Z_{g\,q\bar q,cs}$ is the counterterm of the operators $O^{a \bar \alpha \beta}_{\lambda_g(\lambda_q;\lambda_\ell)}$ in the $cs+$ regime. Eq.~(\ref{eq:H3cs_def}) gives the same result as \eq{H3s_def} due to the identical form of the Wilson lines $X_{n}$ and $Y_{n}$, $V_{n}$ and $Y_{\bar{n}}$, see \eqs{Vn_Xn}{Yn}.

These expressions can be easily adapted to $N$ jets, which only affects the form of \eq{H3s_def} due to the fact that more usoft Wilson lines appear, and to initial-state splittings. Furthermore, also the hard functions with several additional emissions discussed in \subsec{bso} can be computed in the same way.

\subsection{Calculation of $H_s$}
\label{app:Hs}

Here we calculate the hard function for the soft splitting in the $s+$ regime with two hard jets at one-loop order, and extract the two-loop result from the literature. Following \eq{H3s_def} we write
\begin{align}\label{eq:H3s_calc}
H_{s}\Big(\frac{t\, u}{Q^2},\mu\Big) 
& \equiv  \ABS{\frac{Z_{g\,q\bar q,s}}{Z_{q\bar q}}}^2  S^{(\rm bare)} = \frac{Z_{3,s}}{Z_{2}}  \, \sum_k \biggl(\sum_{i+j=k} S^{(i,j)} +S^{(k,\rm ct)}\biggr)
\,,\end{align}
where $S^{(\rm bare)}$ indicates the bare soft matrix element (with renormalized $\alpha_s$). At order $\alpha_s^{k+1}$ this originates from the interference $S^{(i,j)}$ of soft currents with loop corrections of $\mathcal{O}(\alpha_s^{i})$ and $\mathcal{O}(\alpha_s^{j})$ and $k=i+j$. Since $S^{(i,j)}$ are given in terms of the unrenormalized strong coupling, we include the associated counterterm $S^{(k,\rm ct)}$. For convenience we abbreviate $Z_{2}\equiv|Z_{q\bar q}|^2$ and $Z_{3,s}\equiv|Z_{g\,q\bar q,s}|^2$.

In our calculation we employ the 't Hooft-Veltman (HV) scheme~\cite{Hooft:1972fi}, in which the momentum and polarization of the measured external soft gluon is kept in four dimensions, and only unresolved partons in loops obtain nonvanishing components in $d-4$ dimensions. This gives the same results as in conventional dimensional regularization, but is more convenient since $\mathcal{O}(\epsilon)$ corrections do not arise in the tree-level correction.

\begin{figure}
\centering
\includegraphics[height=3.8cm]{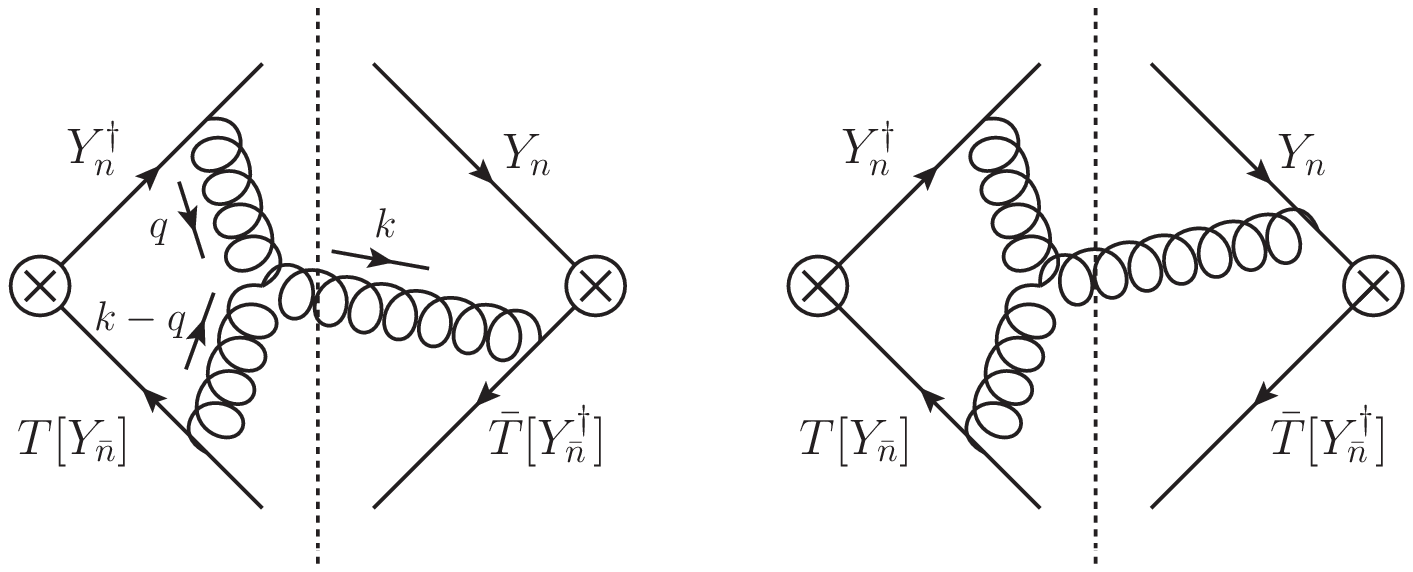}%
\hspace{0.7cm}
\includegraphics[height=3.8cm]{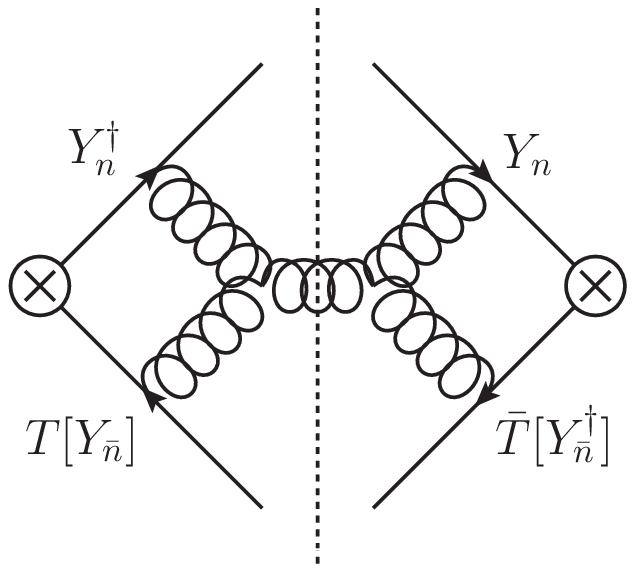}
\caption{Diagrams contribution to $H_{s}$ for $e^+ e^- \to$ 3 jets. The left and middle diagram are the nonvanishing contributions to $S^{(1,0)}$. The right diagram gives $S^{(1,1)}$, entering $H_s$ at two loops.}
\label{fig:softmatching}
\end{figure}

At one loop, only the first two diagrams in \fig{softmatching} contribute, yielding
\begin{align}
S^{(1,0)}
&= - g^4 C_A C_F \Bigl(\frac{\mu^2 e^{\gamma_E}}{4 \pi}\Bigr)^\eps \int\! \frac{\df^4 k}{(2 \pi)^4} \,\frac{1}{k^- -\img \delta} \,(-2\pi \img) \,\delta(k^2) \,\delta(t-Q k^+) \,\delta(u-Q k^-)
\nn \\ & \quad
\times \int\! \frac{\df^d q}{(2 \pi)^d} \, \frac{2k ^- - q^-}{k^--q^-+\img\delta}\, \frac{1}{q^+ +\img\delta}\,\frac{1}{(k-q)^2+\img\delta}\,\frac{1}{q^2+\img\delta}
\,,\end{align}
where we rescaled $\mu^2 \to \mu^2 e^{\gamma_E}/4\pi$ anticipating $\MS$ renormalization.
The integrals can be easily solved using standard methods like Feynman parameters. The final $d$-dimensional result reads
\begin{align}
S^{(1,0)}=-\frac{\alpha_s^2 C_F C_A}{4\pi^2} \,  \frac{1}{t \, u} \biggl(-\frac{t \, u}{Q^2 \mu^2}\biggr)^{-\epsilon}   \frac{e^{\gamma_E \epsilon}\,  \Gamma^2(\epsilon)\, \Gamma^3(1-\epsilon)}{\Gamma(1-2\epsilon)}
\,.\end{align}
We renormalize $\alpha_s$ in the $\MS$-scheme which gives rise to the contribution
\begin{align}
S^{(1,\rm ct)}=-\frac{\alpha_s \beta_0}{4\pi} \, \frac{1}{\epsilon} \, H^{(0)}_{s} \, .
\end{align}
Thus we obtain for the full one-loop hard matching function given in \eq{H3s_oneloop},
\begin{align}
H_{s}^{(1)}\biggl(\frac{t\, u}{Q^2},\mu\biggr)
&= S^{(10)} + S^{(01)} +S^{(1,\rm ct)}+\Bigl(Z_{3,s}^{(1)}-Z_{2}^{(1)}\Bigr)H_{s}^{(0)}
\nn \\
& = - H^{(0)}_{s} \, \frac{\alpha_s C_A}{4 \pi}\biggl[\ln^2\Bigl(\frac{t \, u}{Q^2  \mu^2}\Bigr) -\frac{5\pi^2}{6}\biggr]
\,,\end{align}
where we used the one-loop $\MS$ counterterms
\begin{align}
Z_{2}^{(1)} & = \frac{\alpha_s C_F}{2\pi} \biggl[\frac{2}{\epsilon^2} -\frac{2}{\epsilon}\ln\frac{Q^2}{\mu^2} + \frac{3}{\epsilon}\biggr]
\,, \nn \\
Z_{3,s}^{(1)} & =\frac{\alpha_s}{2\pi} \biggl[\frac{2C_F+C_A}{\epsilon^2}-\frac{2C_F-C_A}{\epsilon}\ln\frac{Q^2}{\mu^2} -
\frac{C_A}{\epsilon}\biggl(\ln \frac{t}{\mu^2} + \ln\frac{u}{\mu^2} \biggr)
+\frac{3C_F+\beta_0/2}{\epsilon} \biggr]
\, . \end{align}

We now extract $H_{s}$ at two loops, which can be written as
\begin{align}
H_{s}^{(2)}
&= S^{(0,2)}+S^{(2,0)}+S^{(1,1)}+S^{(2, \rm ct)}
+ \bigl(S^{(1,0)}+S^{(0,1)}\bigr)\bigl(Z_{3,s}^{(1)}-Z_{2}^{(1)}\bigr)
\nn \\ & \quad
+\Bigl[ Z_{3,s}^{(2)}-Z_{2}^{(2)} - Z_{2}^{(1)} Z_{3,s}^{(1)}+\bigl(Z_{2}^{(1)}\bigr)^2 \Bigr] H_{s}^{(0)} \, .
\end{align}
in the notation of \eq{H3s_calc}.
The interference term between the two-loop and tree-level current has been calculated in refs.~\cite{Badger:2004uk,Duhr:2013msa,Li:2013lsa}. The finite pieces are given by
\begin{align}
S^{(0,2)}+S^{(2,0)} \Big\vert_{\rm finite}
&= H^{(0)}_{s} \, \frac{\alpha_s^2 C_A}{16 \pi^2}\biggl\{C_A\biggl[\frac{2}{3}L^4 +\frac{22}{9}L^3-\biggl(\frac{67}{9}+\frac{10\pi^2}{3} \biggr)L^2
\\ & \quad
+ \biggl(\frac{386}{27}-\frac{121\pi^2}{18}+\frac{22}{3}\zeta_3\biggr)L -\frac{1142}{81} + \frac{737 \pi^2}{108}+\frac{341}{9} \zeta_3+\frac{7\pi^4}{360} \biggr]
\nn \\ & \quad
 \biggl.+ \, T_F n_f \biggl[-\frac{8}{9}L^3+\frac{20}{9}L^2-\biggl(\frac{76}{27}-\frac{22\pi^2}{9}\biggr)L+\frac{130}{81}-\frac{55\pi^2}{27}-\frac{124}{9}\zeta_3\biggr]\biggr\}
\, , \nn \end{align}
with $L=\ln(\frac{t \, u} {Q^2  \mu^2})$ and using still an unrenormalized strong coupling. The only nonvanishing diagram for the interference contribution between the two one-loop currents is given in \fig{softmatching}. The result for the bare correction is given by square of the one-loop contribution,
\begin{align}
S^{(1,1)}=H^{(0)}_{s} \biggl|\frac{\alpha_s C_A}{4\pi} \biggl(-\frac{t \, u}{Q^2 \mu^2}\biggr)^{-\epsilon} \frac{ e^{\gamma_E \epsilon}\,\Gamma^2(\epsilon)\,\Gamma^3(1-\epsilon)}{\Gamma(1-2\epsilon)}\biggr|^2 \, .
\end{align}
Including also the counterterm corrections for the renormalization of $\alpha_s$ in the $\MS$-scheme,
\begin{align}
S^{(2,\rm ct)}=-\frac{\alpha_s^2 \beta_1}{(4\pi)^2} \frac{1}{\epsilon} H^{(0)}_{s}-\frac{\alpha_s \beta_0}{2\pi} \frac{1}{\epsilon} \bigl(S^{(10)}+S^{(01)}\bigr)
\, ,\end{align}
and the cross term between the one-loop counterterm and one-loop contributions to the bare soft matrix element yields
\begin{align}
& S^{(1,1)}+ \bigl(S^{(1,0)}+S^{(0,1)}\bigr)\bigl(Z_{3,s}^{(1)}-Z_{2}^{(1)}\bigr)+S^{(2,\rm ct)}\Big\vert_{\rm finite}
\\ & \quad
=  H^{(0)}_{s} \, \frac{\alpha_s^2 C_A}{16 \pi^2} \biggl\{C_A\biggl[-\frac{1}{6}L^4+ \frac{17\pi^2}{6}L^2 -
 \frac{28}{3}\zeta_3 L+ \frac{11 \pi^4}{180}\biggr]+\beta_0\biggl[-\frac{1}{3}L^3+\frac{5\pi^2}{6}L -\frac{14}{3} \zeta_3\biggr]\biggr\}  \nn 
\, .\end{align}
Summing all contributions gives the final two-loop result,
\begin{align}\label{eq:H3_twoloop}
H_{s}^{(2)} \biggl(\frac{t\, u}{Q^2},\mu\biggr)
&= H^{(0)}_{s} \, \frac{\alpha_s^2 C_A}{16 \pi^2}\biggl\{C_A\biggl[\frac{1}{2}L^4 +\frac{11}{9}L^3-\biggl(\frac{67}{9}+\frac{\pi^2}{2} \biggr)L^2
\\ & \quad
+ \biggl(\frac{386}{27}-\frac{11\pi^2}{3}-2\zeta_3\biggr)L -\frac{1142}{81} + \frac{737 \pi^2}{108}+\frac{187}{9} \zeta_3+\frac{29\pi^4}{360} \biggr]
\nn \\ & \quad
+ T_F n_f \biggl[-\frac{4}{9}L^3+\frac{20}{9}L^2-\biggl(\frac{76}{27}-\frac{4\pi^2}{3}\biggr)L+\frac{130}{81}-\frac{55\pi^2}{27}-\frac{68}{9}\zeta_3\biggr]\biggr\}
\, . \nn\end{align}
As a cross check, we verified that this result agrees with the anomalous dimension in \eq{Cs_anom} using the expressions in \eq{anom_dim}.

\phantomsection
\addcontentsline{toc}{section}{References}
\bibliographystyle{../jhep}
\bibliography{../scetplus}

\providecommand{\href}[2]{#2}\begingroup\raggedright\begin{thebibliography}{10}

\bibitem{Bauer:2000ew}
C.~W. Bauer, S.~Fleming, and M.~E. Luke, {\it {Summing Sudakov logarithms in
  $B\to X_s \gamma$ in effective field theory}},  {\em Phys.~Rev.} {\bf D63}
  (2000) 014006, [\href{http://arXiv.org/abs/hep-ph/0005275}{{\tt
  hep-ph/0005275}}].

\bibitem{Bauer:2000yr}
C.~W. Bauer, S.~Fleming, D.~Pirjol, and I.~W. Stewart, {\it {An Effective field
  theory for collinear and soft gluons: Heavy to light decays}},  {\em
  Phys.~Rev.} {\bf D63} (2001) 114020,
  [\href{http://arXiv.org/abs/hep-ph/0011336}{{\tt hep-ph/0011336}}].

\bibitem{Bauer:2001ct}
C.~W. Bauer and I.~W. Stewart, {\it {Invariant operators in collinear effective
  theory}},  {\em Phys. Lett.} {\bf B516} (2001) 134--142,
  [\href{http://arXiv.org/abs/hep-ph/0107001}{{\tt hep-ph/0107001}}].

\bibitem{Bauer:2001yt}
C.~W. Bauer, D.~Pirjol, and I.~W. Stewart, {\it {Soft collinear factorization
  in effective field theory}},  {\em Phys.~Rev.} {\bf D65} (2002) 054022,
  [\href{http://arXiv.org/abs/hep-ph/0109045}{{\tt hep-ph/0109045}}].

\bibitem{Bauer:2002nz}
C.~W. Bauer, S.~Fleming, D.~Pirjol, I.~Z. Rothstein, and I.~W. Stewart, {\it
  {Hard scattering factorization from effective field theory}},  {\em
  Phys.~Rev.~D} {\bf 66} (2002) 014017,
  [\href{http://arXiv.org/abs/hep-ph/0202088}{{\tt hep-ph/0202088}}].

\bibitem{Beneke:2002ph}
M.~Beneke, A.~P. Chapovsky, M.~Diehl, and T.~Feldmann, {\it {Soft collinear
  effective theory and heavy to light currents beyond leading power}},  {\em
  Nucl. Phys.} {\bf B643} (2002) 431--476,
  [\href{http://arXiv.org/abs/hep-ph/0206152}{{\tt hep-ph/0206152}}].

\bibitem{Bauer:2011uc}
C.~W. Bauer, F.~J. Tackmann, J.~R. Walsh, and S.~Zuberi, {\it {Factorization
  and Resummation for Dijet Invariant Mass Spectra}},  {\em Phys. Rev.} {\bf
  D85} (2012) 074006, [\href{http://arXiv.org/abs/1106.6047}{{\tt
  arXiv:1106.6047}}].

\bibitem{Procura:2014cba}
M.~Procura, W.~J. Waalewijn, and L.~Zeune, {\it {Resummation of
  Double-Differential Cross Sections and Fully-Unintegrated Parton Distribution
  Functions}},  {\em JHEP} {\bf 1502} (2015) 117,
  [\href{http://arXiv.org/abs/1410.6483}{{\tt arXiv:1410.6483}}].

\bibitem{Larkoski:2015zka}
A.~J. Larkoski, I.~Moult, and D.~Neill, {\it {Non-Global Logarithms,
  Factorization, and the Soft Substructure of Jets}},  {\em JHEP} {\bf 09}
  (2015) 143, [\href{http://arXiv.org/abs/1501.04596}{{\tt arXiv:1501.04596}}].

\bibitem{Neill:2015nya}
D.~Neill, {\it {The Edge of Jets and Subleading Non-Global Logs}},
  \href{http://arXiv.org/abs/1508.07568}{{\tt arXiv:1508.07568}}.

\bibitem{Larkoski:2015kga}
A.~J. Larkoski, I.~Moult, and D.~Neill, {\it {Analytic Boosted Boson
  Discrimination}},  \href{http://arXiv.org/abs/1507.03018}{{\tt
  arXiv:1507.03018}}.

\bibitem{Larkoski:2014gra}
A.~J. Larkoski, I.~Moult, and D.~Neill, {\it {Power Counting to Better Jet
  Observables}},  {\em JHEP} {\bf 12} (2014) 009,
  [\href{http://arXiv.org/abs/1409.6298}{{\tt arXiv:1409.6298}}].

\bibitem{Becher:2015hka}
T.~Becher, M.~Neubert, L.~Rothen, and D.~Y. Shao, {\it {An Effective Field
  Theory for Jet Processes}},  \href{http://arXiv.org/abs/1508.06645}{{\tt
  arXiv:1508.06645}}.

\bibitem{Chien:2015cka}
Y.-T. Chien, A.~Hornig, and C.~Lee, {\it {A Soft-Collinear Mode for Jet Cross
  Sections in Soft Collinear Effective Theory}},
  \href{http://arXiv.org/abs/1509.04287}{{\tt arXiv:1509.04287}}.

\bibitem{Seymour:1997kj}
M.~H. Seymour, {\it {Jet shapes in hadron collisions: Higher orders,
  resummation and hadronization}},  {\em Nucl. Phys.} {\bf B513} (1998)
  269--300, [\href{http://arXiv.org/abs/hep-ph/9707338}{{\tt hep-ph/9707338}}].

\bibitem{Cheung:2009sg}
W.~M.-Y. Cheung, M.~Luke, and S.~Zuberi, {\it {Phase Space and Jet Definitions
  in SCET}},  {\em Phys.~Rev.~D} {\bf 80} (2009) 114021,
  [\href{http://arXiv.org/abs/0910.2479}{{\tt arXiv:0910.2479}}].

\bibitem{Ellis:2010rwa}
S.~D. Ellis, C.~K. Vermilion, J.~R. Walsh, A.~Hornig, and C.~Lee, {\it {Jet
  Shapes and Jet Algorithms in SCET}},  {\em JHEP} {\bf 1011} (2010) 101,
  [\href{http://arXiv.org/abs/1001.0014}{{\tt arXiv:1001.0014}}].

\bibitem{Banfi:2010pa}
A.~Banfi, M.~Dasgupta, K.~Khelifa-Kerfa, and S.~Marzani, {\it {Non-global
  logarithms and jet algorithms in high-pT jet shapes}},  {\em JHEP} {\bf 08}
  (2010) 064, [\href{http://arXiv.org/abs/1004.3483}{{\tt arXiv:1004.3483}}].

\bibitem{Kelley:2012kj}
R.~Kelley, J.~R. Walsh, and S.~Zuberi, {\it {Abelian Non-Global Logarithms from
  Soft Gluon Clustering}},  {\em JHEP} {\bf 09} (2012) 117,
  [\href{http://arXiv.org/abs/1202.2361}{{\tt arXiv:1202.2361}}].

\bibitem{vonManteuffel:2013vja}
A.~von Manteuffel, R.~M. Schabinger, and H.~X. Zhu, {\it {The Complete Two-Loop
  Integrated Jet Thrust Distribution In Soft-Collinear Effective Theory}},
  {\em JHEP} {\bf 03} (2014) 139, [\href{http://arXiv.org/abs/1309.3560}{{\tt
  arXiv:1309.3560}}].

\bibitem{Stewart:2010tn}
I.~W. Stewart, F.~J. Tackmann, and W.~J. Waalewijn, {\it {N-Jettiness: An
  Inclusive Event Shape to Veto Jets}},  {\em Phys. Rev. Lett.} {\bf 105}
  (2010) 092002, [\href{http://arXiv.org/abs/1004.2489}{{\tt
  arXiv:1004.2489}}].

\bibitem{Stewart:2015waa}
I.~W. Stewart, F.~J. Tackmann, J.~Thaler, C.~K. Vermilion, and T.~F. Wilkason,
  {\it {XCone: N-jettiness as an Exclusive Cone Jet Algorithm}},  {\em JHEP}
  {\bf 11} (2015) 072, [\href{http://arXiv.org/abs/1508.01516}{{\tt
  arXiv:1508.01516}}].

\bibitem{Thaler:2015xaa}
J.~Thaler and T.~F. Wilkason, {\it {Resolving Boosted Jets with XCone}},
  \href{http://arXiv.org/abs/1508.01518}{{\tt arXiv:1508.01518}}.

\bibitem{Catani:1992tm}
S.~Catani, B.~R. Webber, Y.~L. Dokshitzer, and F.~Fiorani, {\it {Average
  multiplicities in two and three jet e+ e- annihilation events}},  {\em Nucl.
  Phys.} {\bf B383} (1992) 419--441.

\bibitem{Fleming:2007qr}
S.~Fleming, A.~H. Hoang, S.~Mantry, and I.~W. Stewart, {\it {Jets from massive
  unstable particles: Top-mass determination}},  {\em Phys.~Rev.} {\bf D77}
  (2008) 074010, [\href{http://arXiv.org/abs/hep-ph/0703207}{{\tt
  hep-ph/0703207}}].

\bibitem{Bauer:2008dt}
C.~W. Bauer, S.~P. Fleming, C.~Lee, and G.~F. Sterman, {\it {Factorization of
  $e^+ e^-$ Event Shape Distributions with Hadronic Final States in Soft
  Collinear Effective Theory}},  {\em Phys.~Rev.} {\bf D78} (2008) 034027,
  [\href{http://arXiv.org/abs/0801.4569}{{\tt arXiv:0801.4569}}].

\bibitem{Bauer:2006mk}
C.~W. Bauer and M.~D. Schwartz, {\it {Event Generation from Effective Field
  Theory}},  {\em Phys. Rev.} {\bf D76} (2007) 074004,
  [\href{http://arXiv.org/abs/hep-ph/0607296}{{\tt hep-ph/0607296}}].

\bibitem{Baumgart:2010qf}
M.~Baumgart, C.~Marcantonini, and I.~W. Stewart, {\it {Systematic Improvement
  of Parton Showers with Effective Theory}},  {\em Phys.Rev.} {\bf D83} (2011)
  034011, [\href{http://arXiv.org/abs/1007.0758}{{\tt arXiv:1007.0758}}].

\bibitem{Larkoski:2015uaa}
A.~J. Larkoski and I.~Moult, {\it {The Singular Behavior of Jet Substructure
  Observables}},  \href{http://arXiv.org/abs/1510.08459}{{\tt
  arXiv:1510.08459}}.

\bibitem{Manohar:2006nz}
A.~V. Manohar and I.~W. Stewart, {\it {The Zero-Bin and Mode Factorization in
  Quantum Field Theory}},  {\em Phys. Rev.} {\bf D76} (2007) 074002,
  [\href{http://arXiv.org/abs/hep-ph/0605001}{{\tt hep-ph/0605001}}].

\bibitem{Ligeti:2008ac}
Z.~Ligeti, I.~W. Stewart, and F.~J. Tackmann, {\it {Treating the b quark
  distribution function with reliable uncertainties}},  {\em Phys.~Rev.~D} {\bf
  78} (2008) 114014, [\href{http://arXiv.org/abs/0807.1926}{{\tt
  arXiv:0807.1926}}].

\bibitem{Abbate:2010xh}
R.~Abbate, M.~Fickinger, A.~H. Hoang, V.~Mateu, and I.~W. Stewart, {\it {Thrust
  at N$^3$LL with Power Corrections and a Precision Global Fit for
  $\alpha_s(m_Z)$}},  {\em Phys.~Rev.~D} {\bf 83} (2011) 074021,
  [\href{http://arXiv.org/abs/1006.3080}{{\tt arXiv:1006.3080}}].

\bibitem{Thaler:2010tr}
J.~Thaler and K.~Van~Tilburg, {\it {Identifying Boosted Objects with
  N-subjettiness}},  {\em JHEP} {\bf 1103} (2011) 015,
  [\href{http://arXiv.org/abs/1011.2268}{{\tt arXiv:1011.2268}}].

\bibitem{Thaler:2011gf}
J.~Thaler and K.~Van~Tilburg, {\it {Maximizing Boosted Top Identification by
  Minimizing N-subjettiness}},  {\em JHEP} {\bf 02} (2012) 093,
  [\href{http://arXiv.org/abs/1108.2701}{{\tt arXiv:1108.2701}}].

\bibitem{Larkoski:2014uqa}
A.~J. Larkoski, D.~Neill, and J.~Thaler, {\it {Jet Shapes with the Broadening
  Axis}},  {\em JHEP} {\bf 1404} (2014) 017,
  [\href{http://arXiv.org/abs/1401.2158}{{\tt arXiv:1401.2158}}].

\bibitem{Chiu:2011qc}
J.-y. Chiu, A.~Jain, D.~Neill, and I.~Z. Rothstein, {\it {The Rapidity
  Renormalization Group}},  {\em Phys. Rev. Lett.} {\bf 108} (2012) 151601,
  [\href{http://arXiv.org/abs/1104.0881}{{\tt arXiv:1104.0881}}].

\bibitem{Chiu:2012ir}
J.-Y. Chiu, A.~Jain, D.~Neill, and I.~Z. Rothstein, {\it {A Formalism for the
  Systematic Treatment of Rapidity Logarithms in Quantum Field Theory}},  {\em
  JHEP} {\bf 1205} (2012) 084, [\href{http://arXiv.org/abs/1202.0814}{{\tt
  arXiv:1202.0814}}].

\bibitem{Stewart:2010qs}
I.~W. Stewart, F.~J. Tackmann, and W.~J. Waalewijn, {\it {The Quark Beam
  Function at NNLL}},  {\em JHEP} {\bf 09} (2010) 005,
  [\href{http://arXiv.org/abs/1002.2213}{{\tt arXiv:1002.2213}}].

\bibitem{Berger:2010xi}
C.~F. Berger, C.~Marcantonini, I.~W. Stewart, F.~J. Tackmann, and W.~J.
  Waalewijn, {\it {Higgs Production with a Central Jet Veto at NNLL+NNLO}},
  {\em JHEP} {\bf 04} (2011) 092, [\href{http://arXiv.org/abs/1012.4480}{{\tt
  arXiv:1012.4480}}].

\bibitem{Stewart:2012yh}
I.~W. Stewart, F.~J. Tackmann, and W.~J. Waalewijn, {\it {Combining Fixed-Order
  Helicity Amplitudes With Resummation Using SCET}},  {\em PoS} {\bf LL2012}
  (2012) 058, [\href{http://arXiv.org/abs/1211.2305}{{\tt arXiv:1211.2305}}].

\bibitem{Moult:2015aoa}
I.~Moult, I.~W. Stewart, F.~J. Tackmann, and W.~J. Waalewijn, {\it {Employing
  Helicity Amplitudes for Resummation}},
  \href{http://arXiv.org/abs/1508.02397}{{\tt arXiv:1508.02397}}.

\bibitem{Kolodrubetz:2016uim}
D.~W. Kolodrubetz, I.~Moult, and I.~W. Stewart, {\it {Building Blocks for
  Subleading Helicity Operators}},  \href{http://arXiv.org/abs/1601.02607}{{\tt
  arXiv:1601.02607}}.

\bibitem{Ellis:1980wv}
R.~K. Ellis, D.~Ross, and A.~Terrano, {\it {The Perturbative Calculation of Jet
  Structure in $e^+ e^-$ Annihilation}},  {\em Nucl. Phys.} {\bf B178} (1981)
  421.

\bibitem{Bauer:2003pi}
C.~W. Bauer and A.~V. Manohar, {\it {Shape function effects in $B \to X_s \ga$
  and $B \to X_u \ell \bar \nu$ decays}},  {\em Phys. Rev.} {\bf D70} (2004)
  034024, [\href{http://arXiv.org/abs/hep-ph/0312109}{{\tt hep-ph/0312109}}].

\bibitem{Fleming:2003gt}
S.~Fleming, A.~K. Leibovich, and T.~Mehen, {\it {Resumming the color octet
  contribution to $e^{+} e^{-} \to J/\psi$ + $X$}},  {\em Phys. Rev.} {\bf D68}
  (2003) 094011, [\href{http://arXiv.org/abs/hep-ph/0306139}{{\tt
  hep-ph/0306139}}].

\bibitem{Becher:2006qw}
T.~Becher and M.~Neubert, {\it {Toward a NNLO calculation of the $\bar B \to
  X_s \gamma$ decay rate with a cut on photon energy. II. Two-loop result for
  the jet function}},  {\em Phys. Lett.} {\bf B637} (2006) 251--259,
  [\href{http://arXiv.org/abs/hep-ph/0603140}{{\tt hep-ph/0603140}}].

\bibitem{Becher:2009th}
T.~Becher and M.~D. Schwartz, {\it {Direct photon production with effective
  field theory}},  {\em JHEP} {\bf 1002} (2010) 040,
  [\href{http://arXiv.org/abs/0911.0681}{{\tt arXiv:0911.0681}}].

\bibitem{Becher:2010pd}
T.~Becher and G.~Bell, {\it {The gluon jet function at two-loop order}},  {\em
  Phys. Lett.} {\bf B695} (2011) 252--258,
  [\href{http://arXiv.org/abs/1008.1936}{{\tt arXiv:1008.1936}}].

\bibitem{Jouttenus:2011wh}
T.~T. Jouttenus, I.~W. Stewart, F.~J. Tackmann, and W.~J. Waalewijn, {\it {The
  Soft Function for Exclusive N-Jet Production at Hadron Colliders}},  {\em
  Phys. Rev.} {\bf D83} (2011) 114030,
  [\href{http://arXiv.org/abs/1102.4344}{{\tt arXiv:1102.4344}}].

\bibitem{Boughezal:2015eha}
R.~Boughezal, X.~Liu, and F.~Petriello, {\it {$N$-jettiness soft function at
  next-to-next-to-leading order}},  {\em Phys. Rev.} {\bf D91} (2015), no.~9
  094035, [\href{http://arXiv.org/abs/1504.02540}{{\tt arXiv:1504.02540}}].

\bibitem{Almelid:2015jia}
O.~Almelid, C.~Duhr, and E.~Gardi, {\it {Three-loop corrections to the soft
  anomalous dimension in multi-leg scattering}},
  \href{http://arXiv.org/abs/1507.00047}{{\tt arXiv:1507.00047}}.

\bibitem{Manohar:2002fd}
A.~V. Manohar, T.~Mehen, D.~Pirjol, and I.~W. Stewart, {\it {Reparameterization
  invariance for collinear operators}},  {\em Phys. Lett.} {\bf B539} (2002)
  59--66, [\href{http://arXiv.org/abs/hep-ph/0204229}{{\tt hep-ph/0204229}}].

\bibitem{Manohar:2003vb}
A.~V. Manohar, {\it {Deep inelastic scattering as $x\to1$ using soft collinear
  effective theory}},  {\em Phys. Rev.} {\bf D68} (2003) 114019,
  [\href{http://arXiv.org/abs/hep-ph/0309176}{{\tt hep-ph/0309176}}].

\bibitem{Bauer:2003di}
C.~W. Bauer, C.~Lee, A.~V. Manohar, and M.~B. Wise, {\it {Enhanced
  nonperturbative effects in Z decays to hadrons}},  {\em Phys. Rev.} {\bf D70}
  (2004) 034014, [\href{http://arXiv.org/abs/hep-ph/0309278}{{\tt
  hep-ph/0309278}}].

\bibitem{Berends:1987me}
F.~A. Berends and W.~Giele, {\it {Recursive Calculations for Processes with n
  Gluons}},  {\em Nucl. Phys.} {\bf B306} (1988) 759.

\bibitem{Mangano:1990by}
M.~L. Mangano and S.~J. Parke, {\it {Multiparton amplitudes in gauge
  theories}},  {\em Phys. Rept.} {\bf 200} (1991) 301--367,
  [\href{http://arXiv.org/abs/hep-th/0509223}{{\tt hep-th/0509223}}].

\bibitem{Bern:1994zx}
Z.~Bern, L.~J. Dixon, D.~C. Dunbar, and D.~A. Kosower, {\it {One loop n point
  gauge theory amplitudes, unitarity and collinear limits}},  {\em Nucl. Phys.}
  {\bf B425} (1994) 217--260, [\href{http://arXiv.org/abs/hep-ph/9403226}{{\tt
  hep-ph/9403226}}].

\bibitem{Kosower:1999xi}
D.~A. Kosower, {\it {All order collinear behavior in gauge theories}},  {\em
  Nucl. Phys.} {\bf B552} (1999) 319--336,
  [\href{http://arXiv.org/abs/hep-ph/9901201}{{\tt hep-ph/9901201}}].

\bibitem{Kosower:1999rx}
D.~A. Kosower and P.~Uwer, {\it {One loop splitting amplitudes in gauge
  theory}},  {\em Nucl. Phys.} {\bf B563} (1999) 477--505,
  [\href{http://arXiv.org/abs/hep-ph/9903515}{{\tt hep-ph/9903515}}].

\bibitem{Bern:1999ry}
Z.~Bern, V.~Del~Duca, W.~B. Kilgore, and C.~R. Schmidt, {\it {The infrared
  behavior of one loop QCD amplitudes at next-to-next-to leading order}},  {\em
  Phys. Rev.} {\bf D60} (1999) 116001,
  [\href{http://arXiv.org/abs/hep-ph/9903516}{{\tt hep-ph/9903516}}].

\bibitem{Sborlini:2013jba}
G.~F. Sborlini, D.~de~Florian, and G.~Rodrigo, {\it {Double collinear splitting
  amplitudes at next-to-leading order}},  {\em JHEP} {\bf 1401} (2014) 018,
  [\href{http://arXiv.org/abs/1310.6841}{{\tt arXiv:1310.6841}}].

\bibitem{Bern:1998sc}
Z.~Bern, V.~Del~Duca, and C.~R. Schmidt, {\it {The Infrared behavior of one
  loop gluon amplitudes at next-to-next-to-leading order}},  {\em Phys.Lett.}
  {\bf B445} (1998) 168--177, [\href{http://arXiv.org/abs/hep-ph/9810409}{{\tt
  hep-ph/9810409}}].

\bibitem{Catani:2000pi}
S.~Catani and M.~Grazzini, {\it {The soft gluon current at one loop order}},
  {\em Nucl. Phys.} {\bf B591} (2000) 435--454,
  [\href{http://arXiv.org/abs/hep-ph/0007142}{{\tt hep-ph/0007142}}].

\bibitem{Duhr:2013msa}
C.~Duhr and T.~Gehrmann, {\it {The two-loop soft current in dimensional
  regularization}},  {\em Phys. Lett.} {\bf B727} (2013) 452--455,
  [\href{http://arXiv.org/abs/1309.4393}{{\tt arXiv:1309.4393}}].

\bibitem{Li:2013lsa}
Y.~Li and H.~X. Zhu, {\it {Single soft gluon emission at two loops}},  {\em
  JHEP} {\bf 1311} (2013) 080, [\href{http://arXiv.org/abs/1309.4391}{{\tt
  arXiv:1309.4391}}].

\bibitem{Chiu:2008vv}
J.-y. Chiu, R.~Kelley, and A.~V. Manohar, {\it {Electroweak Corrections using
  Effective Field Theory: Applications to the LHC}},  {\em Phys. Rev.} {\bf
  D78} (2008) 073006, [\href{http://arXiv.org/abs/0806.1240}{{\tt
  arXiv:0806.1240}}].

\bibitem{Gardi:2009qi}
E.~Gardi and L.~Magnea, {\it {Factorization constraints for soft anomalous
  dimensions in QCD scattering amplitudes}},  {\em JHEP} {\bf 03} (2009) 079,
  [\href{http://arXiv.org/abs/0901.1091}{{\tt arXiv:0901.1091}}].

\bibitem{Korchemsky:1987wg}
G.~Korchemsky and A.~Radyushkin, {\it {Renormalization of the Wilson Loops
  Beyond the Leading Order}},  {\em Nucl. Phys.} {\bf B283} (1987) 342--364.

\bibitem{Aybat:2006mz}
S.~M. Aybat, L.~J. Dixon, and G.~F. Sterman, {\it {The Two-loop soft anomalous
  dimension matrix and resummation at next-to-next-to leading pole}},  {\em
  Phys.~Rev.~D} {\bf 74} (2006) 074004,
  [\href{http://arXiv.org/abs/hep-ph/0607309}{{\tt hep-ph/0607309}}].

\bibitem{Stewart:2009yx}
I.~W. Stewart, F.~J. Tackmann, and W.~J. Waalewijn, {\it {Factorization at the
  LHC: From PDFs to Initial State Jets}},  {\em Phys. Rev.} {\bf D81} (2010)
  094035, [\href{http://arXiv.org/abs/0910.0467}{{\tt arXiv:0910.0467}}].

\bibitem{Gaunt:2014ska}
J.~R. Gaunt, {\it {Glauber Gluons and Multiple Parton Interactions}},  {\em
  JHEP} {\bf 1407} (2014) 110, [\href{http://arXiv.org/abs/1405.2080}{{\tt
  arXiv:1405.2080}}].

\bibitem{Zeng:2015iba}
M.~Zeng, {\it {Drell-Yan process with jet vetoes: breaking of generalized
  factorization}},  {\em JHEP} {\bf 10} (2015) 189,
  [\href{http://arXiv.org/abs/1507.01652}{{\tt arXiv:1507.01652}}].

\bibitem{Rothstein:2016bsq}
I.~Z. Rothstein and I.~W. Stewart, {\it {An Effective Field Theory for Forward
  Scattering and Factorization Violation}},
  \href{http://arXiv.org/abs/1601.04695}{{\tt arXiv:1601.04695}}.

\bibitem{Jouttenus:2013hs}
T.~T. Jouttenus, I.~W. Stewart, F.~J. Tackmann, and W.~J. Waalewijn, {\it {Jet
  mass spectra in Higgs boson plus one jet at next-to-next-to-leading
  logarithmic order}},  {\em Phys.Rev.} {\bf D88} (2013), no.~5 054031,
  [\href{http://arXiv.org/abs/1302.0846}{{\tt arXiv:1302.0846}}].

\bibitem{Andersen:2009nu}
J.~R. Andersen and J.~M. Smillie, {\it {Constructing All-Order Corrections to
  Multi-Jet Rates}},  {\em JHEP} {\bf 01} (2010) 039,
  [\href{http://arXiv.org/abs/0908.2786}{{\tt arXiv:0908.2786}}].

\bibitem{Giele:1993dj}
W.~T. Giele, E.~W.~N. Glover, and D.~A. Kosower, {\it {Higher order corrections
  to jet cross-sections in hadron colliders}},  {\em Nucl. Phys.} {\bf B403}
  (1993) 633--670, [\href{http://arXiv.org/abs/hep-ph/9302225}{{\tt
  hep-ph/9302225}}].

\bibitem{Dixon:1996wi}
L.~J. Dixon, {\it {Calculating scattering amplitudes efficiently}},
  \href{http://arXiv.org/abs/hep-ph/9601359}{{\tt hep-ph/9601359}}.

\bibitem{Dixon:2013uaa}
L.~J. Dixon, {\it {A brief introduction to modern amplitude methods}},
  \href{http://arXiv.org/abs/1310.5353}{{\tt arXiv:1310.5353}}.

\bibitem{Arnesen:2005nk}
C.~M. Arnesen, J.~Kundu, and I.~W. Stewart, {\it {Constraint equations for
  heavy-to-light currents in SCET}},  {\em Phys. Rev.} {\bf D72} (2005) 114002,
  [\href{http://arXiv.org/abs/hep-ph/0508214}{{\tt hep-ph/0508214}}].

\bibitem{Kang:2015moa}
D.~Kang, O.~Z. Labun, and C.~Lee, {\it {Equality of hemisphere soft functions
  for $e^+e^-$, DIS and $pp$ collisions at $\mathcal{O}(\alpha_s^2)$}},  {\em
  Phys. Lett.} {\bf B748} (2015) 45--54,
  [\href{http://arXiv.org/abs/1504.04006}{{\tt arXiv:1504.04006}}].

\bibitem{Becher:2009qa}
T.~Becher and M.~Neubert, {\it {On the Structure of Infrared Singularities of
  Gauge-Theory Amplitudes}},  {\em JHEP} {\bf 06} (2009) 081,
  [\href{http://arXiv.org/abs/0903.1126}{{\tt arXiv:0903.1126}}]. [Erratum:
  JHEP11,024(2013)].

\bibitem{Campbell:1997hg}
J.~M. Campbell and E.~N. Glover, {\it {Double unresolved approximations to
  multiparton scattering amplitudes}},  {\em Nucl. Phys.} {\bf B527} (1998)
  264--288, [\href{http://arXiv.org/abs/hep-ph/9710255}{{\tt hep-ph/9710255}}].

\bibitem{Catani:1999ss}
S.~Catani and M.~Grazzini, {\it {Infrared factorization of tree level QCD
  amplitudes at the next-to-next-to-leading order and beyond}},  {\em Nucl.
  Phys.} {\bf B570} (2000) 287--325,
  [\href{http://arXiv.org/abs/hep-ph/9908523}{{\tt hep-ph/9908523}}].

\bibitem{DelDuca:1999ha}
V.~Del~Duca, A.~Frizzo, and F.~Maltoni, {\it {Factorization of tree QCD
  amplitudes in the high-energy limit and in the collinear limit}},  {\em Nucl.
  Phys.} {\bf B568} (2000) 211--262,
  [\href{http://arXiv.org/abs/hep-ph/9909464}{{\tt hep-ph/9909464}}].

\bibitem{Kosower:2003bh}
D.~A. Kosower, {\it {Antenna factorization in strongly ordered limits}},  {\em
  Phys. Rev.} {\bf D71} (2005) 045016,
  [\href{http://arXiv.org/abs/hep-ph/0311272}{{\tt hep-ph/0311272}}].

\bibitem{Somogyi:2005xz}
G.~Somogyi, Z.~Trocsanyi, and V.~Del~Duca, {\it {Matching of singly- and
  doubly-unresolved limits of tree-level QCD squared matrix elements}},  {\em
  JHEP} {\bf 0506} (2005) 024, [\href{http://arXiv.org/abs/hep-ph/0502226}{{\tt
  hep-ph/0502226}}].

\bibitem{Catani:1998nv}
S.~Catani and M.~Grazzini, {\it {Collinear factorization and splitting
  functions for next-to-next-to-leading order QCD calculations}},  {\em
  Phys.Lett.} {\bf B446} (1999) 143--152,
  [\href{http://arXiv.org/abs/hep-ph/9810389}{{\tt hep-ph/9810389}}].

\bibitem{Alioli:2012fc}
S.~Alioli, C.~W. Bauer, C.~J. Berggren, A.~Hornig, F.~J. Tackmann, C.~K.
  Vermilion, J.~R. Walsh, and S.~Zuberi, {\it {Combining Higher-Order
  Resummation with Multiple NLO Calculations and Parton Showers in GENEVA}},
  {\em JHEP} {\bf 09} (2013) 120, [\href{http://arXiv.org/abs/1211.7049}{{\tt
  arXiv:1211.7049}}].

\bibitem{Alioli:2013hqa}
S.~Alioli, C.~W. Bauer, C.~Berggren, F.~J. Tackmann, J.~R. Walsh, and
  S.~Zuberi, {\it {Matching Fully Differential NNLO Calculations and Parton
  Showers}},  {\em JHEP} {\bf 06} (2014) 089,
  [\href{http://arXiv.org/abs/1311.0286}{{\tt arXiv:1311.0286}}].

\bibitem{Alioli:2015toa}
S.~Alioli, C.~W. Bauer, C.~Berggren, F.~J. Tackmann, and J.~R. Walsh, {\it
  {Drell-Yan production at NNLL$'+$NNLO matched to parton showers}},  {\em
  Phys.~Rev.~D} {\bf 92} (2015), no.~9 094020,
  [\href{http://arXiv.org/abs/1508.01475}{{\tt arXiv:1508.01475}}].

\bibitem{Hamilton:2012np}
K.~Hamilton, P.~Nason, and G.~Zanderighi, {\it {MINLO: Multi-Scale Improved
  NLO}},  {\em JHEP} {\bf 10} (2012) 155,
  [\href{http://arXiv.org/abs/1206.3572}{{\tt arXiv:1206.3572}}].

\bibitem{Hamilton:2012rf}
K.~Hamilton, P.~Nason, C.~Oleari, and G.~Zanderighi, {\it {Merging H/W/Z + 0
  and 1 jet at NLO with no merging scale: a path to parton shower + NNLO
  matching}},  {\em JHEP} {\bf 05} (2013) 082,
  [\href{http://arXiv.org/abs/1212.4504}{{\tt arXiv:1212.4504}}].

\bibitem{Frederix:2015fyz}
R.~Frederix and K.~Hamilton, {\it {Extending the MINLO method}},
  \href{http://arXiv.org/abs/1512.02663}{{\tt arXiv:1512.02663}}.

\bibitem{Caswell:1974gg}
W.~E. Caswell, {\it {Asymptotic Behavior of Nonabelian Gauge Theories to Two
  Loop Order}},  {\em Phys. Rev. Lett.} {\bf 33} (1974) 244.

\bibitem{Jones:1974mm}
D.~R.~T. Jones, {\it {Two Loop Diagrams in Yang-Mills Theory}},  {\em Nucl.
  Phys.} {\bf B75} (1974) 531.

\bibitem{Idilbi:2005ni}
A.~Idilbi, X.-d. Ji, J.-P. Ma, and F.~Yuan, {\it {Threshold resummation for
  Higgs production in effective field theory}},  {\em Phys.~Rev.~D} {\bf 73}
  (2006) 077501, [\href{http://arXiv.org/abs/hep-ph/0509294}{{\tt
  hep-ph/0509294}}].

\bibitem{Idilbi:2006dg}
A.~Idilbi, X.-d. Ji, and F.~Yuan, {\it {Resummation of threshold logarithms in
  effective field theory for DIS, Drell-Yan and Higgs production}},  {\em Nucl.
  Phys.} {\bf B753} (2006) 42--68,
  [\href{http://arXiv.org/abs/hep-ph/0605068}{{\tt hep-ph/0605068}}].

\bibitem{Hooft:1972fi}
G.~'t~Hooft and M.~Veltman, {\it {Regularization and Renormalization of Gauge
  Fields}},  {\em Nucl. Phys.} {\bf B44} (1972) 189--213.

\bibitem{Badger:2004uk}
S.~Badger and E.~N. Glover, {\it {Two loop splitting functions in QCD}},  {\em
  JHEP} {\bf 0407} (2004) 040, [\href{http://arXiv.org/abs/hep-ph/0405236}{{\tt
  hep-ph/0405236}}].

\end{thebibliography}\endgroup

\end{document}